\documentclass[aps,twocolumn,prb,showpacs,floatfix,superscriptaddress]{revtex4}
\usepackage{amsfonts}
\usepackage{graphicx}
\usepackage{amsmath}
\usepackage{amssymb}

\input{tcilatex}

\begin{document}

\title{Ground-state of fractional and integral quantum Hall systems at $\nu
\leq 1$ and it excitations}
\author{O. G. Balev}
\email[Electronic address:]{ogbalev@ufam.edu.br}
\affiliation{Departamento de F\'{\i}sica, Universidade Federal do Amazonas, 69077-000,
Manaus, Amazonas, Brazil}
\affiliation{Institute of Semiconductor Physics, NAS of Ukraine, Kiev 03650, Ukraine }

\begin{abstract}
Many-body variational ground-state wave function of two-dimensional
electron system (2DES), localized in the main strip (MS)
$L_{x}^{\square} \times L_{y}$ of
the finite width $L_{x}^{\square}=\sqrt{2 \pi m} \ell_{0}$ (and the
periodic boundary condition (PBC) imposed along $x-$direction), is presented at
the fractional and the integral filling factors $\nu=1/m$ for two different
ion backgrounds, giving homogeneous ion density: microscopical uniform ion
background (UIB) and classical ion jellium background (IJB); $\ell_{0}$ is
the magnetic length, $m=2\ell+1$ and $\ell=0, \; 1, \; 2,\ldots$.
It is shown that the ground-state and the lowest excited-state can
correspond to partial crystal-like correlation order among $N$ electrons
of the main region (MR) $L_{x} \times L_{y}$; then the study of
2DES of $N$ electrons within MR is exactly reduced to the treatment
of 2DES of $\tilde{N}=N L_{x}^{\square}/L_{x}$ electrons localized
within MS, with PBC along $x$.
Both for
UIB and IJB, the ground-state manifests the broken symmetry liquid-crystal
state with 2DES density that is periodic along the $y-$ direction, with the
period $L_{x}^{\square}/m$, and independent of $x$.
The
difference between the ground-state energy, per electron, for these two
backgrounds is only due to the difference between the energies of UIB-UIB
and IJB-IJB interactions. For IJB, at $m=3, \;5$, the ground-state has
essentially lower energy per electron than the Laughlin, uniform liquid,
ground-state (the Laughlin model uses IJB); the same holds at $m=1$.
At $m \geq 3$, the compound form of the many-body ground-state wave function
leads to the compound structure of each electron already within MS
(due to important
similarities between the present $m=1$ and $m \geq 3$ states, and to
symplify notations, the term ``compound'' is often used as well at $m=1$):
these compound electrons play important role in the properties of the
many-body excited-states. Obtained compound exciton (without the change of
spin of the excited compound electron) and compound spin-exciton (with the
change of spin of the excited compound electron) states show finite
excitation gaps, for $m=1, \; 3, \; 5, \; 7$. The excited compound electron
(hole) is composed, within MS, from $m$ strongly correlated
quasielectrons (quasiholes) of the total charge $e/m$ ($-e/m$) each; this
charge is fractionally quantized at $m \geq 3$. The activation gap,
experimentally observable from the activation behavior of the direct current
magnetotransport coefficients, is obtained: it is given by the excitation
gap of relevant compound exciton, at $m \geq 3$, and by the gap of pertinent
compound spin-exciton, at $m=1$. Quantized Hall conductance $%
\sigma_{H}=e^{2}/(2 m \pi \hbar)$ is obtained;
it is fractional at $m \geq 3$.
The theory is in good agreement with experiments.
\end{abstract}

\date{December 8, 2007}
\maketitle



\section{INTRODUCTION}

The discoveries of the integer \cite{klitzing1980} and the fractional \cite%
{tsui1982} quantum Hall effects in two-dimensional electron systems (2DES)
of a semiconductor based samples had born strong ongoing interest to the
ground-state of 2DES, and it elementary excitations, in a quantum Hall
regime at a filling factor $\nu \leq 1$ \cite%
{laughlin1983,prang1990,stone1992,chakraborty1995,sarma1997,
yoshioka1983,maki1983,yoshioka83i84,yoshioka84,
levesque1984,haldane1985,morf1986,
macdonald1986,kivelson1986,halperin1986,claro1987,willett1988,
jain1989,dorozhkin1995,mikhailov2001,wexler2004,cabo2004}. Since seminal
work of Laughlin \cite{laughlin1983}, a particular attention in past years
is given to properties of the $\nu=1/3$ and $1/5$ fractional quantum Hall
effect (FQHE) states \cite{prang1990,stone1992,
chakraborty1995,sarma1997,yoshioka1983,maki1983,yoshioka83i84,yoshioka84,
levesque1984,haldane1985,morf1986,macdonald1986,kivelson1986,
halperin1986,claro1987,willett1988,jain1989,dorozhkin1995,mikhailov2001,
wexler2004,cabo2004}. At $\nu=1$, corresponding to fully occupied the lowest
Landau level (LLL), the Laughlin wave function gives the same total energy
per electron as a Hartree-Fock wave function (if the latter is built from
the symmetric gauge the LLL single-particle wave functions it coincides with
the former one, see, e.g., Ref. \onlinecite{mikhailov2001}), \cite%
{laughlin1983,mikhailov2001}
\begin{equation}
\epsilon_{HF}=-\frac{1}{2}\sqrt{\frac{\pi}{2}} \frac{e^{2}}{\varepsilon
\ell_{0}} ,  \notag  \label{1da}
\end{equation}%
where $\ell_{0}=\sqrt{\hbar c/|e|B}$ is the magnetic length and $\varepsilon$
is the background dielectric constant. Point out, for $\nu=1$ the
Hartree-Fock approximation (HFA) study that uses the Landau gauge
single-particle wave functions also gives $\epsilon_{HF}$, %
%
see, e.g., Refs. \onlinecite{yoshioka1983,macdonald1986b}. However, the HFA
result, obtained for $0 <\nu \leq 1$,\cite{yoshioka1983,macdonald1986b} 
\begin{equation}
\epsilon_{HF}(\nu)=-\frac{\nu}{2}\sqrt{\frac{\pi}{2}} \frac{e^{2}}{%
\varepsilon \ell_{0}} ,  \notag  \label{2da}
\end{equation}%
gives for $\nu=1/3$ and $1/5$ the total energy per electron substantially
higher than that of Laughlin model \cite{laughlin1983} and even the energy
of corresponding charge-density wave or Wigner crystal states \cite%
{chakraborty1995,yoshioka1983,maki1983,cabo2004,fertig1997}.

To date understanding is that for $m=3,5$ the Laughlin wave function \cite%
{laughlin1983} gives the best known analytical approximation of exact
many-body ground-state wave function \cite%
{prang1990,stone1992,chakraborty1995,sarma1997,yoshioka1983,
maki1983,yoshioka83i84,yoshioka84,
levesque1984,haldane1985,morf1986,macdonald1986,
kivelson1986,halperin1986,claro1987,willett1988,jain1989,dorozhkin1995,
mikhailov2001,wexler2004,cabo2004}. Point out, there were many attempts
(some of them are quite recent) to find out a ground-state, at $\nu =1/3$,
with an energy lower than in the Laughlin theory,\cite{laughlin1983} see,
e.g., \cite{prang1990,chakraborty1995,kivelson1986,claro1987,
jain1989,mikhailov2001,wexler2004,cabo2004} and references cited therein. In
all pertinent previous theoretical works neutralizing ion background is
treated as a \textquotedblleft classical\textquotedblright\ uniform ion
density, see, e.g., Refs. \cite{laughlin1983,morf1986,mikhailov2001} and 
\cite{bonsall1977,fano1986}, where discrete nature of ions and specific form
of single-ion wave functions does not appear. In addition, in these
theoretical works the 2DES is placed in the 2D-plane of the neutralizing ion
background, e.g., see Refs. \cite%
{laughlin1983,morf1986,mikhailov2001,bonsall1977,fano1986}; we call this
model of neutralizing ion background as classical ion jellium background
(IJB). Besides the IJB model, I treat the other theoretical model of the ion
background (also localized within the plane of 2DES) where I assume, in good
agreement with typical experimental conditions, that electric charge of each
ion is totally localized within the finite square unit cell, $L_{x}^{\square
}\times L_{x}^{\square }$. The latter model also gives exactly homogeneous
ion density; we call this model of the ion background as microscopical
uniform ion background (UIB). The latter model treats the neutralizing ion
background in more correct manner (it excludes, in particular, interaction
of an ion with itself). However, as IJB model typically is used, \cite%
{laughlin1983,morf1986,mikhailov2001,bonsall1977,fano1986} to make
comparison with previous studies we present the results for IJB as well.

At $\nu=1/m$, I present variational many-body wave functions of
ground-states for UIB and IJB as $\Psi _{\tilde{N},\tilde{N}}^{(m),eh}$,
Sec. IV A, and $\Psi _{\tilde{N}}^{(m),JB}$, Sec. IV B, respectively. In
these ground-states the total lowering per electron due to many-body
interactions is given (in units of $e^{2}/\varepsilon \ell_{0}$) by (i)
 $U^{UB}(m)$, Eq. (\ref{92}), and (ii) $U^{JB}(m)$, Eq. (\ref%
{66d}), respectively. In particular, $U^{JB}(m)$ presents, for $m=1,
\; 3, \; 5$, substantially stronger lowering than pertinent total lowering
for the Laughlin variational wave function.\cite{laughlin1983}
In addition, $U^{UB}(m)$ presents, for $m=1, \; 3, \; 5$,
much stronger lowering
than pertinent total lowering $U^{JB}(m)$. For IJB model I obtain
that $U^{JB}(1) \approx -0.66510$, $U^{JB}(3) \approx
-0.42854$, and $U^{JB}(5) \approx -0.33885$ are well below of pertinent
total lowerings for the Laughlin variational wave function that are given as%
\cite{laughlin1983} $\; -\; \sqrt{\pi/8} \approx -0.6267$, $-0.4156 \pm
0.0012$ (notice, more accurate calculations for the Laughlin model show here
the lowering\cite{morf1986} $\;-\;0.410 \pm 0.001$),
and $-0.3340 \pm 0.0028$, respectively.

For treated electron-ion systems many-body effects are essentially related
with $\tilde{N}$ electrons of 2DES localized within the main strip (MS)
$L_{x}^{\square }\times L_{y}$, to which periodic boundary condition (PBC)
along $x-$direction is imposed. Of course, the \textquotedblright
images\textquotedblright\ of MS (cf. with Refs. \cite%
{yoshioka83i84,yoshioka84,chakraborty1995}), periodically repeated with the
period $L_{x}^{\square }$ along $x$-direction of the main region (MR)
$L_{x}\times L_{y}$ ($L_{x,y}\rightarrow \infty $), are taken into account as
well. We assume that within MR there are present $N$ electrons
and $N$ ions such that $N/\tilde{N}=L_{x}/L_{x}^{\square}\rightarrow \infty $%
. It is important that more adequate, physically, sets of single-electron
wave functions are used than previously. These wave functions are localized
mainly within the square unit cell $L_{x}^{\square }\times L_{x}^{\square }$%
, where $L_{x}^{\square }=\sqrt{2\pi \;m}\ell _{0}$; e.g., see Eqs. (\ref{7}%
)-(\ref{16}). This choice helps to reflect the tendency (proven by present
results) of each electron: a) to be present mainly within one such
relatively \textquotedblleft localized\textquotedblright\ (around its centre
point) unit cell and, in addition, b) to occupy all these $\tilde{N}$ unit
cells of MS with equal probability. The latter is achieved by
proper construction of many-body wave functions $\Psi _{\tilde{N},\tilde{N}%
}^{(m),eh}$, $\Psi _{\tilde{N},\tilde{N};(m)}^{i_{0},j_{0};\tilde{n}}$, etc.
Notice, in the present study it is assumed that $\tilde{N}\rightarrow \infty 
$.

It is shown that the ground-state and the lowest excited-state can
correspond to partial crystal-like correlation order,
Eq. (\ref{DA7}), among $N$ electrons
of MR. As a result the study of 2DES of $N$ electrons within MR
is exactly reduced to the treatment of 2DES of
$\tilde{N}$ electrons localized within MS of the finite width (along $x$)
to which PBC is imposed along $x$-direction.
I.e., present study shows that proper PBC
can be totally relevant to symmetry, periodicity, correlations, etc.
properties, e.g., of a sought ground-state. Then it will not lead
to any oversimplification or nonphysical ``boundary effects''.
Present below study of ground-state (e.g., the trial wave function
of ground-state with the energy, per electron, lower than
pertinent energy of Ref. \cite{laughlin1983})
confirms assumption Eq. (\ref{DA7}) and relevant PBC.


Notice that for an infinite MR any specific orientation of $x$- (or
$y$-) axis, within 2D-plane, is not defined until the quantum phase
transition to broken symmetry liquid-crystal state will take place. Point
out, all main formulas of the work are obtained by exact analytical
calculations. In this work I further develop main physical ideas
outlined in Ref. \cite{balev2005}.
All present physical results for the electron-ion system with UIB coincide
with those for the electron-ion system with IJB, except that for IJB the
energies of the ground-state and the excited-states are shifted upwards on
the same value, for given $m$, with respect to the energies of relevant
many-body states for UIB.

The paper is organized as follows. In Sec. II we present many-body
Hamiltonian of the electron-ion system as for UIB, Sec. II A, so for IJB,
Sec. II B. In addition, in Sec. II C we present exact obtaining of
the model Hamiltonians of Secs. II A, II B from first principles and
physical conditions involved.
In Sec. III we introduce complete set of single-body wave
functions. At $\nu=1/m$ and odd integer $m$, in Sec. IV we present two
variational ground-state wave functions: one for UIB, Sec. IV A, and other
for IJB, Sec. IV B. In Sec. V A and Sec. V B we calculate the ground-state
energy of electron-ion system for each of these two ground-state wave
functions. In Sec. V C we give more details and remarks on partial
crystal-like correlation order and energy of ground-state.
In Sec. VI we study excited-states of the ground-states of Secs.
IV A, IV B, and calculate their energies of excitation. In Secs. VI A we
present (both for UIB and IJB) the many-body wave functions both for the
compound exciton and the compound spin-exciton; in addition, we obtain
compound structure for the charge density of these excitons. In Sec. VI B,
at $m \geq 3$, we obtain the energies of the compound excitons and treat
them, both for UIB and IJB. In Sec. VI C, at $m \geq 1$, we obtain the
energies of the compound spin-excitons (both for UIB and IJB) and treat
them. In Sec. VII quantized Hall conductance is calculated for present two
ground-states, at $\nu=1/m$. Finally, we make concluding remarks in Sec.
VIII.

\section{Hamiltonian of the electron-ion system}

\subsection{For a microscopical model of ion background}

First we consider the Hamiltonian of the electron-ion system that will allow
to introduce correctly microscopical ion background, in particular, UIB.

We consider a zero-thickness 2DES localized within the main strip (MS)
of the
finite width $L_{x}^{\square }=\sqrt{2\pi m}\ell _{0}$ ($L_{x}^{\square
}n_{xs}^{\alpha }>x>L_{x}^{\square }(n_{xs}^{\alpha }-1$); where $%
n_{xs}^{\alpha }$ is a finite integer) and of very large (in principle,
infinite) length $L_{y}$ ($L_{y}/2>y>-L_{y}/2$) in the presence of a strong
perpendicular magnetic field $\mathbf{B}=B\hat{\mathbf{z}}$. The Landau
gauge for the vector potential $\mathbf{A}(\mathbf{r})=(-By,0,0)$ is used.
We assume that both $\tilde{N}$ electrons of a 2DES and $\tilde{N}$ ions
(single-charged, with the charge $|e|$) of a neutralizing background are
located, within MS, at the same $(z=0)$-plane (it is $(Z=0)$%
-plane, for ions); i.e., at the same 2D-plane. Further, we assume PBC
only along $x$ direction (cf. with Refs. \cite%
{yoshioka83i84,yoshioka84}, for review see Ref.\cite{chakraborty1995}).
Notice, already there is essential difference from the approach of Refs.
\cite{yoshioka83i84,yoshioka84}, where a rectangular cell is considered and
in both directions PBCs are imposed; however, still
some important analogies between, e.g., the forms of the Coulomb
interaction, properties of matrix elements will be seen. We
assume that the ions are very heavy such that their kinetic energy can be
neglected; similar approximation is widely used in solid-state theory \cite%
{ansel'm1978,madelung1981} as well as in studies of quantum Hall effects,
see, e.g.\cite{laughlin1983,prang1990,mikhailov2001} Then the
many-body Hamiltonian, $\hat{H}_{\tilde{N},\tilde{N}}\equiv \hat{H}_{\tilde{N%
},\tilde{N}}(\mathbf{r}_{1},\ldots ,\mathbf{r}_{\tilde{N}};\mathbf{R}%
_{1},\ldots ,\mathbf{R}_{\tilde{N}})$ of electron-ion system, of $\tilde{N}$
electrons and $\tilde{N}$ ions, is given as %
\begin{equation}
\hat{H}_{\tilde{N},\tilde{N}}=\hat{H}_{0}+V_{ee}+V_{ei}+V_{ii},  \label{1}
\end{equation}%
where the kinetic energy of electrons
\begin{equation}
\hat{H}_{0}=\sum_{i=1}^{\tilde{N}}\hat{h}_{0i}=\frac{1}{2m^{\ast }}%
\sum_{i=1}^{\tilde{N}}[\hat{\mathbf{p}}_{i}-\frac{e}{c}\mathbf{A}(\mathbf{r}%
_{i})]^{2},  \label{2}
\end{equation}%
where $\hat{\mathbf{p}}=-i\hbar \mathbf{\nabla }$ is the in-plane momentum
operator, $m^{\ast }$ the electron effective mass, $\mathbf{r}%
_{i}=(x_{i},y_{i})$ and $\mathbf{R}_{i}=(X_{i},Y_{i})$, $i=1,...,\tilde{N}$,
are the electron and the ion in-plane coordinates. To simplify notations,
here in the ideal single-body Hamiltonian $\hat{h}_{0}$ there is omitted the
Zeeman energy, due to bare g-factor $g_{0}$, as it effect on the studied
states of the system (typically with only the LLL, spin-split, corresponding
to the spin quantum number $\sigma =1$, is occupied) can be easily taken
into account in present treatment when will be needed. For definiteness, we
assume that $g_{0}<0$, as in typical GaAs-based samples, and $B>0$. Further,
in Eq. (\ref{1}) the electron-electron potential 
\begin{eqnarray}
V_{ee} &=&\frac{1}{2}\sum_{i=1}^{\tilde{N}}\sum_{j=1,j\neq i}^{\tilde{N}%
}\sum_{k=-N_{C}}^{N_{C}}\frac{e^{2}}{\varepsilon |\mathbf{r}_{i}-\mathbf{r}%
_{j}-kL_{x}^{\square }\hat{\mathbf{x}}|}  \notag \\
&&+\tilde{N}\sum_{k=1}^{N_{C}}\frac{e^{2}}{\varepsilon L_{x}^{\square }k},
\label{3}
\end{eqnarray}%
where $N_{C}$ is a very large natural number; as it is expected, physical
results will not depend on $N_{C}\rightarrow \infty $. In Eq. (\ref{1}) the
electron-ion potential 
\begin{equation}
V_{ei}=-\sum_{i=1}^{\tilde{N}}\sum_{j=1}^{\tilde{N}}\sum_{k=-N_{C}}^{N_{C}}%
\frac{e^{2}}{\varepsilon |\mathbf{r}_{i}-\mathbf{R}_{j}-kL_{x}^{\square }%
\hat{\mathbf{x}}|},  \label{4}
\end{equation}%
and the ion-ion potential
\begin{eqnarray}
V_{ii} &=&\frac{1}{2}\sum_{i=1}^{\tilde{N}}\sum_{j=1,j\neq i}^{\tilde{N}%
}\sum_{k=-N_{C}}^{N_{C}}\frac{e^{2}}{\varepsilon |\mathbf{R}_{i}-\mathbf{R}%
_{j}-kL_{x}^{\square }\hat{\mathbf{x}}|}  \notag \\
&&+\tilde{N}\sum_{k=1}^{N_{C}}\frac{e^{2}}{\varepsilon L_{x}^{\square }k}.
\label{5}
\end{eqnarray}%
Notice, the modified form of the Coulomb interaction, due to PBC,
that appears in Eqs. (\ref{3})-(\ref{5}) is quite similar with
the one given in Refs. \cite{yoshioka83i84,yoshioka84} see also Ref. \cite%
{chakraborty1995}. Further, it is seen that the second (constant) term in
Eq. (\ref{3}) gives the total contribution due to the interaction of each
electron, within MS, with its images (that appear in \textit{%
other} strips due to PBC); here the final sum over all $%
\tilde{N}$ electrons of MS leads to the factor $\tilde{N}$.
Point out, that the second term in Eq. (\ref{3}) it follows from the first
term in Eq. (\ref{3}) if formally to assume that $j=i$ (i.e., formally
neglecting the important condition $j\neq i$) and excluding $k=0$ term from
the sum over $k$; it clearly should be absent as now it gives
self-interaction of an electron with itself, not with it image. In addition,
if formally to discard PBC then only electrons
within MS are present and respectively in Eq. (\ref{3}): the
second term should be dropped and in the first term the sum over $k$ must be
reduced to only one term, $k=0$. I.e., in this limiting case Eq. (\ref{3})
reduces to the correct form of the electron-electron potential.

Point out, the form of the electron-electron interaction Eq. (\ref{3}),
modified due to PBC, can be proven by a detailed
consideration. The latter is mainly omitted as it final result Eq. (\ref{3})
is rather natural, as we have shown above. Notice, such detailed
consideration should exclude double counting of the interactions between an $%
i-$ electron, of MS, with all other $j \neq i$ electrons of the
main strip and their images; the same is valid for the interactions of this $%
i-$ electron with its images. In particular, it should be kept in mind that $%
\tilde{N}$ electrons of MS represent only very small fraction of
the $N$ electrons of the main region and for any other strip $n_{xs}^{\beta}
\neq n_{xs}^{\alpha}$ the electrons of such $\beta-$strip will interact with
electrons of the $\alpha-$strip (i.e., our MS) as with their
images in the $\alpha-$strip.

As it should be, the form of the ion-ion potential Eq. (\ref{5}) is totally
analogous to the one of the electron-electron potential Eq. (\ref{3}). The
form of the electron-ion interaction Eq. (\ref{4}), modified due to
PBC, can be proven by a detailed consideration; the
latter is omitted as it final result Eq. (\ref{4}) is quite natural. Notice,
this consideration takes into account, without double counting, as
interactions of any electron of MS with all ions and their
images so interactions of any ion of MS with all electrons and
their images. Point out, if formally to discard PBC
then only electrons and ions within MS are present and
in Eq. (\ref{4}) in the sum over $k$ should be left only one term, $k=0$;
i.e., Eq. (\ref{4}) in this limiting case reduces to the correct form of the
electron-electron potential.

Further, for implicit limit $N_{C} \rightarrow \infty$ (assuming that
MS is repeated $N/\tilde{N}= L_{x}/L_{x}^{\square} \rightarrow
\infty $ times within MR $L_{x} \times L_{y}$), it is
easy to see that the Hamiltonian $\hat{H}_{\tilde{N},\tilde{N}}$, defined by
Eqs. (\ref{1})-(\ref{5}), is periodic with the period $L_{x}^{\square}$
along any of it $2\tilde{N}$ variables $x_{i}$ and $X_{j}$, where $%
i,j=1,\cdots,\tilde{N}$. Then a many-body wave function that describes a
state pertinent to the Hamiltonian $\hat{H}_{\tilde{N},\tilde{N}}$, Eq. (\ref%
{1}), should satisfy the same property, i.e., to be periodic with the period 
$L_{x}^{\square}$ along any of $2\tilde{N}$ variables $x_{i}$ and $X_{j}$.
Point out, present problem has the translational symmetry along the $x$ axis
very similar with pertinent translational symmetry of Refs. \cite%
{yoshioka83i84,yoshioka84} (i.e., along the $y$ axis in Refs. \cite%
{yoshioka83i84,yoshioka84}, due to another form of the Landau gauge used by 
\cite{yoshioka83i84,yoshioka84}).

As a good approximation of typical experimental conditions, in present study
we will assume for UIB model that each ion is located totally (as, e.g., in
Sec. IV A) in a square unit cell $L_{x}^{\square} \times L_{x}^{\square}$,
such that $L_{x}^{\square} L_{y}/(L_{x}^{\square})^{2}=\tilde{N}$. Further,
for a ground-state we will assume that any particular electron tends to be
with equal probability in all these $\tilde{N}$ unit cells. This in turn
leads to some important conditions for an optimal set of single-electron
wave functions.

\subsection{For classical jellium model of ion background}

For widely used\cite{laughlin1983,morf1986,mikhailov2001} model of the
classical ion jellium background (IJB), we need to modify the Hamiltonian
Eq. (\ref{1}). Then the relevant many-electron Hamiltonian of electron-ion
system, $\hat{H}_{\tilde{N}}^{JB}$, is given as 
\begin{equation}
\hat{H}_{\tilde{N}}^{JB}=\hat{H}_{0}+V_{ee}+V_{eb}+V_{bb},  \label{1d}
\end{equation}%
where $\hat{H}_{0}$ and $V_{ee}$ are defined by Eqs. (\ref{2}) and (\ref{3}%
), respectively. Here the electron-ion system, localized within MS
$L_{x}^{\square }n_{xs}^{\alpha }>x>L_{x}^{\square }(n_{xs}^{\alpha
}-1)$, consists: i) from $\tilde{N}$ electrons, of MS,
interacting with IJB of MR and ii) from
the uniform positive charge density of the ion background, $|e|n_{b}$,
localized within MS that interacts with IJB of the main region.
Respectively, in Eq. (\ref{1d}) we have, cf. with Refs. \cite%
{morf1986,mikhailov2001}, that 
\begin{equation}
V_{eb}=-\sum_{i=1}^{\tilde{N}}
\int_{MR}d\mathbf{R}\frac{e^{2}n_{b}(\mathbf{R}%
)}{\varepsilon |\mathbf{r}_{i}-\mathbf{R}|},  \label{2d}
\end{equation}%
where the subscript \textquotedblleft MR\textquotedblright\ shows that
integration is carried out over MR, $L_{x}\times L_{y}$, and
\begin{equation}
V_{bb}=\frac{1}{2}\int_{MS}d\mathbf{R}\int_{MR}d\mathbf{R}^{\prime }\frac{%
e^{2}n_{b}(\mathbf{R})n_{b}(\mathbf{R}^{\prime })}{\varepsilon |\mathbf{R}-%
\mathbf{R}^{\prime }|},  \label{3d}
\end{equation}%
where the subscript \textquotedblleft MS\textquotedblright\ shows that
integration is carried out over MS. Notice, it is assumed that
within the main region $n_{b}(\mathbf{R})=const(\mathbf{R})=n_{b}$ and $%
\int_{MS}d\mathbf{R}n_{b}=\tilde{N}$.

Assuming that MS is repeated $N/\tilde{N}= L_{x}/L_{x}^{\square}
\rightarrow \infty$ times within MR, it is
easy to see that the Hamiltonian $\hat{H}_{\tilde{N}}^{JB}$, defined by Eqs.
(\ref{1d})-(\ref{3d}), (\ref{2}), (\ref{3}), is periodic with the period $%
L_{x}^{\square}$ along any of it variable $x_{i}$, $i=1,\cdots,\tilde{N}$.
Then a many-body wave function that describes a state pertinent to the
Hamiltonian $\hat{H}_{\tilde{N}}^{JB}$, Eq. (\ref{1d}), should satisfy the
same property, i.e., to be periodic with the period $L_{x}^{\square}$ along
any of $\tilde{N}$ variables $x_{i}$.

\subsection{Hamiltonian $\hat{H}_{\tilde{N}}^{JB}$: straightforward
obtaining from first principles}

For definiteness, here we will present straightforward
justification, starting from first principles,
of the model Hamiltonian $\hat{H}_{\tilde{N}}^{JB}$.
Point out, very similar treatment will justify
$\hat{H}_{\tilde{N},\tilde{N}}$ model Hamiltonian.

Here we will start with the same many-electron Hamiltonian,
$\hat{H}(\textbf{r}_{1},\ldots,\textbf{r}_{N})$, for
2DES of $N$ electrons as in \cite{laughlin1983,morf1986},
only for the Landau vector potencial gauge
$\textbf{A}=-By\hat{\textbf{x}}$. We assume that $N$ electrons
are localized in MR; $N/L_{x}L_{y}=\nu/2\pi \ell_{0}^{2}$.
As in \cite{laughlin1983,morf1986}, the IJB-model
of neutralizing ion background is assumed. Eigenstates
$\Psi(\textbf{r}_{1},\ldots,\textbf{r}_{N})$ of the Hamiltonian
$\hat{H}(\textbf{r}_{1},\ldots,\textbf{r}_{N}) \equiv \hat{H}^{MR}$
and their energies are determined by
\begin{equation}
\hat{H}^{MR}
\Psi(\textbf{r}_{1},\ldots,\textbf{r}_{N})=E_{N}
\Psi(\textbf{r}_{1},\ldots,\textbf{r}_{N}) ,
\label{DA1}
\end{equation}%
here
\begin{equation}
\hat{H}^{MR}=\hat{H}_{0}^{MR}+V_{ee}^{MR}+V_{eb}^{MR}+V_{bb}^{MR},
\label{DA2}
\end{equation}%
where
\begin{equation}
\hat{H}_{0}^{MR}=\frac{1}{2m^{\ast }}\sum_{i=1}^{N}
[\hat{\mathbf{p}}_{i}-\frac{e}{c}\mathbf{A}(\mathbf{r}_{i})]^{2},
\label{DA3}
\end{equation}%
\begin{equation}
V_{ee}^{MR}=\frac{1}{2}\sum_{i=1}^{N}\sum_{j=1,j\neq i}^{N}
\frac{e^{2}}{\varepsilon |\mathbf{r}_{i}-\mathbf{r}_{j}|} ,
\label{DA4}
\end{equation}%
further,
\begin{equation}
V_{eb}^{MR}=-\sum_{i=1}^{N}\int_{MR}d\mathbf{R}
\frac{e^{2}n_{b}(\mathbf{R})}{\varepsilon
|\mathbf{r}_{i}-\mathbf{R}|},  \label{DA5}
\end{equation}%
and
\begin{equation}
V_{bb}^{MR}=\frac{1}{2}\int_{MR}d\mathbf{R}\int_{MR}
d\mathbf{R}^{\prime }\frac{e^{2}n_{b}(\mathbf{R})
n_{b}(\mathbf{R}^{\prime })}{\varepsilon |\mathbf{R}-%
\mathbf{R}^{\prime }|}.
\label{DA6}
\end{equation}%
We can assume that 2DES, with IJB, is located within the ribbon of
width $L_{y}$ bent into loop of radius $L_{x}/2\pi$.
Then Born-Carman PBCs
$\textbf{r}_{i} \pm L_{x}\hat{\textbf{x}}=\textbf{r}_{i}$ are
holded, where $i=1,\dots,N$.
It is seen that the area of MR per electron, $L_{x}L_{y}/N=
(L_{x}^{\square})^{2}$, where $L_{x}^{\square}=\sqrt{2 m \pi}\ell_{0}$.
Then the strip of the width $L_{x}^{\square}$, along $x$-direction,
and of the length $L_{y}$ contains $\tilde{N}=L_{y}/L_{x}^{\square}$
of the (square) unit cells,
$L_{x}^{\square} \times L_{x}^{\square}$. The integer number
of such strips within MR is
given as $n_{xs}^{\max}=L_{x}/L_{x}^{\square}=N/\tilde{N}$;
for definiteness, odd (as $\tilde{N}$).

Further, we assume that for Eq. (\ref{DA1}) the ground-state and,
at the least, the lowest excited-state  correspond to partial
crystal-like correlation order among $N$ electrons of MR as
\begin{eqnarray}
&&\textbf{r}_{1+k_{1} \tilde{N}}=\textbf{r}_{1}+k_{1} L_{x}^{\square}
\hat{\textbf{x}}, \;\;\;
\textbf{r}_{2+k_{2} \tilde{N}}=\textbf{r}_{2}+k_{2} L_{x}^{\square}
\hat{\textbf{x}},
\notag \\
&& \cdots, \;\;\;
\textbf{r}_{\tilde{N}+k_{\tilde{N}} \tilde{N}}=
\textbf{r}_{\tilde{N}}+k_{\tilde{N}} L_{x}^{\square} \hat{\textbf{x}} ,
\label{DA7}
\end{eqnarray}%
where $k_{i}=1, 2,\ldots, n_{xs}^{\max}$;
$n_{xs}^{\max}=L_{x}/L_{x}^{\square}$.
Then using Eq. (\ref{DA7}) in Eqs. (\ref{DA2})-(\ref{DA5}) it is easy
to see that exact many-body
Hamiltonian $\hat{H}^{MR}$ becomes:
i) dependent only on $\tilde{N}$ $\mathbf{r}_{i}=(x_{i},y_{i})$
and ii) periodic, with period $L_{x}^{\square}$,
on any $x_{i}$; $i=1,\ldots,\tilde{N}$.
Hence, for a wave function in Eq. (\ref{DA1}) the properties (i) and (ii)
also will be valid. Then the study (e.g., calculation of the
total energy per electron, etc.) within MR of Eq. (\ref{DA1}),
for 2DES of $N$ electrons with many-electron wave functions
orthonormal within MR, can be exactly reduced to the treatment of the
Schr\"{o}dinger equation for 2DES of $\tilde{N}$ ``compound'' electrons
within MS as
\begin{equation}
\hat{H}_{\tilde{N}}^{JB}(\textbf{r}_{1},\ldots,\textbf{r}_{\tilde{N}})
\Psi_{\tilde{N}}=E_{\tilde{N}}^{JB}
\Psi_{\tilde{N}}(\textbf{r}_{1},\ldots,\textbf{r}_{\tilde{N}}) ,
\label{DA8}
\end{equation}%
with many-electron wave functions orthonormal within MS; also for
pertinent discussion see Sec. V.C.
Notice, here we formally obtain
that $N_{C}=(n_{xs}^{\max}-1)/2$ in Eq. (\ref{3}). However,
as physical results will be practically independent of $N_{C}$ for
$N_{C} \gg 1$, we can assume $N_{C}$ in Eq. (\ref{3})
very large, however, such that
$N_{C} \ll (n_{xs}^{\max}-1)/2 \rightarrow \infty$; notice, the same
conditions on $N_{C}$ will be obtained for $\hat{H}_{\tilde{N},\tilde{N}}$
model Hamiltonian.
Conditions Eq. (\ref{DA7}) can be thought as assumed physical
constaints that are justified only if they will lead to a
ground-state with the energy, per electron, lower than obtained in
Ref. \cite{laughlin1983}.

I.e., we assume that there are some low
energy eigenstates for which holds partial crystal-like correlation
order, Eq. (\ref{DA7}); due to the latter some many-body correlations
are implicitly included in the Hamiltonian Eq. (\ref{1d}).
Present below study of the low energy eigenstates (in
particular, the trial wave function of ground-state with the energy
lower than the energy for the Laughlin's trial wave function of
ground-state) confirms this physical assumption,
Eq. (\ref{DA7}).

\section{Single-body wave functions}

Let us for MS ($L_{x}^{\square}n_{xs}^{\alpha}
>x>L_{x}^{\square}(n_{xs}^{\alpha}-1)$; $L_{y}/2> y >-L_{y}/2$) introduce
normalized solutions of the single-electron Schr\"{o}dinger ($%
\omega_{c}=|e|B/m^{\ast} c$) equation
\begin{equation}
\hat{h}_{0} \psi_{n_{\alpha}; k_{x\alpha}}^{L_{x}^{\square}}(\mathbf{r})=
\hbar \omega_{c} (n_{\alpha}+1/2) \psi_{n_{\alpha};
k_{x\alpha}}^{L_{x}^{\square}}(\mathbf{r}),  \label{6}
\end{equation}%
that satisfy PBC ($y_{0}(k_{x\alpha})=%
\ell_{0}^{2} k_{x\alpha}$; $k_{x\alpha}=(2\pi/L_{x}^{\square})
n_{ys}^{\alpha}$), of the form
\begin{eqnarray}
&&\psi_{n_{\alpha};k_{x\alpha}}^{L_{x}^{\square}}(\mathbf{r}) \equiv
\psi_{n_{\alpha}; n_{ys}^{\alpha}}^{L_{x}^{\square}}(\mathbf{r})  \notag \\
&&= e^{ik_{x\alpha} x} \Psi_{n_{\alpha}}(y-y_{0}(k_{x\alpha})) /\sqrt{%
L_{x}^{\square}} ,  \label{7}
\end{eqnarray}%
where $\Psi_{n}(y)$ is the harmonic oscillator function, $%
n_{ys}^{\alpha}=0,\pm 1,\ldots, \pm(\tilde{N}_{L}-1)/2$; $\tilde{N}_{L}$ is
the odd integer such that $(2\pi/L_{x}^{\square}) \tilde{N}_{L}
\ell_{0}^{2}=L_{y}$. As we see $n_{ys}^{\alpha}$ gives the number to a
``bare'' cell of the width $\Delta y_{0}=2\pi \ell_{0}^{2}/L_{x}^{\square}$,
the quantum of $y_{0}(k_{x\alpha})$, and of the length $L_{x}^{\square}$,
along $x$. So within each ``bare'' cell $L_{x}^{\square} \times \Delta y_{0}$
there is only one quantized $y_{0}(k_{x\alpha})$ (or $k_{x\alpha}$) at the $%
n_{\alpha}-$th Landau level; i.e., there is only one state Eq. (\ref{7}) per
``bare'' cell at the latter level. Respectively, the total number of the
bare cells, per Landau level, within MS is equal to $\tilde{N}%
_{L}$.

It can be shown that, within MS, the set of single-body wave
functions Eq. (\ref{7}) is complete. In addition, they are orthonormal
within the strip as 
\begin{eqnarray}
&&\int_{L_{x}^{\square}(n_{xs}^{\alpha}-1)}^{L_{x}^{\square}n_{xs}^{\alpha}}
dx \int_{-\infty}^{\infty} dy \;
\psi_{n_{\beta};k_{x\beta}}^{L_{x}^{\square} \ast}(\mathbf{r})
\psi_{n_{\alpha};k_{x\alpha}}^{L_{x}^{\square}}(\mathbf{r})  \notag \\
&&\equiv \langle \psi _{n_{\beta };k_{x\beta }}^{L_{x}^{\square }}| \psi
_{n_{\alpha };k_{x\alpha}}^{L_{x}^{\square }}\rangle =
\delta_{n_{\beta},n_{\alpha}} \delta_{k_{x\beta},k_{x\alpha}} ,  \label{8}
\end{eqnarray}%
which can be also rewritten in the equivalent form by formally changing in
Eq. (\ref{8}) of $k_{x\alpha} \rightarrow n_{ys}^{\alpha}$, and $k_{x\beta}
\rightarrow n_{ys}^{\beta}$. If formally to assume $L_{x}^{\square}=L_{x}$,
Eq. (\ref{8}) reduces to well known result \cite{landau1965} for ``usual''
wave functions, $\psi_{n_{\alpha};k_{x\alpha}}^{L_{x}}(\mathbf{r})$.

Now looking for an optimal set of single-electron wave functions Eq. (\ref{7}%
) at $\nu \leq 1$ and taking into account that the total ion charge within
the unit cell $L_{x}^{\square} \times L_{x}^{\square}$ is equal to $|e|$, we
assume that the total electron charge within the unit cell must be exactly
equal to $e(<0)$. In some way this square unit cell is ``dressed'' by one
electron charge $e$. We call it the unit cell: these cells cannot be
confused with the ``bare'' cells. It follows that 
\begin{equation}
L_{x}^{\square}=\frac{L_{y}}{\tilde{N}}= \frac{L_{y}}{\nu \tilde{N}_{L}}.
\label{9}
\end{equation}%
It is natural to assume that $\tilde{N}$ is fixed for a given sample while
the filling factor $\nu$ can obtain different values. Then $L_{x}^{\square}$
is independent of $\nu$ (or $B$) for the given system.

Using Eqs. (\ref{9}), we obtain 
\begin{equation}
\frac{L_{x}^{\square }}{\Delta y_{0}}=\frac{1}{\nu },  \label{10}
\end{equation}%
so the quantum $\Delta y_{0}\neq L_{x}^{\square }$ for $\nu <1$. More
importantly, with the help of Eq. (\ref{10}) it is seen that within each
unit cell can appear only an odd integer number, $m=1,3,5,\ldots $, of the
quantized oscillator centres $y_{0}(k_{x\alpha })$ of the states Eq. (\ref{7}%
) at the given $n_{\alpha }$-th Landau level. Indeed, for an even integer
number $m=m_{0}$ of the states per unit cell only $m_{0}-1$ states from them
have $y_{0}(k_{\alpha })=\ell _{0}^{2}k_{x\alpha }$ inside the unit cell. As
exactly at the boundaries of unit cells, separating them at $y=\pm
L_{x}^{\square }/2,\pm (1+1/2)L_{x}^{\square }/2,\pm (2+1/2)L_{x}^{\square
}/2,\ldots $, there is one state (i.e., $y_{0}(k_{x\alpha })$) per such
boundary. So the case of even $m$ is a special one and it is not treated in
this work. Then the left hand side (LHS) of Eq. (\ref{10}) must be an odd
integer, $m=2\ell +1$, and it follows that 
\begin{equation}
\frac{1}{\nu }=m,  \label{11}
\end{equation}%
where $\ell =0,1,2,\ldots $.

According to PBC, the wave functions Eq. (\ref{7}%
) are periodic along $x$ with period $L_{x}^{\square}$, cf. with Refs. \cite%
{yoshioka83i84,yoshioka84}. So if multiply Eq. (\ref{6}), from the left, by $%
\psi _{n_{\beta };k_{x\beta }}^{L_{x}^{\square }\ast }(\mathbf{r})$ and then
integrate over $\mathbf{r}$ (as $\int_{L_{x}^{\square}(n_{xs}^{%
\alpha}-1)}^{L_{x}^{\square} n_{xs}^{\alpha}}dx \int_{-\infty }^{\infty
}dy...$) within MS, it follows that
\begin{eqnarray}
&&\langle \psi _{n_{\beta };k_{x\beta }}^{L_{x}^{\square }}|\hat{h}_{0}|\psi
_{n_{\alpha };k_{x\alpha}}^{L_{x}^{\square }}\rangle  \notag \\
&=&\hbar \omega _{c}(n_{\alpha }+\frac{1}{2})\delta _{n_{\beta },n_{\alpha
}}\delta _{k_{x\beta },k_{x\alpha}},  \label{12}
\end{eqnarray}%
where the right-hand side (RHS) have been obtained by using Eq. (\ref{8}).

Notice, from Eqs. (\ref{10}), (\ref{11}) it follows that 
\begin{equation}
L_{x}^{\square }=\sqrt{2\pi m}\;\ell _{0},  \label{13}
\end{equation}%
this form is useful for present study. Notice, as $L_{x}^{\square }$ is
actually independent of $m$, a superscript ($m$) in $L_{x}^{\square}$ is not
used.

Further, as for $\nu=1/m$ there are $m=2\ell+1$ quantized values of $%
y_{0}(k_{x i})$ (or $k_{x i}$) within an $i-$th unit cell and each of them
has a particular position within the unit cell, we separate all $\tilde{N}%
_{L}$ states Eq. (\ref{7}), of the $n_{\alpha}-$th Landau level, into the $m$
sets of wave functions. Within any such $n-$th set of states (here $n$ can
obtain $m$ different values) the difference $[y_{0}(k_{x
j}^{(n)})-y_{0}(k_{x i}^{(n)})]=k L_{x}^{\square}$, where $k$ is an integer.
Here $j$($i$) is the number of a unit cell; it can be any integer from $1$
to $\tilde{N}$. Point out, this $i-$number unambiguously defines the $i-$th
unit cell, among all $\tilde{N}$ unit cells of MS. The
superscript in $k_{x i}^{(n)}$ is given to distinguish the $k_{x i}$
pertinent to the $n-$th set of states; the subscript (superscript) "$i$" in $%
k_{x i}$, $n_{ys}^{(i)}$, etc. indicates belonging to the $i-$th unit cell.
For definiteness, we will choose the values of $n$ as follows: $n=0,...,\pm
\ell$. In particular, for $m=1$ it follows $\ell=0$ and $n=0$, for $m=3$ it
follows $\ell=1$ and $n=0,\pm 1$. We define $k_{x i}^{(n)}$, for $m=2\ell+1$%
, as follows
\begin{equation}
k_{x i}^{(0)}=(2\pi m/L_{x}^{\square}) n_{ys}^{(i)},\ldots, k_{x i}^{(\pm
\ell)}=k_{x i}^{(0)} \pm 2\pi \ell/L_{x}^{\square} ,  \label{14}
\end{equation}%
where $n_{ys}^{(i)}=0, \pm 1,\ldots, \pm (\tilde{N}-1)/2$, and $\tilde{N}=%
\tilde{N}_{L}/m$. It is seen that for the given $n-$th set the total number
of different $k_{x i}^{(n)}$ within MS is equal to $\tilde{N}$,
as it should be. So all $m$ sets of $k_{x i}^{(n)}$ give altogether $\tilde{N%
}_{L}=m \times \tilde{N}$ different values, the same as for the $k_{x\alpha}$
in Eq. (\ref{7}). Point out that the choice of $k_{x i}^{(n)}$ in the form
Eq. (\ref{14}) is quite natural as here: i) all $\ell_{0}^{2} k_{x
i}^{(n)}=y_{0}(k_{x i}^{(n)})$ are symmetrical with respect of the $y$%
-centre of MS $y=0$; ii) the smallest $|k_{x i}^{(n)}|$ is given
by $k_{x i}^{(0)}=0$, for $n_{ys}^{(i)}=0$; iii) within an $i-$th unit cell
all it $m$ states have $k_{x i}^{(n)}$ symmetric with the respect of $k_{x
i}^{(0)}$, the centre of this cell; iv) this choice leads to symmetric and
more homogeneous electron charge density within a unit cell, along $y-$%
direction.

To simplify writing, we will use notation $\psi _{n_{\alpha }; k_{x\alpha }}(%
\mathbf{r})\equiv \psi _{n_{\alpha }; k_{x\alpha }}^{L_{x}^{\square }}(%
\mathbf{r})$. Widely used below wave functions Eq. (\ref{7}) of the $%
n_{\alpha }=0$ Landau level $\psi _{0;k_{x\alpha }}(\mathbf{r})\equiv \psi
_{0;n_{ys}^{\alpha }}(\mathbf{r})$ we denote, at $\nu =1/m$, as well as 
\begin{equation}
\varphi _{k_{xi}^{(n)}}^{(m)}(\mathbf{r})\equiv \psi _{0;k_{xi}^{(n)}}(%
\mathbf{r}),  \label{15}
\end{equation}%
where $i=1,2,\ldots,\tilde{N}$ is the number of a unit cell and $%
n=0,\dots,\pm \ell$ the "set" number; they unambiguously define $%
k_{xi}^{(n)} $. For these wave functions Eq. (\ref{8}) reduces to
\begin{eqnarray}
&&\int_{L_{x}^{\square}(n_{xs}^{\alpha}-1)}^{L_{x}^{\square}n_{xs}^{\alpha}}
dx \int_{-\infty}^{\infty} dy \; \varphi_{k_{xj}^{(k)}}^{(m)\ast }(\mathbf{r}%
)\;\varphi _{k_{xi}^{(n)}}^{(m)}(\mathbf{r})  \notag \\
&&\equiv \langle \varphi_{k_{xj}^{(k)}}^{(m)}|
\varphi_{k_{xi}^{(n)}}^{(m)}\rangle=\; \delta
_{k,n}\;\delta_{k_{xj}^{(n)},k_{xi}^{(n)}}\;,  \label{16}
\end{eqnarray}%
i.e., the single-body wave functions of the same $n-$th set are orthonormal
and they are orthogonal to any wave function from another set $k\neq n$.
From Eq. (\ref{12}) we have 
\begin{equation}
\langle \varphi _{k_{xj}^{(k)}}^{(m)}|\hat{h}_{0}|
\varphi_{k_{xi}^{(n)}}^{(m)}\rangle =\frac{\hbar \omega _{c}}{2} \;\delta
_{k,n}\; \delta _{k_{xj}^{(n)},k_{xi}^{(n)}}.  \label{17}
\end{equation}%
Point out, Eq. (\ref{17}) is very similar to pertinent result of Refs. \cite%
{yoshioka83i84,yoshioka84}, for their finite rectangular main cell.

\section{Ground-state wave function of electron-ion system at $\protect\nu%
=1/m$}

At $\nu =1/m$ ($m=2\ell +1$; $\ell=0, 1,\dots$), we look for the
ground-state many-body wave function of the electron-ion system Eq. (\ref{1}%
), Sec. IV A, and Eq. (\ref{1d}), Sec. IV B. In Sec. IV A we consider
electron-ion system for UIB with
the total wave function $\Psi _{\tilde{N},\tilde{N}}^{(m),eh}(\mathbf{r}%
_{1},\ldots, \mathbf{r}_{\tilde{N}};\mathbf{R}_{1},\ldots,\mathbf{R}_{\tilde{%
N}})$, that corresponds to totally homogeneous ion density $n_{io}^{eh}$,
Eq. (\ref{32}). It is important to point out that for UIB model (in
difference from typically used\cite{laughlin1983,morf1986,mikhailov2001} IJB
model) it is absent, e.g., self-interaction of an ion with itself. In Sec.
IV B we treat the ground-state, $\Psi _{\tilde{N}}^{(m),JB}(\mathbf{r}%
_{1},\ldots, \mathbf{r}_{\tilde{N}})$, of electron-ion system for IJB model.
To simplify notations, arguments $\mathbf{r}_{i},\mathbf{R}_{i}$ in the wave
functions $\Psi_{\tilde{N},\tilde{N}}^{(m)}$, $\Psi _{\tilde{N},\tilde{N}%
}^{(m),eh}$, etc. are often suppressed.

\subsection{Ground-state, $\Psi _{\tilde{N},\tilde{N}}^{(m),eh}$, for UIB
model. Compound electrons.}

Now we will consider a ground-state, $\Psi _{\tilde{N},\tilde{N}}^{(m),eh}$,
of electron-ion system which gives exactly homogeneous ion density in MS.
Notice, due to PBC here the ion
density is homogeneous for the whole MR.
We assume that $\Psi _{\tilde{N},%
\tilde{N}}^{(m),eh}$ has the \textquotedblleft compound\textquotedblright\
form
\begin{eqnarray}
&&\Psi _{\tilde{N},\tilde{N}}^{(m),eh}(\mathbf{r}_{1},\ldots ,\mathbf{r}_{%
\tilde{N}};\mathbf{R}_{1},\ldots ,\mathbf{R}_{\tilde{N}})=\dprod%
\limits_{i=1}^{\tilde{N}}\phi _{n_{ys}^{(i)}}(\mathbf{R}_{i})  \notag \\
&&\;\;\;\;\;\;\;\;\;\;\;\;\;\;\;\;\;\;\times \sum_{n=-\ell }^{\ell
}C_{n}(m)\Psi _{\tilde{N}}^{n,(m)}(\mathbf{r}_{1},\ldots ,\mathbf{r}_{\tilde{%
N}}),  \label{18}
\end{eqnarray}%
where 
\begin{equation}
|C_{n}(m)|^{2}=1/m,  \label{19}
\end{equation}%
the \textquotedblleft partial\textquotedblright\ many-electron wave function
$\Psi _{\tilde{N}}^{n,(m)}(\mathbf{r}_{1},\mathbf{r}_{2},\ldots ,\mathbf{r}_{%
\tilde{N}})$ (or the $n-$th set many-electron wave function) is an $\tilde{N}%
-$dimensional Slater determinant of the wave functions Eq. (\ref{15}) given
as 
\begin{equation}
\Psi _{\tilde{N}}^{n,(m)}=\frac{1}{\sqrt{\tilde{N}!}}%
\begin{vmatrix}
\varphi _{k_{x1}^{(n)}}^{(m)}(\mathbf{r}_{1}) & \cdots & \varphi
_{k_{x1}^{(n)}}^{(m)}(\mathbf{r}_{\tilde{N}}) \\
\varphi _{k_{x2}^{(n)}}^{(m)}(\mathbf{r}_{1}) & \cdots & \varphi
_{k_{x2}^{(n)}}^{(m)}(\mathbf{r}_{\tilde{N}}) \\
\vdots & \ddots & \vdots \\ 
\varphi _{k_{x\tilde{N}}^{(n)}}^{(m)}(\mathbf{r}_{1}) & \cdots & \varphi
_{k_{x\tilde{N}}^{(n)}}^{(m)}(\mathbf{r}_{\tilde{N}})%
\end{vmatrix}%
,  \label{20}
\end{equation}%
i.e., in the \textquotedblleft Hartree-Fock\textquotedblright -alike form.
Point out that all $m$ many-electron wave functions Eq. (\ref{20}) form the
orthonormal assembly as 
\begin{equation}
\langle \Psi _{\tilde{N}}^{k,(m)}(\mathbf{r}_{1},\ldots ,\mathbf{r}_{\tilde{N%
}})|\Psi _{\tilde{N}}^{n,(m)}(\mathbf{r}_{1},\ldots ,\mathbf{r}_{\tilde{N}%
})\rangle =\delta _{k,n}.  \label{21}
\end{equation}%
It is readily seen that due to PBC satisfied by
the single-electron wave functions, Eqs. (\ref{7}), (\ref{15}), the
many-body wave function, Eq. (\ref{18}), and the \textquotedblleft
partial\textquotedblright\ many-electron wave functions, Eq. (\ref{20}), are
periodic with period $L_{x}^{\square }$ with respect to any $x_{i}$; $%
i=1,\ldots ,\tilde{N}$.

Further, in Eq. (\ref{18}) the ``partial'' many-ion wave function $%
\prod\limits_{i=1}^{\tilde{N}} \phi _{n_{ys}^{(i)}}(\mathbf{R}_{i})$ is
given in the ``Hartree''-alike form, where a ``single-ion'' wave function $%
\phi_{n_{ys}^{(i)}}(\mathbf{R})$, localized in the $i-$th unit cell of the
main strip, is introduced as follows. For both $X \in
(L_{x}^{\square}(n_{xs}^{\alpha}-1),L_{x}^{\square}n_{xs}^{\alpha})$ and $Y
\in (L_{x}^{\square}(n_{ys}^{(i)}-1/2),L_{x}^{\square} (n_{ys}^{(i)}+1/2))$,
we have 
\begin{equation}
|\phi_{n_{ys}^{(i)}}(\mathbf{R})|^{2}= 1/(L_{x}^{\square})^{2} ,  \label{22}
\end{equation}%
if $Y$ is outside of the $i-$th unit cell then $\phi_{n_{ys}^{(i)}}(\mathbf{R%
}) \equiv 0$. The set of these single-body wave functions is orthonormal,
within MS, as we have
\begin{eqnarray}
\int_{L_{x}^{\square}(n_{xs}^{\alpha}-1)}^{L_{x}^{\square}n_{xs}^{\alpha}}
dX \int_{-\infty}^{\infty} &&dY \; \phi_{n_{ys}^{(i)}}^{\ast}(\mathbf{R})
\phi_{n_{ys}^{(j)}}(\mathbf{R})  \notag \\
&&= \delta_{n_{ys}^{(i)},n_{ys}^{(j)}} \equiv \delta_{i,j} .  \label{23}
\end{eqnarray}%
Point out, PBC is also applied to the single-ion
wave functions $\phi_{n_{ys}^{(i)}}(\mathbf{R})$; i.e., the many-body wave
function Eq. (\ref{18}) is periodic with period $L_{x}^{\square}$ with
respect to any $X_{i}$ as well. It is seen that wave function Eq. (\ref{18})
is normalized, $\langle \Psi _{\tilde{N},\tilde{N}}^{(m),eh}| \Psi_{\tilde{N}%
,\tilde{N}}^{(m),eh}\rangle=1$. We also will need to use a shorter notation
for the integral over MS (given, e.g., in Eqs. (\ref{16}), (\ref%
{23})) as $\int d \mathbf{R} \ldots=
\int_{L_{x}^{\square}(n_{xs}^{\alpha}-1)}^{L_{x}^{\square}n_{xs}^{\alpha}}
dX \int_{-\infty}^{\infty} dY \ldots$.

Point out that, due to the quantized according to Eq. (\ref{19})
contributions from the partial many-electron wave functions Eq. (\ref{20}),
for $m \geq 3$ the compound form of the ground-state wave function
Eq. (\ref{18}) leads to the compound structure of each electron within MS.
In particular, this compound structure of the electrons plays
important role in the treatment of
excited-states of the present ground-state, as it is shown in Sec. VI.

Now we consider the electron charge density $\rho _{el}^{eh}(\mathbf{r}%
)=en^{eh}(\mathbf{r})$ in the state Eq. (\ref{18}), in MS. We
have
\begin{equation}
n^{eh}(\mathbf{r})=\langle \Psi _{\tilde{N},\tilde{N}}^{(m),eh}| \sum_{j=1}^{%
\tilde{N}}\delta (\mathbf{r}-\mathbf{r}_{j})| \Psi _{\tilde{N},\tilde{N}%
}^{(m),eh}\rangle ,  \label{24}
\end{equation}%
where in the RHS integration is taken over all $\mathbf{r}_{i}$ and $\mathbf{%
R}_{i}$. Notice, the matrix element of Eq. (\ref{24}) cannot be mixed with
the matrix elements of Eqs. (\ref{16}), (\ref{17}). Using Eqs. (\ref{16}), (%
\ref{18})-(\ref{20}), from Eq. (\ref{24}) it follows 
\begin{equation}
n^{eh}(\mathbf{r})=\frac{1}{m} \sum_{ n=-\ell }^{\ell }\langle \Psi _{\tilde{%
N}}^{n,(m)}|\sum_{j=1}^{\tilde{N}} \delta (\mathbf{r}-\mathbf{r}_{j})|\Psi_{%
\tilde{N}}^{n,(m)}\rangle ,  \label{25}
\end{equation}%
where arguments $\mathbf{r}_{i}$ in the $n-$th set many-electron wave
function $\Psi_{\tilde{N}}^{n,(m)}$, over which the integration holds in the
RHS of Eq. (\ref{25}), are suppressed. Eq. (\ref{25}), after using Eq. (\ref%
{20}), gives 
\begin{eqnarray}
n^{eh}(\mathbf{r})&=&\frac{1}{m}\sum_{n=-\ell }^{\ell }\sum_{i=1}^{\tilde{N}%
} |\varphi_{k_{xi}^{(n)}}^{(m)}(\mathbf{r})|^{2}=\frac{1}{\sqrt{2\pi }%
m^{3/2}\ell _{0}}  \notag \\
&&\times \sum_{n=-\ell }^{\ell }\sum_{k_{xi}^{(n)}} \Psi
_{0}^{2}(y-y_{0}(k_{xi}^{(n)})) ,  \label{26}
\end{eqnarray}%
where the $k_{xi}^{(n)}$ are given by Eq. (\ref{14}). Due to PBC,
using Eq. (\ref{26}) we conclude that both in MS and MR
$n^{eh}(\mathbf{r})\equiv n^{eh}(y)$, i.e., it is independent of $x$.

To study $n^{eh}(y)$, we apply to Eq. (\ref{26}) the Fourier transformation
over $y$, $n(q_{y})=\int_{-\infty }^{\infty }dy\;n(y)\exp (-iq_{y}y)$; we
obtain 
\begin{eqnarray}
n^{eh}(q_{y}) &=&\frac{e^{-q_{y}^{2}\ell _{0}^{2}/4}}{\sqrt{2\pi }%
m^{3/2}\ell _{0}}\sum_{n=-\ell }^{\ell }e^{-i\sqrt{2\pi /m}\;n\;q_{y}\ell
_{0}}  \notag \\
&&\times \sum_{n_{ys}^{(i)}=-(\tilde{N}-1)/2}^{(\tilde{N}-1)/2}e^{-i\sqrt{%
2\pi m}\;n_{ys}^{(i)}\;q_{y}\ell _{0}},  \label{27}
\end{eqnarray}%
where in the RHS, using $\tilde{N}\rightarrow \infty $ assumed in the
present study, the last sum is given by $\sum_{k=-\infty }^{\infty }\exp (-i%
\sqrt{2\pi m}kq_{y}\ell _{0})=\sum_{k=-\infty }^{\infty }\exp
(-ikq_{y}L_{x}^{\square })$. Using in the latter Poisson's summation formula
\cite{hilbert} we obtain (it is well known result) that 
\begin{equation}
\sum_{k=-\infty }^{\infty }\exp [-ikq_{y}L_{x}^{\square }]=\frac{2\pi }{%
L_{x}^{\square }}\sum_{M=-\infty }^{\infty }\delta (q_{y}+M\;\frac{2\pi }{%
L_{x}^{\square }}).  \label{28}
\end{equation}%
By making use of Eq. (\ref{28}) in Eq. (\ref{27}), we calculate
\begin{eqnarray}
n^{eh}(q_{y}) &=&\frac{1}{m^{2}\ell _{0}^{2}}\sum_{M=-\infty }^{\infty
}e^{-q_{y}^{2}\ell _{0}^{2}/4}\delta (q_{y}+\sqrt{\frac{2\pi }{m}}\frac{M}{%
\ell _{0}})  \notag \\
&&\times \sum_{n=-\ell }^{\ell }\exp (i2\pi Mn/m)  \notag \\
&=&\frac{1}{m\ell _{0}^{2}}\sum_{k=-\infty }^{\infty }e^{-q_{y}^{2}\ell
_{0}^{2}/4}\delta (q_{y}+\frac{\sqrt{2\pi m}}{\ell _{0}}k),  \label{29}
\end{eqnarray}%
where it is used (remind, $m=2\ell +1$) that i) $\sum_{n=-\ell }^{\ell }\exp
(i2\pi Mn/m)=m$ for $M=0,\pm m,\pm 2m,...$ and ii) this sum is equal to zero
for any other $M$ (by definition, $M$ is an integer); then after using the
notation $M=m\times k$, where $k=0,\pm 1,\pm 2,...$, we readily arrive to
the final form in Eq. (\ref{29}). Applying inverse Fourier transformation to
Eq. (\ref{29}), $n(y)=(1/2\pi )\int_{-\infty }^{\infty }dq_{y}\;n(q_{y})\exp
(iq_{y}y)$, we obtain 
\begin{equation}
n^{eh}(y)=\frac{1}{2\pi m\ell _{0}^{2}}[1+2\sum_{k=1}^{\infty }e^{-\pi
mk^{2}/2}\cos (\frac{\sqrt{2\pi m}}{\ell _{0}}ky)].  \label{30}
\end{equation}

Now we consider the ion charge density $\rho_{io}^{eh}(\mathbf{r})=-
en_{io}^{eh}(\mathbf{r})$ in the state Eq. (\ref{18}). The ion density,
within MS, is given as
\begin{eqnarray}
&&n_{io}^{eh}(\mathbf{r})= \langle \Psi _{\tilde{N},\tilde{N}%
}^{(m),eh}|\sum_{j=1}^{\tilde{N}}\delta (\mathbf{r}-\mathbf{R}_{j})|\Psi _{%
\tilde{N},\tilde{N}}^{(m),eh}\rangle  \notag \\
&&=\sum_{n_{ys}^{(i)}=-(\tilde{N}-1)/2}^{(\tilde{N}-1)/2}
|\phi_{n_{ys}^{(i)}}(\mathbf{r})|^{2} ,  \label{31}
\end{eqnarray}%
where, using Eq. (\ref{22}), it follows that $n_{io}^{eh}(\mathbf{r}) \equiv
const(x,y)=n_{io}^{eh}$, has the form 
\begin{equation}
n_{io}^{eh}=\frac{1}{(L_{x}^{\square})^{2}}=\frac{1}{2\pi m \ell_{0}^{2}} .
\label{32}
\end{equation}%
I.e., in MS (and, due to PBC, in
the main region $L_{x} \times L_{y}$) the ion density Eq. (\ref{32}) is
spatially homogeneous and independent of $m$.
Point out that $\sum_{n_{ys}^{(i)}=-(%
\tilde{N}-1)/2}^{(\tilde{N}-1)/2} \equiv \sum_{i=1}^{\tilde{N}}$ and $%
i=\{n_{ys}^{(i)}\}$; equivalently $"i"$ can be understood as $%
i=\{k_{xi}^{(0)}\}$.

Relative inhomogeneity $\delta \tilde{n}^{eh}$ of $n^{eh}(y)$, Eq. (\ref{30}%
), is very well approximated by the amplitude of $k=1$ oscillating term,
i.e., $\delta \tilde{n}^{eh} \approx 2 \times \exp(-\pi m/2)$. We have that
for $m=1, \; 3,$ and 5 $\delta \tilde{n}^{eh} \approx 0.416, \; 1.8 \times
10^{-2},$ and $7.8 \times 10^{-4}$. So the relative inhomogeneity of the
electron density $n^{eh}(y)$, Eq. (\ref{30}), pertinent to the homogeneous
ion background Eq. (\ref{32}), is not very small only for $m=1$, while for $%
m \geq 3$ the inhomogeneity is very small.

Point out that for the electron-ion system described by the wave function
Eq. (\ref{18}), the electron charge density does not exactly cancels the ion
charge density, $\rho_{el}^{eh}(\mathbf{r})+\rho_{io}^{eh}(\mathbf{r}) \neq
0 $. I.e., the system is not exactly electrically neutral, even though
within each unit cell the total electron charge, $e$, exactly cancels the
total ion charge, $-e$.

\subsection{Ground-state, $\Psi_{\tilde{N}}^{(m),JB}$, for IJB model.
Compound electrons.}

Now we consider a ground-state, $\Psi _{\tilde{N}}^{(m),JB}$, of
electron-ion system for typically used form of the ion background, i.e., the
continuous homogeneous one that we call as IJB. It is exactly homogeneous as
well within MS. From the Hamiltonian Eq. (\ref{1d}) it is clear
that here PBC is applied only to the electrons
coordinates and their functions. Based on Eq. (\ref{18}), it is natural to
assume that $\Psi _{\tilde{N}}^{(m),JB}$ has the \textquotedblleft
compound\textquotedblright\ form as follows 
\begin{eqnarray}
&&\Psi _{\tilde{N}}^{(m),JB}(\mathbf{r}_{1},\ldots ,\mathbf{r}_{\tilde{N}})=
\notag \\
&&\;\;\;\;\;\;\;\;\;\;\;\;\;\;\;\;\;\;\sum_{n=-\ell }^{\ell }C_{n}(m)\Psi _{%
\tilde{N}}^{n,(m)}(\mathbf{r}_{1},\ldots , \mathbf{r}_{\tilde{N}}).
\label{18d}
\end{eqnarray}%
It is readily seen that due to PBC satisfied by
the single-electron wave functions, Eqs. (\ref{7}), (\ref{15}), the
many-body wave function, Eq. (\ref{18d}), is periodic with period $%
L_{x}^{\square }$ with respect to any $x_{i}$; $i=1,\ldots ,\tilde{N}$.
Respectively, the electron density in the state Eq. (\ref{18d}), within the
main strip,
\begin{equation}
n^{JB}(\mathbf{r})=\langle \Psi _{\tilde{N}}^{(m),JB}| \sum_{j=1}^{\tilde{N}%
}\delta (\mathbf{r}-\mathbf{r}_{j})| \Psi _{\tilde{N}}^{(m),JB} \rangle ,
\label{24d}
\end{equation}%
coincides with $n^{eh}(\mathbf{r})$, see Eqs. (\ref{25})-(\ref{30}); i.e.,
$n^{JB}(\mathbf{r}) \equiv n^{JB}(y)=n^{eh}(y)$ both in MS and MR.

For the ion density, $n_{b}$, it follows that $n_{b}=n_{io}^{eh}$.

\section{Ground-state energy of electron-ion system at $\protect\nu=1/m$}

In Sec. IV A, for UIB model Eqs. (\ref{1})-(\ref{5}), we study the energy of
the ground-state Eq. (\ref{18}). In Sec. IV B, for IJB model Eqs. (\ref{1d}%
)-(\ref{3d}), (\ref{2}), (\ref{3}), we treat the energy of the ground-state
Eq. (\ref{18d}). As only for the latter model of the ion background we can
directly compare the total lowering per electron due to many-body
interactions (it includes any electron-electron, electron-ion and ion-ion
contributions) with pertinent total lowering for the Laughlin variational
wave function \cite{laughlin1983}. As we will show, the difference between
the ground-state energy per electron $U^{UB}(m)$, for UIB, and $U^{JB}(m)$,
for IJB, is related only with the difference between the contributions from
the ion-ion interaction $V_{ii}$, Eq. (\ref{5}), and $V_{bb}$,
Eq. (\ref{3d}), respectively.
In Sec. V C we present additional remarks on partial crystal-like
correlation and energy of ground-state.

Point out, all analytical results obtained in Sec. V are exact.

\subsection{Energy of ground-state $\Psi _{\tilde{N},\tilde{N}}^{(m),eh}$,
for UIB}

Using the Hamiltonian Eq. (\ref{1}), we calculate the total energy of
electron-ion system in the state Eq. (\ref{18}) as 
\begin{equation}
E_{\tilde{N}}^{(m),eh}=\langle \Psi _{\tilde{N},\tilde{N}}^{(m),eh}| \hat{H}%
_{\tilde{N},\tilde{N}} |\Psi_{\tilde{N},\tilde{N}}^{(m),eh}\rangle ,
\label{40}
\end{equation}%
where in the RHS for the kinetic energy term
\begin{equation}
\langle \Psi _{\tilde{N},\tilde{N}}^{(m),eh}|\hat{H}_{0} | \Psi _{\tilde{N},%
\tilde{N}}^{(m),eh}\rangle=\tilde{N}E_{kin}  \label{41}
\end{equation}%
we obtain (remind, all matrix elements should be calculated within MS)
the kinetic energy per electron (or per electron-ion pair) as
\begin{eqnarray}
&&E_{kin}= \frac{1}{m \tilde{N}} \sum_{n=-\ell }^{\ell } \langle \Psi _{%
\tilde{N}}^{n,(m)}|\hat{H}_{0}|\Psi_{\tilde{N}}^{n,(m)}\rangle  \notag \\
&&=\frac{1}{m \tilde{N}^{2}} \sum_{n=-\ell }^{\ell } \sum_{j=1}^{\tilde{N}}
\sum_{i=1}^{\tilde{N}} \langle \varphi _{k_{xi}^{(n)}}^{(m)}(\mathbf{r}_{j})|%
\hat{h}_{0j}|\varphi _{k_{xi}^{(n)}}^{(m)}(\mathbf{r}_{j})\rangle  \notag \\
&&=\frac{1}{m \tilde{N}} \sum_{n=-\ell }^{\ell } \sum_{j=1}^{\tilde{N}} 
\frac{\hbar \omega_{c}}{2}=\frac{\hbar \omega_{c}}{2} .  \label{42}
\end{eqnarray}%
In Eq. (\ref{42}) the RHS of the first line contains usual in the
Hartree-Fock theory \cite{ansel'm1978,madelung1981} matrix elements, on the
``Hartree-Fock''-alike many-electron wave functions Eq. (\ref{20}); this RHS
it follows straightforwardly from the LHS, quite similar with transition
from Eq. (\ref{24}) to Eq. (\ref{25}). Then transition to the second line of
Eq. (\ref{42}) is analogous to well known one in the Hartee-Fock theory \cite%
{ansel'm1978,madelung1981}. Notice, in Eq. (\ref{42}) index ``j''
distinguishes electrons. Further, the second line in Eq. (\ref{42}) is
simplified by using Eq. (\ref{17}). Point out, from Sec. IV as well as Eq. (%
\ref{42}) it is seen that within the subspace of the $n-$th set of
single-electron states (from which the $n-$th set many-electron wave
function Eq. (\ref{20}) is constructed) an $j-$th electron is equally
distributed (present) over (in) all these $\tilde{N}$ states. Point out, the
kinetic energy $E_{kin}=\hbar \omega_{c}/2$ coincides with the pertinent
result of Refs. \cite{yoshioka83i84,yoshioka84}, for their finite
rectangular main cell.

In the RHS of Eq. (\ref{40}) the term related with ion-ion interaction, we
call it also UIB-UIB interaction, is given as

\begin{equation}
\langle \Psi _{\tilde{N},\tilde{N}}^{(m),eh}|V_{ii}| \Psi _{\tilde{N},\tilde{%
N}}^{(m),eh}\rangle= \tilde{N}(E_{ii}^{a}+E_{ii}^{b}),  \label{43}
\end{equation}%
where, due to the second term in the RHS of Eq. (\ref{5}), 
\begin{equation}
E_{ii}^{a}=\sum_{k=1}^{N_{C}}\frac{e^{2}}{\varepsilon L_{x}^{\square} k}
\label{44}
\end{equation}%
and, due to the first term in the RHS of Eq. (\ref{5}),
\begin{eqnarray}
E_{ii}^{b}&=& \frac{1}{2\tilde{N}}\sum_{i_{1}=1}^{\tilde{N}} \sum_{j_{1}=1,
j_{1} \neq i_{1}}^{\tilde{N}} \sum_{k=-N_{C}}^{N_{C}}  \notag \\
&& \times \langle \Psi _{\tilde{N},\tilde{N}}^{(m),eh}| \frac{e^{2}}{%
\varepsilon |\mathbf{R}_{i_{1}}- \mathbf{R}_{j_{1}}-k L_{x}^{\square} \hat{%
\mathbf{x}}|} |\Psi _{\tilde{N},\tilde{N}}^{(m),eh}\rangle  \notag \\
&& =\frac{1}{2\tilde{N}} \sum_{i_{1}=1}^{\tilde{N}} \sum_{j_{1}=1, j_{1}
\neq i_{1}}^{\tilde{N}} \sum_{k=-N_{C}}^{N_{C}} \langle \dprod\limits_{i=1}^{%
\tilde{N}}\phi _{n_{ys}^{(i)}}(\mathbf{R}_{i})|  \notag \\
&&\times \frac{e^{2}}{\varepsilon |\mathbf{R}_{i_{1}}-\mathbf{R}_{j_{1}}- k
L_{x}^{\square} \hat{\mathbf{x}}|} |\dprod\limits_{j=1}^{\tilde{N}}\phi
_{n_{ys}^{(j)}}(\mathbf{R}_{j}) \rangle .  \label{45}
\end{eqnarray}%
In Eq. (\ref{45}) we have used Eqs. (\ref{18})-(\ref{21}). Finally, by
standard transformations \cite{ansel'm1978,madelung1981}, Eq. (\ref{45}) is
rewritten as 
\begin{eqnarray}
E_{ii}^{b}&=& \frac{1}{2\tilde{N}} \sum_{k=-N_{C}}^{N_{C}} \sum_{i=1}^{%
\tilde{N}} \sum_{j=1, j \neq i}^{\tilde{N}} \int \int \frac{e^{2} d \mathbf{R%
} d \mathbf{R}^{\prime}}{\varepsilon |\mathbf{R}-\mathbf{R}^{\prime}- k
L_{x}^{\square} \hat{\mathbf{x}}|}  \notag \\
&&\;\;\; \times |\phi _{n_{ys}^{(i)}}(\mathbf{R})|^{2} \; \;
|\phi_{n_{ys}^{(j)}}(\mathbf{R}^{\prime})|^{2} .  \label{46}
\end{eqnarray}

In the RHS of Eq. (\ref{40}) the term related with electron-ion interaction,
we call it also electron-UIB interaction, is given as
\begin{equation}
\langle \Psi _{\tilde{N},\tilde{N}}^{(m),eh}|V_{ei}| \Psi _{\tilde{N},\tilde{%
N}}^{(m),eh}\rangle= \tilde{N}E_{ei},  \label{47}
\end{equation}%
where
\begin{eqnarray}
&&E_{ei}= - \frac{1}{\tilde{N}}\sum_{i_{1}=1}^{\tilde{N}} \sum_{j_{1}=1}^{%
\tilde{N}} \sum_{k=-N_{C}}^{N_{C}} \langle \Psi _{\tilde{N},\tilde{N}%
}^{(m),eh}| \frac{e^{2}/\varepsilon}{ |\mathbf{r}_{i_{1}}- \mathbf{R}%
_{j_{1}}-k L_{x}^{\square} \hat{\mathbf{x}}|}  \notag \\
&& \times |\Psi _{\tilde{N},\tilde{N}}^{(m),eh}\rangle = - \frac{1}{m \tilde{%
N}} \sum_{n=-\ell}^{\ell} \sum_{i_{1}=1}^{\tilde{N}} \sum_{j_{1}=1}^{\tilde{N%
}} \sum_{k=-N_{C}}^{N_{C}}  \notag \\
&& \times \langle \Psi _{\tilde{N}}^{n,(m)}(\mathbf{r}_{1},\ldots)
\dprod\limits_{i=1}^{\tilde{N}}\phi _{n_{ys}^{(i)}}(\mathbf{R}_{i})| \frac{%
e^{2}/\varepsilon}{ |\mathbf{r}_{i_{1}}-\mathbf{R}_{j_{1}}- k
L_{x}^{\square} \hat{\mathbf{x}}|}  \notag \\
&&\times | \Psi _{\tilde{N}}^{n,(m)}(\mathbf{r}_{1},\ldots)
\dprod\limits_{j=1}^{\tilde{N}}\phi _{n_{ys}^{(j)}}(\mathbf{R}_{j}) \rangle .
\label{48}
\end{eqnarray}%
In Eq. (\ref{48}) we have used Eqs. (\ref{18})-(\ref{20}) and (cf. with Eq. (%
\ref{21})) the property
\begin{eqnarray}
&& \langle \Psi _{\tilde{N}}^{k,(m)}(\mathbf{r}_{1},\ldots)| \frac{e^{2}}{%
\varepsilon |\mathbf{r}_{i}-\mathbf{R}_{j}- k L_{x}^{\square} \hat{\mathbf{x}%
}|} |\Psi _{\tilde{N}}^{n,(m)}(\mathbf{r}_{1},\ldots) \rangle  \notag \\
&&= \delta_{k,n} \; \langle \Psi _{\tilde{N}}^{n,(m)}| \frac{e^{2}}{%
\varepsilon |\mathbf{r}_{i}-\mathbf{R}_{j}- k L_{x}^{\square} \hat{\mathbf{x}%
}|} |\Psi _{\tilde{N}}^{n,(m)} \rangle ,  \label{49}
\end{eqnarray}
which is valid for $\tilde{N} \geq 2$. By making calculations analogous to
ones of the HFA and the Hartree approximation \cite{ansel'm1978,madelung1981}%
, from Eq. (\ref{48}) we obtain 
\begin{eqnarray}
E_{ei}&=&-\frac{1}{m \tilde{N}} \sum_{n=-\ell}^{\ell} \sum_{i=1}^{\tilde{N}}
\sum_{j=1}^{\tilde{N}} \sum_{k=-N_{C}}^{N_{C}} \int d \mathbf{r} \int d 
\mathbf{R}  \notag \\
&&\times \frac{e^{2}/\varepsilon}{|\mathbf{r}- \mathbf{R}-k L_{x}^{\square} 
\hat{\mathbf{x}}|} |\varphi_{k_{xi}^{(n)}}^{(m)}(\mathbf{r})|^{2}
|\phi_{n_{ys}^{(j)}}(\mathbf{R})|^{2} .  \label{50}
\end{eqnarray}

In the RHS of Eq. (\ref{40}) the term related with electron-electron
interaction 
\begin{equation}
\langle \Psi _{\tilde{N},\tilde{N}}^{(m),eh}|V_{ee}|\Psi _{\tilde{N},\tilde{N%
}}^{(m),eh}\rangle =\tilde{N}(E_{ee}^{a}+E_{ee}^{b}),  \label{51}
\end{equation}%
where, due to the second term in the RHS of Eq. (\ref{3}), 
\begin{equation}
E_{ee}^{a}=\sum_{k=1}^{N_{C}}\frac{e^{2}}{\varepsilon L_{x}^{\square }k}
\label{52}
\end{equation}%
and, due to the first term in the RHS of Eq. (\ref{3}), 
\begin{eqnarray}
E_{ee}^{b} &=&\frac{1}{2\tilde{N}}\sum_{i=1}^{\tilde{N}}\sum_{j=1,j\neq i}^{%
\tilde{N}}\sum_{k=-N_{C}}^{N_{C}}  \notag \\
&&\times \langle \Psi _{\tilde{N},\tilde{N}}^{(m),eh}|\frac{e^{2}}{%
\varepsilon |\mathbf{r}_{i}-\mathbf{r}_{j}-kL_{x}^{\square }\hat{\mathbf{x}}|%
}|\Psi _{\tilde{N},\tilde{N}}^{(m),eh}\rangle  \notag \\
&=&\frac{1}{2m\tilde{N}}\sum_{n=-\ell }^{\ell }\sum_{i=1}^{\tilde{N}%
}\sum_{j=1,j\neq i}^{\tilde{N}}\sum_{k=-N_{C}}^{N_{C}}  \notag \\
&&\times \langle \Psi _{\tilde{N}}^{n,(m)}|\frac{e^{2}}{\varepsilon |\mathbf{%
r}_{i}-\mathbf{r}_{j}-kL_{x}^{\square }\hat{\mathbf{x}}|}|\Psi _{\tilde{N}%
}^{n,(m)}\rangle .  \label{53}
\end{eqnarray}%
In Eq. (\ref{53}) we have used Eqs. (\ref{18})-(\ref{20}) and (cf. with Eq. (%
\ref{49})) the property 
\begin{eqnarray}
&&\langle \Psi _{\tilde{N}}^{k,(m)}(\mathbf{r}_{1},\ldots )|\frac{e^{2}}{%
\varepsilon |\mathbf{r}_{i}-\mathbf{r}_{j}-kL_{x}^{\square }\hat{\mathbf{x}}|%
}|\Psi _{\tilde{N}}^{n,(m)}(\mathbf{r}_{1},\ldots )\rangle  \notag \\
&=&\delta _{k,n}\;\langle \Psi _{\tilde{N}}^{n,(m)}|\frac{e^{2}}{\varepsilon
|\mathbf{r}_{i}-\mathbf{r}_{j}-kL_{x}^{\square }\hat{\mathbf{x}}|}|\Psi _{%
\tilde{N}}^{n,(m)}\rangle ,  \label{54}
\end{eqnarray}%
which is valid for $\tilde{N}\geq 3$ (remind, in the present study it is
assumed that $\tilde{N}\rightarrow \infty $). As in the RHS of Eq. (\ref{53}%
) the matrix elements are calculated on many-electron wave functions Eq. (%
\ref{20}) of \textquotedblleft Hartree-Fock\textquotedblright -alike form,
by making typical HFA calculations \cite{ansel'm1978,madelung1981} of the
matrix elements, we rewrite Eq. (\ref{53}) as 
\begin{equation}
E_{ee}^{b}=E_{ee}^{di}+E_{ee}^{xc},  \label{55}
\end{equation}%
where the direct-alike (or the Hartree-alike) contribution 
\begin{eqnarray}
E_{ee}^{di} &=&\frac{1}{2m\tilde{N}}\sum_{n=-\ell }^{\ell }\sum_{i=1}^{%
\tilde{N}}\sum_{j=1,j\neq i}^{\tilde{N}}\sum_{k=-N_{C}}^{N_{C}}\int d\mathbf{%
r}\int d\mathbf{r}^{\prime }  \notag \\
&&\times \frac{e^{2}/\varepsilon }{|\mathbf{r}-\mathbf{r}^{\prime
}-kL_{x}^{\square }\hat{\mathbf{x}}|}|\varphi _{k_{xi}^{(n)}}^{(m)}(\mathbf{r%
})|^{2}|\varphi _{k_{xj}^{(n)}}^{(m)}(\mathbf{r}^{\prime })|^{2},  \label{56}
\end{eqnarray}%
and the exchange-alike (or the Fock-alike) contribution 
\begin{eqnarray}
E_{ee}^{xc} &=&-\frac{1}{2m\tilde{N}}\sum_{n=-\ell }^{\ell }\sum_{i=1}^{%
\tilde{N}}\sum_{j=1,j\neq i}^{\tilde{N}}\sum_{k=-N_{C}}^{N_{C}}\int d\mathbf{%
r}\int d\mathbf{r}^{\prime }  \notag \\
&&\times \frac{e^{2}/\varepsilon }{|\mathbf{r}-\mathbf{r}^{\prime
}-kL_{x}^{\square }\hat{\mathbf{x}}|}\;\;\varphi _{k_{xi}^{(n)}}^{(m)\ast }(%
\mathbf{r})\;\varphi _{k_{xj}^{(n)}}^{(m)}(\mathbf{r})  \notag \\
&&\times \;\varphi _{k_{xi}^{(n)}}^{(m)}(\mathbf{r}^{\prime })\;\;\varphi
_{k_{xj}^{(n)}}^{(m)\ast }(\mathbf{r}^{\prime }).  \label{57}
\end{eqnarray}%
Notice, the direct-alike term, Eq. (\ref{56}), is always positive and the
exchange-alike term, Eq. (\ref{57}), is always negative. Point out that
strictly speaking these two terms cannot be called as the direct and the
exchange ones because some correlations are already taken into account in
the form of many-body electron-ion wave function $\Psi _{\tilde{N},\tilde{N}%
}^{(m),eh}$.

Point out, in final expressions of Eqs. (\ref{46}), (\ref{56}), (\ref{57})
the sums over $i$ and $j$ have the additional condition, $j \neq i$;
however, it is important that this condition in Eq. (\ref{50}) is absent.
Remind that $\int d \mathbf{r}\ldots=
\int_{L_{x}^{\square}(n_{xs}^{\alpha}-1)}^{L_{x}^{\square}n_{xs}^{\alpha}}
dx \int_{-\infty}^{\infty} dy \ldots $.

The sum of direct and direct-alike terms 
\begin{equation}
E^{da}=E_{ee}^{a}+E_{ii}^{a}+E_{ii}^{b}+E_{ee}^{di}+E_{ei} ,  \label{58}
\end{equation}%
we rewrite (by, quite obvious, exact transformations) as follows 
\begin{equation}
E^{da}=E_{1}^{(m)}+E_{2}^{(m)}+E_{3}^{(m)} ,  \label{59}
\end{equation}%
where $E_{ee}^{a}$, $E_{ii}^{a}$, and the "diagonal" terms , $j=i$, from $%
E_{ei}$, Eq. (\ref{50}), for $k \neq 0$, have contributed to
\begin{eqnarray}
E_{1}^{(m)}&=&\frac{2e^{2}}{\varepsilon L_{x}^{\square}} \sum_{k=1}^{N_{C}}%
\frac{1}{k} -\frac{1}{m \tilde{N}} \sum_{n=-\ell }^{\ell } \sum_{i=1}^{%
\tilde{N}} \sum_{k=-N_{C}; k \neq 0}^{N_{C}} \int d \mathbf{r} \int d 
\mathbf{R}  \notag \\
&&\times \frac{e^{2}}{\varepsilon |\mathbf{r}- \mathbf{R}-k L_{x}^{\square} 
\hat{\mathbf{x}}|} |\varphi_{k_{xi}^{(n)}}^{(m)}(\mathbf{r})|^{2}
|\phi_{n_{ys}^{(i)}}(\mathbf{R})|^{2} ,  \label{60}
\end{eqnarray}%
further, the only left, however, very important "diagonal" term, $j=i$, from 
$E_{ei}$, Eq. (\ref{50}), for $k=0$, have contributed to
\begin{eqnarray}
E_{2}^{(m)}&=&-\frac{1}{m \tilde{N}} \sum_{n=-\ell }^{\ell } \sum_{i=1}^{%
\tilde{N}} \int d \mathbf{r} \int d \mathbf{R}  \notag \\
&&\times \frac{e^{2}}{\varepsilon |\mathbf{r}- \mathbf{R}|}
|\varphi_{k_{xi}^{(n)}}^{(m)}(\mathbf{r})|^{2} |\phi_{n_{ys}^{(i)}}(\mathbf{R%
})|^{2} ,  \label{61}
\end{eqnarray}%
and $E_{ii}^{b}$, $E_{ee}^{di}$ and all the rest "nondiagonal" terms, $j
\neq i$, from $E_{ei}$, Eq. (\ref{50}), for any $k$ from $-N_{C}$ to $N_{C}$%
, have given
\begin{eqnarray}
E_{3}^{(m)}&=&\frac{1}{2m \tilde{N}} \sum_{n=-\ell }^{\ell } \sum_{i=1}^{%
\tilde{N}} \sum_{j=1, j \neq i}^{\tilde{N}} \sum_{k=-N_{C}}^{N_{C}} \int d 
\mathbf{r} \int d \mathbf{r}^{\prime}  \notag \\
&&\times \frac{e^{2}/\varepsilon}{ |\mathbf{r}-\mathbf{r}^{\prime}- k
L_{x}^{\square} \hat{\mathbf{x}}|} [|\phi_{n_{ys}^{(i)}}(\mathbf{r})|^{2}-
|\varphi_{k_{xi}^{(n)}}^{(m)}(\mathbf{r})|^{2}]  \notag \\
&& \times [|\phi_{n_{ys}^{(j)}}(\mathbf{r}^{\prime})|^{2}-
|\varphi_{k_{xj}^{(n)}}^{(m)}(\mathbf{r}^{\prime})|^{2}] .  \label{62}
\end{eqnarray}

As an $i-$th term in the sums over $i$ in Eqs. (\ref{60})-(\ref{62}) is
independent of $i=n_{ys}^{(i)}$, these equations can be rewritten as 
\begin{eqnarray}
E_{1}^{(m)}&=&\frac{2e^{2}}{\varepsilon L_{x}^{\square}} \sum_{k=1}^{N_{C}}%
\frac{1}{k} -\frac{1}{m} \sum_{n=-\ell }^{\ell } \sum_{k=-N_{C}; k \neq
0}^{N_{C}} \int d \mathbf{r} \int d \mathbf{R}  \notag \\
&&\times \frac{e^{2}}{\varepsilon |\mathbf{r}- \mathbf{R}-k L_{x}^{\square} 
\hat{\mathbf{x}}|} |\varphi_{k_{xi}^{(n)}}^{(m)}(\mathbf{r})|^{2}
|\phi_{n_{ys}^{(i)}}(\mathbf{R})|^{2} ,  \label{63}
\end{eqnarray}%
\begin{eqnarray}
E_{2}^{(m)}&=&-\frac{1}{m} \sum_{n=-\ell }^{\ell } \int d \mathbf{r} \int d
\mathbf{R}  \notag \\
&&\times \frac{e^{2}}{\varepsilon |\mathbf{r}- \mathbf{R}|}
|\varphi_{k_{xi}^{(n)}}^{(m)}(\mathbf{r})|^{2} |\phi_{n_{ys}^{(i)}}(\mathbf{R%
})|^{2} ,  \label{64}
\end{eqnarray}%
and 
\begin{eqnarray}
E_{3}^{(m)}&=&\frac{1}{2m} \sum_{n=-\ell }^{\ell } \sum_{j=1, j \neq i}^{%
\tilde{N}} \sum_{k=-N_{C}}^{N_{C}} \int d \mathbf{r} \int d \mathbf{r}%
^{\prime}  \notag \\
&&\times \frac{e^{2}/\varepsilon}{ |\mathbf{r}-\mathbf{r}^{\prime}- k
L_{x}^{\square} \hat{\mathbf{x}}|} [|\phi_{n_{ys}^{(i)}}(\mathbf{r})|^{2}-
|\varphi_{k_{xi}^{(n)}}^{(m)}(\mathbf{r})|^{2}]  \notag \\
&& \times [|\phi_{n_{ys}^{(j)}}(\mathbf{r}^{\prime})|^{2}-
|\varphi_{k_{xj}^{(n)}}^{(m)}(\mathbf{r}^{\prime})|^{2}] .  \label{65}
\end{eqnarray}%
As it can be expected, in particular, we will see that in Eqs. (\ref{60})-(%
\ref{65}) their RHS are actually independent of $n_{xs}^{\alpha}$.

Point out, it easy to see that a weak logarithmic divergence of the first
sum over $k$ in the RHS of Eq. (\ref{63}), that appears due to $k\gg 1$, is
exactly canceled by the main contributions to the second sum over the $k$ in
the RHS of Eq. (\ref{63}), at $|k|\gg 1$. Such that the total sum over $k$
in the RHS of Eq. (\ref{63}) will quickly converge with the increase of $|k|$%
. Indeed, for $|k|\rightarrow \infty $ in the second term of the RHS of Eq. (%
\ref{63}) as $|\mathbf{r}-\mathbf{R}|\alt L_{x}^{\square }$, due to
properties of the factors given by the squares of the single-electron and
the single-ion wave functions involved, we have in a very good approximation
that $e^{2}/(\varepsilon |\mathbf{r}-\mathbf{R}-kL_{x}^{\square }\hat{%
\mathbf{x}}|)\approx e^{2}/(\varepsilon |k|L_{x}^{\square })$. Using the
latter constant expression in the second term of Eq. (\ref{63}) leads to the
mutual canceling with the relevant $k-$term of the first contribution to the
RHS of Eq. (\ref{63}). By similar considerations, it is easy to see that the
sum over $k$ in the RHS of Eq. (\ref{65}) is quickly convergent. As a
result, we can in Eqs. (\ref{63}), (\ref{65}) and, respectively, (\ref{60}),
(\ref{62}) to assume that $N_{C}\rightarrow \infty $; it is in agreement
with pertinent discussion in Sec. II.

To calculate the RHS of Eqs. (\ref{63}), (\ref{64}) we will use that 
\begin{equation}
\frac{e^{2}}{\varepsilon |\mathbf{r}-\mathbf{R}|} =\frac{e^{2}}{2\pi
\varepsilon} \int_{-\infty}^{\infty} d q_{x} \int_{-\infty}^{\infty} d q_{y}
\frac{e^{i \mathbf{q} (\mathbf{r}-\mathbf{R})}}{ \sqrt{q_{x}^{2}+q_{y}^{2}+%
\delta^{2}/\ell_{0}^{2}}} ,  \label{66}
\end{equation}%
where, e.g., cf. with \cite{kittel1987}, a dimensionless $\delta \rightarrow
0$ and it is implicit that $L_{\mu} \rightarrow \infty$ such that $%
\ell_{0}/(\delta \times L_{\mu}) \rightarrow 0$, $\mu=x, y$. Respectively,
we have that $L_{x}^{\square}/(\delta \times L_{\mu}) \rightarrow 0$ as only
finite $m$ are treated. In addition, we will use the matrix elements 
\begin{eqnarray}
&&\int d \mathbf{r} \; e^{i \mathbf{q} \mathbf{r}} \; |\varphi
_{k_{xi}^{(n)}}^{(m)}(\mathbf{r})|^{2} =\exp(i q_{y}k_{xi}^{(n)}
\ell_{0}^{2}-q_{y}^{2} \ell_{0}^{2}/4)  \notag \\
&&\times \frac{\exp(i q_{x} L_{x}^{\square} n_{xs}^{\alpha})}{ i q_{x}
L_{x}^{\square} } [1-\exp(-i q_{x} L_{x}^{\square})] ,  \label{67}
\end{eqnarray}%
and 
\begin{eqnarray}
&&\int d \mathbf{R} \; e^{-i \mathbf{q} \mathbf{R}} \; |\phi _{n_{ys}^{(i)}}(%
\mathbf{R})|^{2} = e^{-iq_{y}L_{x}^{\square} n_{ys}^{(i)}} \; S_{m}(q_{y}
\ell_{0})  \notag \\
&&\times \frac{\exp(-i q_{x} L_{x}^{\square} n_{xs}^{\alpha})}{-i q_{x}
L_{x}^{\square} } [1-\exp(i q_{x} L_{x}^{\square})] ,  \label{68}
\end{eqnarray}%
where$^{\cite{balev2005}}$ $S_{m}(q_{\mu} \ell_{0})= \sin(q_{\mu}
L_{x}^{\square}/2)/(q_{\mu} L_{x}^{\square}/2)= \sin(\sqrt{\pi m/2} \;
q_{\mu} \ell_{0})/(\sqrt{\pi m/2} \; q_{\mu} \ell_{0})$; it is even
function. In particular, in the LHS of Eq. (\ref{67}) i) the integral over $%
x $ is calculated as $\int_{L_{x}^{\square}(n_{xs}^{\alpha}-1)}^{L_{x}^{%
\square}n_{xs}^{\alpha}} e^{i q_{x}x} \;dx/L_{x}^{\square}= e^{iq_{x}
L_{x}^{\square} n_{xs}^{\alpha}} [1-e^{-iq_{x} L_{x}^{\square}}]/(i q_{x}
L_{x}^{\square})$ and ii) the integral over $y$ is calculated as \cite%
{balev1997} $\int_{-\infty}^{\infty} dy \; e^{i q_{y} y} \;
\Psi_{0}^{2}(y-y_{0}(k_{xi}^{(n)}))= \exp(-\frac{1}{4} q_{y}^{2}
\ell_{0}^{2}+i q_{y}k_{xi}^{(n)} \ell_{0}^{2})$.

After using of Eqs. (\ref{66})-(\ref{68}), we rewrite Eq. (\ref{64}) as 
\begin{equation}
E_{2}^{(m)}=\frac{e^{2}}{\varepsilon \ell _{0}} F_{1}^{C}(m) ,  \label{69}
\end{equation}%
where (this, always negative, function is introduced in Ref. $^{\cite%
{balev2005}}$), for $\xi=q_{x} \ell_{0}$ and $\eta=q_{y}\ell_{0}$, we have
that 
\begin{eqnarray}
F^{C}_{1}(m)&=&-\;\frac{2}{\pi} \int_{0}^{\infty} d \xi \int_{0}^{\infty} d
\eta \frac{ e^{-\eta^{2}/4}}{\sqrt{\xi^{2}+\eta^{2}}} f_{m}(\eta)  \notag \\
&&\times S_{m}(\eta) \; S_{m}^{2}(\xi) ,  \label{70}
\end{eqnarray}%
where, as there is no any finite contribution (or divergence) in the RHS for 
$\delta \rightarrow 0$, we have neglected by $\delta^{2}$ in the factor $%
(\xi^{2}+\eta^{2}+\delta^{2})^{-1/2}$. Here (cf. with Ref. \cite{balev2005}) 
$f_{1}(\eta)=1$ and, for $m \geq 3$, the even function
\begin{equation}
f_{m}(\eta)=\frac{1}{m}\left[1+2\sum_{n=1}^{\ell} \cos(\sqrt{\frac{2\pi}{m}}
\; n \; \eta ) \right] .  \label{71}
\end{equation}%
From Eq. (\ref{70}) we calculate: $F^{C}_{1}(1) \approx -1.184787$, $%
F^{C}_{1}(3) \approx - 0.665565$, $F^{C}_{1}(5) \approx -0.518796$, and $%
F^{C}_{1}(7) \approx -0.440366$; notice, these values for $F^{C}_{1}(m)$
were previously obtained in Ref. \cite{balev2005}.

Using Eqs. (\ref{66})-(\ref{68}), we rewrite Eq. (\ref{63}) as
\begin{equation}
E_{1}^{(m)}=\frac{e^{2}}{\varepsilon \ell _{0}}D(m),  \label{72}
\end{equation}%
where 
\begin{equation}
D(m)=\sum_{k=1}^{\infty }D_{1}^{(m)}(k),  \label{73}
\end{equation}%
and
\begin{eqnarray}
D_{1}^{(m)}(k) &=&\sqrt{\frac{2}{m\pi }}\frac{1}{k}-\frac{4}{\pi }%
\int_{0}^{\infty }d\xi \int_{0}^{\infty }d\eta \frac{f_{m}(\eta )}{\sqrt{\xi
^{2}+\eta ^{2}}}  \notag \\
&&\times e^{-\eta ^{2}/4}S_{m}(\eta )\;S_{m}^{2}(\xi )\;\cos [k\sqrt{2\pi m}%
\;\xi ].  \label{74}
\end{eqnarray}%
Point out that the sum in the RHS of Eq. (\ref{73}) is rapidly convergent
as, e.g., $D_{1}^{(m)}(2)/D_{1}^{(m)}(1)<0.1$ and $%
D_{1}^{(m)}(5)/D_{1}^{(m)}(2)<0.1$, for $m=1,3,5,7$; see also the paragraph
above Eq. (\ref{66}). From Eqs. (\ref{73}), (\ref{74}) we calculate: $%
D(1)\approx -0.10661$ (for this precision it is enough to include the first
seventeen terms in the sum of Eq. (\ref{73})), $D(3)\approx -0.05386$ (for
this precision it is enough to include the first fourteen terms of the sum), 
$D(5)\approx -0.04282$ (for this precision it is enough to include the first
fourteen terms of the sum), and $D(7)\approx -0.03678$ (for this precision
it is enough to include the first twelve terms of the sum).

Using Eqs. (\ref{66})-(\ref{68}), we rewrite Eq. (\ref{65}) as
\begin{eqnarray}
E_{3}^{(m)}&=&\frac{e^{2}}{4 \pi \varepsilon} \int_{-\infty}^{\infty} d
q_{x} \int_{-\infty}^{\infty} d q_{y} \frac{ g_{m}(q_{y} \ell_{0})}{\sqrt{%
q_{x}^{2}+q_{y}^{2}+\delta^{2}/\ell_{0}^{2}}}  \notag \\
&& \times S_{m}^{2}(q_{x}\ell_{0}) \; \; [\sum_{k=-\infty}^{\infty} e^{-ik
q_{x} L_{x}^{\square}}]  \notag \\
&& \times \{\sum_{m_{y}=-\infty, m_{y} \neq 0}^{\infty} e^{-im_{y} q_{y}
L_{x}^{\square}}\} ,  \label{75}
\end{eqnarray}%
where (this even function is introduced in Ref. \cite{balev2005}) 
\begin{equation}
g_{m}(\eta)=S_{m}^{2}(\eta)+e^{-\eta^{2}/2} -2 e^{-\eta^{2}/4} \;
f_{m}(\eta)\; S_{m}(\eta) .  \label{76}
\end{equation}%
Notice, $g_{1}(\eta)=[\exp(-\eta^{2}/4)-S_{1}(\eta)]^{2}$. Further, using
Eq. (\ref{28}) we express in the RHS of Eq. (\ref{75}) the product of the
sums in the square and the curly brackets as
\begin{eqnarray}
&&\frac{2 \pi}{L_{x}^{\square}} \sum_{M_{x}=-\infty}^{\infty}
\delta(q_{x}+M_{x} \; \frac{2 \pi}{L_{x}^{\square}})  \notag \\
&& \times \{ [\frac{2 \pi}{L_{x}^{\square}} \sum_{M_{y}=-\infty}^{\infty}
\delta(q_{y}+M_{y} \; \frac{2 \pi}{L_{x}^{\square}}) ] -1\} .  \label{77}
\end{eqnarray}%
Using Eq. (\ref{77}) in Eq. (\ref{76}) and calculating the integrals with
the help of delta-functions we obtain
\begin{equation}
E_{3}^{(m)}=E_{3}^{(m),a}+E_{3}^{(m),b} ,  \label{78}
\end{equation}%
where 
\begin{eqnarray}
E_{3}^{(m),a}&=&\frac{e^{2}}{2 \varepsilon L_{x}^{\square}}
\sum_{M_{x}=-\infty}^{\infty} \sum_{M_{y}=-\infty}^{\infty} S_{m}^{2}(\sqrt{%
2\pi/m} \; M_{x})  \notag \\
&& \times \frac{ g_{m}(\sqrt{2\pi/m} \; M_{y})}{\sqrt{%
M_{x}^{2}+M_{y}^{2}+(m/2\pi) \delta^{2}}} ,  \label{79}
\end{eqnarray}%
and 
\begin{eqnarray}
E_{3}^{(m),b}&=&-\frac{e^{2}}{\varepsilon L_{x}^{\square}}
\sum_{M_{x}=-\infty}^{\infty} S_{m}^{2}(\sqrt{2\pi/m} \; M_{x})
\int_{0}^{\infty} d q_{y}  \notag \\
&& \times \frac{ g_{m}(q_{y} \ell_{0})}{\sqrt{(2\pi/L_{x}^{\square})^{2}
M_{x}^{2}+q_{y}^{2}+ \delta^{2}/\ell_{0}^{2}}} .  \label{80}
\end{eqnarray}%
Taking into account that $S_{m}(\sqrt{2\pi/m} M)= \sin(\pi M)/(\pi M)=0$ for 
$M \neq 0$ and, in addition, that $S_{m}(0)=1$ for $M=0$, we obtain that in
the RHS of Eqs. (\ref{79}), (\ref{80}) only the $M_{x}=0$ term, of the sums
over $M_{x}$, is contributed. Then Eq. (\ref{80}) obtains the form 
\begin{equation}
E_{3}^{(m),b}=-\frac{e^{2}}{\sqrt{2 \pi m} \; \varepsilon \ell_{0}}
\int_{0}^{\infty} d \eta \frac{ g_{m}(\eta)}{\eta} ,  \label{81}
\end{equation}%
where it is used that under the integral in $g_{m}(\eta)/\sqrt{%
\eta^{2}+\delta^{2}}$ it is safe to neglect by $\delta^{2} \rightarrow 0$
as, for $\eta \ll 1$, $g_{m}(\eta)/\eta \propto \eta$ at $m=3, 5, \ldots$
and $g_{m}(\eta)/\eta \propto \eta^{3}$ at $m=1$. Further, in the sum over $%
M_{y}$ in the RHS of Eq. (\ref{79}) it is easy to see that $M_{y}=0$ term
vanish as it is $\propto g_{m}(0)/(\sqrt{m/2\pi}\;\delta) =0$, where $%
g_{m}(0)=0$ and $\delta \neq 0$. Then finally we rewrite Eq. (\ref{79}) as
\begin{eqnarray}
E_{3}^{(m),a}&=&\frac{e^{2}}{\sqrt{2 \pi m} \; \varepsilon \ell_{0}}
\sum_{k=1}^{\infty} \frac{1}{k} \; e^{-\pi \; k^{2}/m} ,  \label{82}
\end{eqnarray}%
where it is used that $g_{m}(\sqrt{2\pi/m} \; k)=\exp(-\pi \; k^{2}/m)$, for
$k \neq 0$. Then using Eqs. (\ref{81}), (\ref{82}) we rewrite Eq. (\ref{78})
in the form
\begin{equation}
E_{3}^{(m)}=\frac{e^{2}}{\varepsilon \ell_{0}} \Delta \tilde{F}^{C}_{1}(m) ,
\label{83}
\end{equation}%
where 
\begin{equation}
\Delta \tilde{F}^{C}_{1}(m)=\frac{1}{\sqrt{2 \pi \; m}} (
\sum_{k=1}^{\infty} \frac{1}{k} \; e^{-\pi \; k^{2}/m} - \int_{0}^{\infty} d
\eta \frac{ g_{m}(\eta)}{\eta}) .  \label{84}
\end{equation}%
From Eqs. (\ref{84}) simple numerical calculations give that: $\Delta \tilde{%
F}^{C}_{1}(1) \approx -0.0021047$, $\Delta \tilde{F}^{C}_{1}(3) \approx
-0.0812376$, $\Delta \tilde{F}^{C}_{1}(5) \approx -0.0654775$, and $\Delta 
\tilde{F}^{C}_{1}(7) \approx -0.0552258$. Now, Eqs. (\ref{59}), (\ref{69})-(%
\ref{74}), (\ref{83})-(\ref{84}) give for $E^{da}(m)=(e^{2}/\varepsilon
\ell_{0})U^{da}_{C}(m)$ the analytical expression, $E^{da}(m) \equiv E^{da}$%
. Then above given numerical results shows that $U^{da}_{C}(1) \approx
-1.29350$, $U^{da}_{C}(3) \approx -0.80066$, $U^{da}_{C}(5) \approx -0.62709$%
, and $U^{da}_{C}(7) \approx -0.53237$.

To calculate $E_{ee}^{xc}$, in the RHS of Eq. (\ref{57}) we use that an $i-$%
th term in the sum over $i$ is independent of $i=n_{ys}^{(i)}$. I.e., in the
RHS of Eq. (\ref{57}) $\tilde{N}^{-1}\sum_{i=1}^{\tilde{N}}A_{i}\rightarrow
A_{i}$. Then Eq. (\ref{57}) can be rewritten, by using Eq. (\ref{66}), as
\begin{eqnarray}
E_{ee}^{xc} &=&-\frac{e^{2}}{4m\pi \varepsilon }\sum_{n=-\ell }^{\ell
}\sum_{j=1;j\neq i}^{\tilde{N}}\sum_{k=-\infty }^{\infty }\int_{-\infty
}^{\infty }dq_{x}\int_{-\infty }^{\infty }dq_{y}  \notag \\
&&\times \frac{e^{-ikq_{x}L_{x}^{\square }}}{\sqrt{q_{x}^{2}+q_{y}^{2}+%
\delta ^{2}/\ell _{0}^{2}}}\;|M(\mathbf{q};k_{xi}^{(n)},k_{xj}^{(n)})|^{2},
\label{85}
\end{eqnarray}%
where the matrix element 
\begin{eqnarray}
&&M(\mathbf{q};k_{xi}^{(n)},k_{xj}^{(n)})=\int d\mathbf{r}e^{i\mathbf{q}%
\mathbf{r}}\varphi _{k_{xi}^{(n)}}^{(m)\ast }(\mathbf{r})\varphi
_{k_{xj}^{(n)}}^{(m)}(\mathbf{r})  \notag \\
&=&\exp
\{[2iq_{y}(k_{xj}^{(n)}+k_{xi}^{(n)})-(k_{xj}^{(n)}-k_{xi}^{(n)})^{2}-q_{y}^{2}]\ell _{0}^{2}/4\}
\notag \\
&&\times \frac{\exp [i(q_{x}+k_{xj}^{(n)}-k_{xi}^{(n)})L_{x}^{\square
}n_{xs}^{\alpha }]}{i(q_{x}+k_{xj}^{(n)}-k_{xi}^{(n)})L_{x}^{\square }}
\notag \\
&&\times \lbrack 1-\exp (-i(q_{x}+k_{xj}^{(n)}-k_{xi}^{(n)})L_{x}^{\square
})].  \label{86}
\end{eqnarray}%
After using Eq. (\ref{86}) in Eq. (\ref{85}), it follows that the factor $|M(%
\mathbf{q};k_{xi}^{(n)},k_{xj}^{(n)})|^{2}$ in Eq. (\ref{85}) is independent
of $n_{xs}^{\alpha }$ and $k_{xi}^{(n)}$. However, it is dependent on $%
\mathbf{q}$ and $\Delta k=k_{xj}^{(n)}-k_{xi}^{(n)}=(\sqrt{2\pi m}/\ell
_{0})(n_{ys}^{(j)}-n_{ys}^{(i)})$; the latter is independent of $n$.
Introducing $m_{y}=n_{ys}^{(j)}-n_{ys}^{(i)}$, we have that $\Delta k=(\sqrt{%
2\pi m}/\ell _{0})m_{y}$, where $m_{y}\neq 0$. It is important to point out
that from above it follows that in the RHS of Eq. (\ref{85}) the $n-$th term
of the sum over $n$ is independent of $n$. Further, using Eq. (\ref{28}) we
readily rewrite Eq. (\ref{85}) as follows 
\begin{eqnarray}
E_{ee}^{xc} &=&-\frac{e^{2}}{\varepsilon L_{x}^{\square }}%
\sum_{M_{x}=-\infty }^{\infty }\sum_{m_{y}=-\infty ;m_{y}\neq 0}^{\infty
}e^{-\pi mm_{y}^{2}}  \notag \\
&&\times \frac{\sin ^{2}[\pi (M_{x}-mm_{y})]}{[\pi (M_{x}-mm_{y})]^{2}}%
\;\int_{0}^{\infty }dq_{y}  \notag \\
&&\times \frac{e^{-q_{y}^{2}\ell _{0}^{2}/2}}{\sqrt{(2\pi /L_{x}^{\square
})^{2}M_{x}^{2}+q_{y}^{2}+\delta ^{2}/\ell _{0}^{2}}},  \label{87}
\end{eqnarray}%
where the integral over $q_{x}$ is already carried out, using obtained
delta-functions. It is readily seen that, due to $m_{y}\neq 0$, the term $%
M_{x}=0$ does not contribute to the RHS of Eq. (\ref{87}). I.e., the sum
over all $M_{x}$ can be substituted only by the sum over $M_{x}\neq 0$,
i.e., by $M_{x}=\pm 1,\pm 2,\ldots $. Further, in the RHS of Eq. (87) we
obtain that 
\begin{equation}
\frac{\sin ^{2}[\pi (M_{x}-mm_{y})]}{[\pi (M_{x}-mm_{y})]^{2}}=\delta
_{M_{x},mm_{y}},  \label{88}
\end{equation}%
i.e., reduces to the Kronecker delta symbol. Then using Eq. (\ref{88}) in
Eq. (\ref{87}) we readily obtain 
\begin{equation}
E_{ee}^{xc}=-\frac{e^{2}}{\varepsilon \ell _{0}}F_{2}(m),  \label{89}
\end{equation}%
where 
\begin{equation}
F_{2}(m)=\sqrt{\frac{2}{m\pi }}\sum_{k=1}^{\infty }e^{-\pi
mk^{2}}\int_{0}^{\infty }d\eta \frac{e^{-\eta ^{2}/2}}{\sqrt{\eta ^{2}+2\pi
mk^{2}}}.  \label{90}
\end{equation}%
Notice, the sum over $k$ in the RHS of Eq. (\ref{90}) is very rapidly
convergent already for $m=1$: in the $k=1$ term $\approx 0.01618$ the $k=2$
term $\approx 6.83\times 10^{-7}$. Numerical calculations (obviously, quite
simple) give that $F_{2}(1)\approx 0.016183$, $F_{2}(3)\approx 1.0475\times
10^{-5}$, $F_{2}(5)\approx 1.18\times 10^{-8}$, and $F_{2}(7)\approx
1.6\times 10^{-11}$.

Then using Eqs. (\ref{40})-(\ref{44}), (\ref{47}), (\ref{51}), (\ref{52}), (%
\ref{55}), (\ref{58}), (\ref{59}), (\ref{69})-(\ref{74}), (\ref{83}), (\ref%
{84}), (\ref{89}), (\ref{90}) we can rewrite Eq. (\ref{40}) as 
\begin{equation}
E_{\tilde{N}}^{(m),eh}=\frac{\hbar \omega_{c}}{2} \tilde{N} + \frac{e^{2}
\tilde{N}}{\varepsilon \ell_{0}} \; U^{UB}(m) ,  \label{91}
\end{equation}%
where $U^{UB}(m)=D(m)+F_{1}^{C}(m)+\Delta \tilde{F}_{1}^{C}(m)-F_{2}(m)$.
Notice that $U^{UB}(m)$ is negative, due to many-body interactions, and
gives lowering of the total energy per electron in the units of $%
e^{2}/\varepsilon \ell_{0}$, i.e., $U^{UB}(m)=[E_{\tilde{N}}^{(m),eh}/\tilde{%
N}-\hbar \omega_{c}/2]/ (e^{2}/\varepsilon \ell_{0})$. Finally, we calculate 
\begin{eqnarray}
&& U^{UB}(1) \approx -1.30968, \;\; \; U^{UB}(3) \approx -0.80067  \notag \\
&& U^{UB}(5) \approx -0.62709, \; \; \; U^{UB}(7) \approx -0.53237 ,
\label{92}
\end{eqnarray}%
where after the decimal point only first five digits are kept. Even though
the values of $U^{UB}(1)$, $U^{UB}(3)$ and $U^{UB}(5)$ given by Eq. (\ref{92}%
) are much lower than pertinent total lowering at $\nu=1, \;1/3$, and $1/5$
for the Laughlin variational wave function\cite{laughlin1983} (i.e., $-\sqrt{%
\pi/8} \approx - 0.6267$, $-0.4156 \pm 0.0012$, and $-0.3340 \pm 0.0028$,
respectively), the comparison of these results by using their face value is
not too useful. In particular, due to the absence in $U^{UB}(m)$ of any
contribution from the self-interaction of an ion with itself while for IJB
model used in Ref. \cite{laughlin1983} (and many others studies) pertinent
contribution is present.

It is interesting that the results Eq. (\ref{92}) are quite close to
pertinent numerical results obtained, for $U^{C}(m)$, in Ref. \cite%
{balev2005} within framework that, however, contains some
oversimplifications the influence of which it is difficult to estimate
beforehand.

\subsection{Energy of ground-state $\Psi _{\tilde{N}}^{(m),JB}$, for IJB}

Using the Hamiltonian Eq. (\ref{1d}), we calculate the total energy of
electron-ion system (matrix elements should be calculated within the main
strip) in the state Eq. (\ref{18d}) as 
\begin{equation}
E_{\tilde{N}}^{(m),JB}=\langle \Psi _{\tilde{N}}^{(m),JB}| \hat{H}_{\tilde{N}%
}^{JB} |\Psi_{\tilde{N}}^{(m),JB}\rangle ,  \label{40d}
\end{equation}%
where in the RHS for the kinetic energy term, cf. with Eqs. (\ref{41})-(\ref%
{42}), it follows
\begin{equation}
\langle \Psi _{\tilde{N}}^{(m),JB}|\hat{H}_{0} | \Psi _{\tilde{N}%
}^{(m),JB}\rangle=\tilde{N} \frac{\hbar \omega_{c}}{2} .  \label{41d}
\end{equation}

In the RHS of Eq. (\ref{40d}) the term related with the $V_{bb}$, Eq. (\ref%
{3d}), interaction (we call it also IJB-IJB interaction) is given as 
\begin{equation}
\langle \Psi _{\tilde{N}}^{(m),JB}|V_{bb}|\Psi _{\tilde{N}}^{(m),JB}\rangle
\equiv V_{bb}=\tilde{N}(E_{bb}^{I}+E_{bb}^{II}),  \label{43d}
\end{equation}%
where it follows (now in the RHS of Eq. (\ref{43d}) all integrations we
reduce to MS) that
\begin{eqnarray}
E_{bb}^{I}&=& \frac{1}{2\tilde{N}} \sum_{i=1}^{\tilde{N}} \sum_{j=1}^{\tilde{%
N}} \int \int \frac{e^{2} d \mathbf{R} d \mathbf{R}^{\prime}}{\varepsilon |%
\mathbf{R}-\mathbf{R}^{\prime}|}  \notag \\
&&\;\;\; \times |\phi _{n_{ys}^{(i)}}(\mathbf{R})|^{2} \; \;
|\phi_{n_{ys}^{(j)}}(\mathbf{R}^{\prime})|^{2} ,  \label{44d}
\end{eqnarray}%
and 
\begin{eqnarray}
E_{bb}^{II}&=& \frac{1}{2\tilde{N}} \sum_{k=-N_{C}, k \neq 0}^{N_{C}}
\sum_{i=1}^{\tilde{N}} \sum_{j=1}^{\tilde{N}} \int \int \frac{e^{2} d 
\mathbf{R} d \mathbf{R}^{\prime}}{\varepsilon |\mathbf{R}-\mathbf{R}%
^{\prime}- k L_{x}^{\square} \hat{\mathbf{x}}|}  \notag \\
&&\;\;\; \times |\phi _{n_{ys}^{(i)}}(\mathbf{R})|^{2} \; \;
|\phi_{n_{ys}^{(j)}}(\mathbf{R}^{\prime})|^{2} ,  \label{45d}
\end{eqnarray}%
where we have used that, e.g., within MS given by Eq. (\ref{31})
$n_{io}^{eh}(\mathbf{R})=n_{b}(\mathbf{R})$, according to Sec. IV B. Point
out, in Eqs. (\ref{44d}), (\ref{45d}) the integrals are carried out within
MS as, e.g., instead of $\int_{MS} d \mathbf{R}$ we write $\int
d \mathbf{R}$.

In the RHS of Eq. (\ref{40d}) the term related with electron-ion
interaction, we call it also electron-IJB interaction, is given as
\begin{equation}
\langle \Psi _{\tilde{N}}^{(m),JB}|V_{eb}| \Psi _{\tilde{N}%
}^{(m),JB}\rangle= \tilde{N}E_{eb},  \label{47d}
\end{equation}%
where using Eqs. (\ref{2d}), (\ref{31}) and $n_{b}(\mathbf{R})=n_{io}^{eh}(%
\mathbf{R})$ we calculate that 
\begin{eqnarray}
E_{eb}&=&-\frac{1}{m \tilde{N}} \sum_{n=-\ell}^{\ell} \sum_{i=1}^{\tilde{N}}
\sum_{j=1}^{\tilde{N}} \sum_{k=-N_{C}}^{N_{C}} \int d \mathbf{r} \int d 
\mathbf{R}  \notag \\
&&\times \frac{e^{2}/\varepsilon}{|\mathbf{r}- \mathbf{R}-k L_{x}^{\square} 
\hat{\mathbf{x}}|} |\varphi_{k_{xi}^{(n)}}^{(m)}(\mathbf{r})|^{2}
|\phi_{n_{ys}^{(j)}}(\mathbf{R})|^{2} ,  \notag  \label{50a}
\end{eqnarray}
and comparing with Eq. (\ref{50}), it follows that
\begin{equation}
E_{eb}=E_{ei} .  \label{50d}
\end{equation}

In the RHS of Eq. (\ref{40d}) the term related with electron-electron
interaction
\begin{eqnarray}
\langle \Psi _{\tilde{N}}^{(m),JB}|V_{ee}|\Psi _{\tilde{N}}^{(m),JB}\rangle
& \equiv & \langle \Psi _{\tilde{N},\tilde{N}}^{(m),eh}| V_{ee}|\Psi _{%
\tilde{N},\tilde{N}}^{(m),eh}\rangle  \notag \\
&& =\tilde{N}(E_{ee}^{a}+E_{ee}^{b}),  \label{51d}
\end{eqnarray}%
where $E_{ee}^{a}$ and $E_{ee}^{b}$ are calculated in Sec. V A.

As in the RHS of Eqs. (\ref{44d}), (\ref{45d}) and (\ref{46}) the $i-$th
term is independent of $i=n_{ys}^{(i)}$, it follows that 
\begin{eqnarray}
E_{bb}^{I}&=& \frac{1}{2} \sum_{j=1}^{\tilde{N}} \int \int \frac{e^{2} d 
\mathbf{R} d \mathbf{R}^{\prime}}{\varepsilon |\mathbf{R}-\mathbf{R}%
^{\prime}|}  \notag \\
&&\;\;\; \times |\phi _{n_{ys}^{(i)}}(\mathbf{R})|^{2} \; \;
|\phi_{n_{ys}^{(j)}}(\mathbf{R}^{\prime})|^{2} ,  \label{52d}
\end{eqnarray}%
\begin{eqnarray}
E_{bb}^{II}&=& \frac{1}{2} \sum_{k=-N_{C}, k \neq 0}^{N_{C}} \sum_{j=1}^{%
\tilde{N}} \int \int \frac{e^{2} d \mathbf{R} d \mathbf{R}^{\prime}}{%
\varepsilon |\mathbf{R}-\mathbf{R}^{\prime}- k L_{x}^{\square} \hat{\mathbf{x%
}}|}  \notag \\
&&\;\;\; \times |\phi _{n_{ys}^{(i)}}(\mathbf{R})|^{2} \; \;
|\phi_{n_{ys}^{(j)}}(\mathbf{R}^{\prime})|^{2} ,  \label{53d}
\end{eqnarray}%
and
\begin{eqnarray}
E_{ii}^{b}&=& \frac{1}{2} \sum_{k=-N_{C}}^{N_{C}} \sum_{j=1, j \neq i}^{%
\tilde{N}} \int \int \frac{e^{2} d \mathbf{R} d \mathbf{R}^{\prime}}{%
\varepsilon |\mathbf{R}-\mathbf{R}^{\prime}- k L_{x}^{\square} \hat{\mathbf{x%
}}|}  \notag \\
&&\;\;\; \times |\phi _{n_{ys}^{(i)}}(\mathbf{R})|^{2} \; \;
|\phi_{n_{ys}^{(j)}}(\mathbf{R}^{\prime})|^{2} .  \label{46d}
\end{eqnarray}

Further, using Eqs. (\ref{40}), (\ref{40d}) and above results, we have that 
\begin{equation}
(E_{\tilde{N}}^{(m),JB}-E_{\tilde{N}}^{(m),eh})/\tilde{N}
=E_{bb}^{I}+E_{bb}^{II}-E_{ii}^{a}-E_{ii}^{b} ,  \label{54d}
\end{equation}%
First, in the RHS of Eq. (\ref{54d}) we rewrite $E_{bb}^{I}-E_{ii}^{b}$, by
separating in $E_{ii}^{b}$ the $k=0$ term and mutually canceling it with all
$j \neq i$ terms in $E_{bb}^{I}$, as follows 
\begin{eqnarray}
E_{bb}^{I}&-&E_{ii}^{b}=\frac{1}{2} \int \int \frac{e^{2} d \mathbf{R} d 
\mathbf{R}^{\prime}}{\varepsilon |\mathbf{R}-\mathbf{R}^{\prime}|} |\phi
_{n_{ys}^{(i)}}(\mathbf{R})|^{2} |\phi_{n_{ys}^{(i)}}(\mathbf{R}%
^{\prime})|^{2}  \notag \\
&&-\frac{1}{2} \sum_{k=-N_{C};k \neq 0}^{N_{C}} \sum_{j=1, j \neq i}^{\tilde{%
N}} \int \int \frac{e^{2} d \mathbf{R} d \mathbf{R}^{\prime}}{\varepsilon |%
\mathbf{R}-\mathbf{R}^{\prime}- k L_{x}^{\square} \hat{\mathbf{x}}|}  \notag
\\
&&\times |\phi _{n_{ys}^{(i)}}(\mathbf{R})|^{2} \; |\phi_{n_{ys}^{(j)}}(%
\mathbf{R}^{\prime})|^{2} .  \label{55d}
\end{eqnarray}%
Now it is easy to see from Eqs. (\ref{55d}) and (\ref{53d}) that 
\begin{eqnarray}
&&E_{bb}^{I}-E_{ii}^{b}+E_{bb}^{II}= \frac{1}{2} \int \int \frac{e^{2} d 
\mathbf{R} d \mathbf{R}^{\prime}}{\varepsilon |\mathbf{R}-\mathbf{R}%
^{\prime}|} |\phi _{n_{ys}^{(i)}}(\mathbf{R})|^{2}  \notag \\
&&\times|\phi_{n_{ys}^{(i)}}(\mathbf{R}^{\prime})|^{2}+ \frac{1}{2}
\sum_{k=-N_{C};k \neq 0}^{N_{C}} \int \int \frac{e^{2} d \mathbf{R} d
\mathbf{R}^{\prime}}{\varepsilon |\mathbf{R}-\mathbf{R}^{\prime}- k
L_{x}^{\square} \hat{\mathbf{x}}|}  \notag \\
&&\times |\phi _{n_{ys}^{(i)}}(\mathbf{R})|^{2} |\phi_{n_{ys}^{(i)}}(\mathbf{%
R}^{\prime})|^{2} ,  \label{56d}
\end{eqnarray}%
as the second term in the RHS of Eq. (\ref{55d}) mutually cancels all terms
in the RHS of Eq. (\ref{53d}) in the sum over $j$ except one, $j=i$. Then
using Eq. (\ref{56d}) and Eq. (\ref{44}) in Eq. (\ref{54d}), we obtain that
\begin{equation}
(E_{\tilde{N}}^{(m),JB}-E_{\tilde{N}}^{(m),eh})/\tilde{N} =\Delta
E_{JB,UB}^{I}+\Delta E_{JB,UB}^{II} ,  \label{57d}
\end{equation}%
where
\begin{eqnarray}
\Delta E_{JB,UB}^{I}&=& \frac{1}{2} \int \int \frac{e^{2} d \mathbf{R} d 
\mathbf{R}^{\prime}}{\varepsilon |\mathbf{R}-\mathbf{R}^{\prime}|}  \notag \\
&&\times |\phi _{n_{ys}^{(i)}}(\mathbf{R})|^{2} |\phi_{n_{ys}^{(i)}}(\mathbf{%
R}^{\prime})|^{2} ,  \label{58d}
\end{eqnarray}%
and
\begin{eqnarray}
\Delta E_{JB,UB}^{II}&=&\frac{1}{2} \sum_{k=-N_{C};k \neq 0}^{N_{C}} (\int
\int \frac{e^{2} d \mathbf{R} d \mathbf{R}^{\prime}}{\varepsilon |\mathbf{R}-%
\mathbf{R}^{\prime}- k L_{x}^{\square} \hat{\mathbf{x}}|}  \notag \\
&&\times |\phi _{n_{ys}^{(i)}}(\mathbf{R})|^{2} |\phi_{n_{ys}^{(i)}}(\mathbf{%
R}^{\prime})|^{2} -\frac{e^{2}}{\varepsilon L_{x}^{\square} \;k}) .
\label{59d}
\end{eqnarray}%
Using Eqs. (\ref{66}), (\ref{68}) in Eqs. (\ref{58d}) and (\ref{59d}), we
calculate that 
\begin{equation}
\Delta E_{JB,UB}^{I}=\frac{e^{2}}{\varepsilon \ell_{0}} F_{JB,UB}^{I}(m) ,
\label{60d}
\end{equation}%
where 
\begin{eqnarray}
F_{JB,UB}^{I}(m)&=&\frac{1}{\pi} \int_{0}^{\infty} d \xi \int_{0}^{\infty} d
\eta \frac{1}{\sqrt{\xi^{2}+\eta^{2}}}  \notag \\
&&\times S_{m}^{2}(\eta) \; S_{m}^{2}(\xi) ,  \label{61d}
\end{eqnarray}%
and
\begin{equation}
\Delta E_{JB,UB}^{II}=\frac{e^{2}}{\varepsilon \ell_{0}} F_{JB,UB}^{II}(m) .
\label{62d}
\end{equation}%
We obtain that
\begin{equation}
F_{JB,UB}^{II}(m)=\sum_{k=1}^{\infty} D_{II}^{(m)}(k) ,  \label{63d}
\end{equation}%
where 
\begin{eqnarray}
D_{II}^{(m)}(k)&=& \frac{2}{\pi} \int_{0}^{\infty} d \xi \int_{0}^{\infty} d
\eta \frac{1}{\sqrt{\xi^{2}+\eta^{2}}} \cos(k\; \sqrt{2 m \pi}\; \xi)  \notag
\\
&&\times S_{m}^{2}(\eta) \; S_{m}^{2}(\xi)-\frac{1}{\sqrt{2 m \pi}} \frac{1}{%
k} .  \label{64d}
\end{eqnarray}%
Point out that the sum in the RHS of Eq. (\ref{63d}) is rapidly convergent
as, e.g., $D_{II}^{(m)}(2)/D_{II}^{(m)}(1)<0.1$ and $%
D_{II}^{(m)}(5)/D_{II}^{(m)}(2)<0.1$, for $m=1,3,5,7$.

From Eqs. (\ref{63d}), (\ref{64d}) we calculate: $F_{JB,UB}^{II}(1)\approx
0.05151$ (for this precision it is enough to include the first eighteen
terms in the sum of Eq. (\ref{73})), $F_{JB,UB}^{II}(3)\approx 0.02972$ (for
this precision it is enough to include the first fifteen terms of the sum), $%
F_{JB,UB}^{II}(5)\approx 0.02301$ (for this precision it is enough to
include the first fourteen terms of the sum), and $F_{JB,UB}^{II}(7)\approx
0.01944$ (for this precision it is enough to include the first thirteen
terms of the sum). Further, from Eq. (\ref{61d}) we calculate: $%
F_{JB,UB}^{I}(1)\approx 0.593068$, $F_{JB,UB}^{I}(3)\approx 0.342408$, $%
F_{JB,UB}^{I}(5)\approx 0.265228$, and $F_{JB,UB}^{I}(7)\approx 0.224158$.
As the result we have that $F_{JB,UB}^{I}(1)+F_{JB,UB}^{II}(1)\approx 0.64458
$, $F_{JB,UB}^{I}(3)+F_{JB,UB}^{II}(3)\approx 0.37213$, $%
F_{JB,UB}^{I}(5)+F_{JB,UB}^{II}(5)\approx 0.28824$, and $%
F_{JB,UB}^{I}(7)+F_{JB,UB}^{II}(7)\approx 0.24360$. Now using Eqs. (\ref{91}%
)-(\ref{92}) and Eqs. (\ref{57d})-(\ref{64d}), we obtain the lowering of the
total energy per electron (in the units of $e^{2}/\varepsilon \ell _{0}$)
for IJB model, $U^{JB}(m)$, as 
\begin{equation}
U^{JB}(m)=U^{UB}(m)+F_{JB,UB}^{I}(m)+F_{JB,UB}^{II}(m).  \label{65d}
\end{equation}%
Finally, we calculate 
\begin{eqnarray}
&&U^{JB}(1)\approx -0.66510,\;\;\;U^{JB}(3)\approx -0.42854  \notag \\
&&U^{JB}(5)\approx -0.33885,\;\;\;U^{JB}(7)\approx -0.28877.  \label{66d}
\end{eqnarray}%
Given by Eq. (\ref{66d}) values of $U^{JB}(1)$, $U^{JB}(3)$ and $U^{JB}(5)$
are substantially lower than pertinent total lowering at $\nu =1,\;1/3$, and
$1/5$ for the Laughlin variational wave function, i.e., \cite{laughlin1983} $%
-\sqrt{\pi /8}\approx -0.6267$, $-0.4156\pm 0.0012$, and $-0.3340\pm 0.0028$%
, respectively. Point out that comparison of $U^{JB}(m)$ with the results of
Ref. \cite{laughlin1983} is totally justified. As for the IJB model the ion
background is totally equivalent to the model of ion background used in Ref.
\cite{laughlin1983}. Notice, more accurate numerical calculations for the
Laughlin trial function, e.g., at $\nu =1/3$ show the lowering\cite{morf1986}
$(-0.410\pm 0.001)$.

\subsection{Remarks on partial crystal-like correlation order
and energy of ground-state}

Here, for IJB, we will make additional remarks on partial
crystal-like correlation order Eq. (\ref{DA7}) and some effects
of it modification on the energy of ground-state.

First, it is easy to see
that using Eq. (\ref{DA7}) in Eq. (\ref{DA2}) we obtain that,
due assumed partial crystal-like correlation order among $N$ electrons
of MR, the Hamiltonian $\hat{H}^{MR}$, Eq. (\ref{DA2}), obtains the
form $n_{xs}^{\max} \times \hat{H}_{\tilde{N}}^{JB}$. The latter is
i) dependent only on $\tilde{N}$ ``compound'' electrons
radius vectors and ii) periodic, with the period $L_{x}^{\square}$,
on any $x_{i}$, $i=1,\ldots,\tilde{N}$; so far these radius vectors
are defined within whole MR. Then we can rewrite
Eq. (\ref{DA1}) as follows
\begin{equation}
n_{xs}^{\max} \hat{H}_{\tilde{N}}^{JB}
\tilde{\Psi}(\textbf{r}_{1},\ldots,\textbf{r}_{\tilde{N}})=E_{N}
\tilde{\Psi}(\textbf{r}_{1},\ldots,\textbf{r}_{\tilde{N}}) ,
\label{DA9}
\end{equation}%
where $\Psi(\textbf{r}_{1},\ldots,\textbf{r}_{N})$, after using
Eq. (\ref{DA7}), is transformed to
$\tilde{\Psi}(\textbf{r}_{1},\ldots,\textbf{r}_{\tilde{N}})$.
Now from Eq. (\ref{DA9}) it is clear that its wave functions
$\tilde{\Psi}(\textbf{r}_{1},\ldots,\textbf{r}_{\tilde{N}})$ have
the same properties, (i) and (ii), as the Hamiltonian
$\hat{H}_{\tilde{N}}^{JB}(\textbf{r}_{1},\ldots,\textbf{r}_{\tilde{N}})$.
So we can assume the ground-state trial wave function
$\tilde{\Psi}(\textbf{r}_{1},\ldots,\textbf{r}_{\tilde{N}})$,
normalized within MR, as
\begin{equation}
\tilde{\Psi}(\textbf{r}_{1},\ldots,\textbf{r}_{\tilde{N}})=
\frac{1}{(n_{xs}^{\max})^{\tilde{N}/2}}
\Psi_{\tilde{N}}^{(m),JB}(\textbf{r}_{1},\ldots,\textbf{r}_{\tilde{N}}) ,
\label{DA10}
\end{equation}%
where
$\Psi_{\tilde{N}}^{(m),JB}(\textbf{r}_{1},\ldots,\textbf{r}_{\tilde{N}})$
is given by Eq. (\ref{18d}) and $\textbf{r}_{i}$ are defined within
whole MR. It is clear that
$\tilde{\Psi}(\textbf{r}_{1},\ldots,\textbf{r}_{\tilde{N}})$ is periodic
in MR, with period $L_{x}^{\square}$, over any its $x_{i}$. It is easy to
see that
\begin{equation}
\int_{MR} d\textbf{r}_{1}\ldots\int_{MR} d\textbf{r}_{\tilde{N}}
|\tilde{\Psi}(\textbf{r}_{1},\ldots,\textbf{r}_{\tilde{N}})|^{2}=1 ,
\label{DA11}
\end{equation}%
indeed, due to its periodicity we have that
\begin{equation}
\int_{MR} d\textbf{r}_{i}
|\tilde{\Psi}(\textbf{r}_{1},\ldots,\textbf{r}_{\tilde{N}})|^{2}=
n_{xs}^{\max} \int_{MS} d\textbf{r}_{i}
|\tilde{\Psi}(\textbf{r}_{1},\ldots,\textbf{r}_{\tilde{N}})|^{2}  .
\label{DA12}
\end{equation}%
Then using Eq. (\ref{DA12}) in Eq. (\ref{DA11}), its LHS we rewrite as
\begin{eqnarray}
&&(n_{xs}^{\max})^{\tilde{N}}
\int_{MS} d\textbf{r}_{1}\ldots\int_{MS} d\textbf{r}_{\tilde{N}}
|\tilde{\Psi}(\textbf{r}_{1},\ldots,\textbf{r}_{\tilde{N}})|^{2}
\notag \\
&&=\int_{MS} d\textbf{r}_{1}\ldots\int_{MS} d\textbf{r}_{\tilde{N}}
|\Psi_{\tilde{N}}^{(m),JB}|^{2}=1 ,
\label{DA13}
\end{eqnarray}%
i.e., the normalization of the wave function
$\tilde{\Psi}(\textbf{r}_{1},\ldots,\textbf{r}_{\tilde{N}})$,
Eq. (\ref{DA10}), within MR is reduced to the normalization
of the wave function $\Psi_{\tilde{N}}^{(m),JB}$ within MS.

Further, from Eq. (\ref{DA9}) the ground-state energy of $N$
electrons within MR, for IJB, is given as
\begin{equation}
E_{N}=n_{xs}^{\max}\int_{MR} d\textbf{r}_{1}\ldots\int_{MR}
d\textbf{r}_{\tilde{N}} \tilde{\Psi}^{\ast}
\hat{H}_{\tilde{N}}^{JB} \tilde{\Psi} ,
\label{DA14}
\end{equation}%
where, using the periodicity of the Hamiltonian
$\hat{H}_{\tilde{N}}^{JB}(\textbf{r}_{1},\ldots,\textbf{r}_{\tilde{N}})$
and property similar to Eq. (\ref{DA12}), we calculate
\begin{equation}
\frac{E_{N}}{n_{xs}^{\max}}= (n_{xs}^{\max})^{\tilde{N}}
\int_{MS} d\textbf{r}_{1}\ldots\int_{MS}
d\textbf{r}_{\tilde{N}} \tilde{\Psi}^{\ast}
\hat{H}_{\tilde{N}}^{JB} \tilde{\Psi} ,
\label{DA15}
\end{equation}%
or
\begin{equation}
\frac{E_{N}}{n_{xs}^{\max}}= E_{\tilde{N}}^{(m),JB},
\label{DA16}
\end{equation}%
where, in agreement with with Eq. (\ref{40d}), we have
\begin{equation}
E_{\tilde{N}}^{(m),JB}=\int_{MS} d\textbf{r}_{1}\ldots\int_{MS}
d\textbf{r}_{\tilde{N}} \Psi_{\tilde{N}}^{(m),JB \ast}
\hat{H}_{\tilde{N}}^{JB} \Psi_{\tilde{N}}^{(m),JB} .
\label{DA17}
\end{equation}%

So it is shown that the ground-state can
correspond to partial crystal-like correlation order,
Eq. (\ref{DA7}), among $N$ electrons
of MR. Then the study of 2DES of $N$ electrons within MR
is exactly reduced to the treatment of 2DES of
$\tilde{N}$ electrons localized within MS
to which PBC is imposed along $x$-direction.
In addition, similar justification can be applied for
the lowest excited-state; i.e., the excited-states
that we study in Sec. VI.

In Sec. VIII, Concluding Remarks, we will discuss effect
on the energy of ground-state, treated in Sec. V, of the change
of the period $L_{x}^{\square}$ in Eqs. (\ref{DA7})-(\ref{DA8})
by period $L_{x}^{ap}=\eta_{a} \times L_{x}^{\square}$
of arbitrary value \cite{balevunpub}. The
area of unit cell is fixed as it is equal to the area
of MR per electron.
It is seen that $\eta_{a} \ll 1$ correspond
to very short period of PBC (and much stronger
crystal-like correlation order than for $\eta_{a}=1$)
while $\eta_{a} \rightarrow \infty$ correspond to
(practical) absence of both the PBC and the crystal-like
correlation order.

\section{Excited-states of the ground-state $\Psi_{\tilde{N},\tilde{N}%
}^{(m),eh}$ and of the ground-state $\Psi_{\tilde{N}}^{(m),JB}$}

In this section we are looking for an excited-state of the lowest energy,
with respect to the ground-state $\Psi _{\tilde{N},\tilde{N}}^{(m),eh}$, Eq.
(\ref{18}); i.e., for the quantum Hall system with UIB at $\nu=1/m$ (remind
that $m=2\ell+1$, $\ell=0, 1, 2, \ldots$). The treatment should include
excitations that occur: i) as without the change of spin of the excited
compound electron (the compound exciton), ii) so with the change of spin of
the excited compound electron (the compound spin-exciton). The treatment of
the compound spin-exciton, at $m=1, \; 3, \; 5,\dots$, should be partly
different from the treatment of the compound exciton, at $m=3,\; 5, \ldots$.
Indeed, 
in $\Psi_{\tilde{N}}^{n,(m)}(\mathbf{r}_{1},...,\mathbf{r}_{\tilde{N}})$,
Eq. (\ref{20}), it is implicit that each single-electron wave function $%
\varphi_{k_{xi}^{(n)}}^{(m)}(\mathbf{r}_{j})$ is multiplied by the spin wave
function $|\sigma>=\psi_{\sigma}(\sigma_{j})=\delta_{\sigma,\sigma_{j}}$,
where spin eigenvalue $\sigma=1$ is pertinent to spin up LLL. Hence, for the
treatment of the compound spin-exciton (in particular, the excited-state at $%
\nu=1$) we will need to take into account as well, at the least partly, the
subspace of the lowest (empty) spin down ($\sigma=-1$) Landau level: with
the same spatial single-electron wave functions and the spin wave function $%
|-1>=\psi_{-1}(\sigma_{j})$.

It follows that all results obtained for the excited-states of the
ground-state $\Psi_{\tilde{N},\tilde{N}}^{(m),eh}$, for UIB, are very simply
extended on the excited-states of the ground-state $\Psi_{\tilde{N}}^{(m),JB}
$, for IJB. In particular, for IJB the energy of any excited-state, counted
from the energy of the ground-state, $\Psi_{\tilde{N}}^{(m),JB}$, coincides
with the energy of relevant excited-state for UIB, counted from the energy
of its ground-state, $\Psi_{\tilde{N},\tilde{N}}^{(m),eh}$. I.e., here the
only difference is that for IJB model the energies of the ground-state and
of its excited-states are shifted upwards on the same value (for given $m$)
with respect to the relevant energies for UIB model.

\subsection{Compound exciton and compound spin-exciton. Compound electrons
and compound hole}

We assume, for $m \geq 3$, the compound exciton wave function $\Psi _{\tilde{%
N},\tilde{N};(m)}^{i_{0},j_{0};\tilde{n}}(\mathbf{r}_{1},..., \mathbf{r}_{%
\tilde{N}};\mathbf{R}_{1},...,\mathbf{R}_{\tilde{N}})$ of the ground-state $%
\Psi_{\tilde{N},\tilde{N}}^{(m),eh}$, Eq. (\ref{18}), in the following (cf.
with Eq. (19) of Ref. \cite{balev2005}) form 
\begin{eqnarray}
\Psi_{\tilde{N},\tilde{N};(m)}^{i_{0},j_{0};\tilde{n}}&=&
\dprod\limits_{i=1}^{\tilde{N}}\phi_{n_{ys}^{(i)}}(\mathbf{R}_{i})
\sum_{n=-\ell}^{\ell } C_{n}(m)  \notag \\
&&\times \Phi_{\tilde{N},(m);n}^{i_{0},j_{0};\tilde{n}} (\mathbf{r}%
_{1},\ldots,\mathbf{r}_{\tilde{N}}) ,  \label{114}
\end{eqnarray}%
where 
an excited ``partial'' many-electron wave function $\Phi_{\tilde{N}%
,(m);n}^{i_{0},j_{0};\tilde{n}}$ it follows from the ground-state
``partial'' many-electron wave function $\Psi _{\tilde{N}}^{n,(m)}$, Eq. (%
\ref{20}), after changing in its $\tilde{N}-$dimensional Slater determinant
of the $i_{0}-$th row $\varphi_{k_{xi_{0}}^{(n)}}^{(m)}(\mathbf{r}_{1}),
\varphi_{k_{xi_{0}}^{(n)}}^{(m)}(\mathbf{r}_{2}),\cdots,
\varphi_{k_{xi_{0}}^{(n)}}^{(m)}(\mathbf{r}_{\tilde{N}})$ by the determinant
row of the following form 
\begin{equation}
\varphi _{k_{xj_{0}}^{(n+\tilde{n})}}^{(m)}(\mathbf{r}_{1}),
\varphi_{k_{xj_{0}}^{(n+\tilde{n})}}^{(m)}(\mathbf{r}_{2}),
\cdots,\varphi_{k_{xj_{0}}^{(n+\tilde{n})}}^{(m)}(\mathbf{r}_{\tilde{N}}),
\label{115}
\end{equation}%
where $\tilde{n}=\pm 1,\ldots, \pm \ell$; $\ell=(m-1)/2$. In particular, we
have that $\tilde{n} \neq 0$, for assumed $m \geq 3$. In Eq. (\ref{115}) the
implicit spin wave function $|1>=\psi_{1}(\sigma_{j})$ is omitted and, using
Eq. (\ref{14}), we have that
\begin{equation}
k_{xj_{0}}^{(n+\tilde{n})}=\frac{2\pi \; m}{L_{x}^{\square}} \left[%
n_{ys}^{(j_{0})} + \frac{n+\tilde{n}}{m} \right] ,  \label{116}
\end{equation}%
where, remind, $n=0, \pm 1,\ldots, \pm \ell$. As only the values $\tilde{n}%
=\pm 1,\ldots, \pm \ell$ are allowed, we have that $k_{xj_{0}}^{(n+\tilde{n}%
)} \neq k_{xi}^{(n)}$, for any possible $i=1,\ldots,\tilde{N}$. We assume
that the $i_{0}$-th unit cell defined by $n_{ys}^{(i_{0})}$ (where $m$
quasihole excitations appear that constitute the compound hole) as well as
the $j_{0}-$th unit cell defined by $n_{ys}^{(j_{0})}$ (where $m$
quasielectron excitations are mainly localized that constitute the excited
compound electron) there are well inside of MS. I.e., they are
far from the edges, along $y-$direction, of MS. Point out that
for the compound exciton Eq. (\ref{114}) the compound electron and the
compound hole have the same spin; i.e., the spin of the compound electron is
not changed in a process of the excitation.

We assume the compound spin-exciton wave function, $\Psi_{\tilde{N},\tilde{N}%
;(m)}^{i_{0},j_{0};\tilde{n},s}$, of the ground-state Eq. (\ref{18}) in the
following form 
\begin{eqnarray}
\Psi_{\tilde{N},\tilde{N};(m)}^{i_{0},j_{0};\tilde{n},s}&=&
\dprod\limits_{i=1}^{\tilde{N}}\phi_{n_{ys}^{(i)}}(\mathbf{R}_{i})
\sum_{n=-\ell}^{\ell } C_{n}(m)  \notag \\
&&\times \Phi_{\tilde{N},(m);n}^{i_{0},j_{0};\tilde{n},s} (\mathbf{r}%
_{1},\ldots,\mathbf{r}_{\tilde{N}}) ,  \label{114s}
\end{eqnarray}%
that (in difference from the compound exciton excited-state, Eq. (\ref{114}%
)) is valid as for $m \geq 3$ so for $m=1$. In Eq. (\ref{114s}) the
spin-excited partial many-electron wave function $\Phi_{\tilde{N}%
,(m);n}^{i_{0},j_{0};\tilde{n},s}$ it follows from the ``usual'' excited
partial many-electron wave function $\Phi_{\tilde{N},(m);n}^{i_{0},j_{0};%
\tilde{n}}$ after in the latter in the determinant row Eq. (\ref{115}): i)
the implicit spin up wave function $|1>=\psi_{1}(\sigma_{j})$ is substituted
by spin down one, $|-1>=\psi_{-1}(\sigma_{j})$; ii) further, here we have
that $\tilde{n}=0, \pm 1, \ldots,\pm \ell$ are allowed (i.e., in addition to
the values of $\tilde{n}$ allowed for the compound exciton, for the compound
spin-exciton it is allowed $\tilde{n}=0$ at any $m=1, \; 3, \; 5,\dots$).
Notice, for the compound spin-exciton Eq. (\ref{116}) is also valid; in
particular, for $m=1$ as $n=\tilde{n}=0$ we have that the second term in its
square brackets, $\propto (n+\tilde{n})$, is zero. For definiteness, we call
the spin-exciton Eq. (\ref{114s}) as compound one as well for $m=1$, even
though here there is no any compound structure of electrons or the hole
resembling pertinent structure for $m=3, \; 5,\ldots$.

Point out, in Eqs. (\ref{114})-(\ref{114s}) we have that $%
n_{ys}^{j_{0}i_{0}}=n_{ys}^{(j_{0})}-n_{ys}^{(i_{0})}$ can obtain any finite
integer value (i.e., $n_{ys}^{j_{0}i_{0}}=0, \pm 1, \pm 2, \dots$) as for $m
\geq 3$ so for $m=1$.

Point out, the excited-state wave functions $\Psi _{\tilde{N},\tilde{N}%
;(m)}^{i_{0},j_{0};\tilde{n}}$, Eq. (\ref{114}), and $\Psi _{\tilde{N},%
\tilde{N};(m)}^{i_{0},j_{0};\tilde{n},s}$, Eq. (\ref{114s}), are periodic
with the period $L_{x}^{\square}$ along any of $2\tilde{N}$ variables $x_{i}$
and $X_{j}$; remind, the same properties have the ground-state wave function 
$\Psi_{\tilde{N},\tilde{N}}^{(m),eh}$, Eq. (\ref{18}). Point out that 
\begin{equation}
\langle \Psi _{\tilde{N},\tilde{N};(m)}^{i_{0},j_{0};\tilde{n}} |\Psi_{%
\tilde{N},\tilde{N}}^{(m),eh}\rangle= \langle \Psi _{\tilde{N},\tilde{N}%
;(m)}^{i_{0},j_{0};\tilde{n},s} |\Psi_{\tilde{N},\tilde{N}%
}^{(m),eh}\rangle=0 ,  \label{117}
\end{equation}%
i.e., as required\cite{macdonald1986,landau1965} the excited-state wave
functions $\Psi _{\tilde{N},\tilde{N};(m)}^{i_{0},j_{0};\tilde{n}}$, $\Psi _{%
\tilde{N},\tilde{N};(m)}^{i_{0},j_{0};\tilde{n},s}$ are orthogonal to the
ground-state wave function $\Psi_{\tilde{N},\tilde{N}}^{(m),eh}$. These
excited-state wave functions are orthonormalized as it is seen, e.g., that
\begin{eqnarray}
\langle \Psi _{\tilde{N}, \tilde{N};(m)}^{i_{0}^{(1)},j_{0}^{(1)};\tilde{n}%
^{(1)}}| \Psi _{\tilde{N},\tilde{N};(m)}^{i_{0}^{(2)},j_{0}^{(2)};\tilde{n}%
^{(2)}} \rangle&=&\delta_{i_{0}^{(1)},i_{0}^{(2)}} \;\;
\delta_{j_{0}^{(1)},j_{0}^{(2)}}  \notag \\
&& \times \delta_{\tilde{n}^{(1)},\tilde{n}^{(2)}}.  \label{117b}
\end{eqnarray}%
Eq. (\ref{117b}) is also valid if in its LHS to substitute $\Psi _{\tilde{N},%
\tilde{N};(m)}^{i_{0}^{(i)},j_{0}^{(i)}; \tilde{n}^{(i)}}$ by $\Psi _{\tilde{%
N},\tilde{N};(m)}^{i_{0}^{(i)},j_{0}^{(i)}; \tilde{n}^{(i)},s}$, $i=1,\; 2$.

It is seen that the compound exciton wave function $\Psi _{\tilde{N},\tilde{N%
};(m)}^{i_{0},j_{0};\tilde{n}}$ describes an exciton like excitation of the
compound (composite) structure, cf. with Refs. \cite%
{morf1986,macdonald1986,bychkov1981,kallin1984,sondhi1993}. In particular,
the form of the exciton charge (cf. with Eqs. (\ref{18}), (\ref{24})-(\ref%
{26}), (\ref{114})) density, $\delta\rho^{(m)}_{i_{0},j_{0};\tilde{n}}(%
\mathbf{r})$, given as 
\begin{eqnarray}
&&\delta\rho^{(m)}_{i_{0},j_{0};\tilde{n}}(\mathbf{r})= e [\langle \Psi _{%
\tilde{N},\tilde{N};(m)}^{i_{0},j_{0};\tilde{n}}| \sum_{j=1}^{\tilde{N}%
}\delta (\mathbf{r}-\mathbf{r}_{j})| \Psi_{\tilde{N},\tilde{N}%
;(m)}^{i_{0},j_{0};\tilde{n}} \rangle  \notag \\
&&- \langle \Psi _{\tilde{N},\tilde{N}}^{(m),eh}|\sum_{j=1}^{\tilde{N}}
\delta (\mathbf{r}-\mathbf{r}_{j})|\Psi _{\tilde{N},\tilde{N}}^{(m),eh}
\rangle ] ,  \label{118}
\end{eqnarray}%
is readily reduced to the form
\begin{equation}
\delta\rho^{(m)}_{i_{0},j_{0};\tilde{n}}(\mathbf{r})=\sum_{n=-\ell}^{\ell}
\delta\rho^{(m),n}_{i_{0},j_{0};\tilde{n}}(\mathbf{r}) ,  \label{118b}
\end{equation}%
where the charge density of the $n-$th quasiexciton 
\begin{equation}
\delta\rho^{(m),n}_{i_{0},j_{0};\tilde{n}}(\mathbf{r})= \frac{e}{m}%
|\varphi_{k_{xj_{0}}^{(n+\tilde{n})}}^{(m)}(\mathbf{r})|^{2}- \frac{e}{m}%
|\varphi_{k_{xi_{0}}^{(n)}}^{(m)}(\mathbf{r})|^{2} .  \label{119}
\end{equation}%
I.e., for $m \geq 3$, the compound exciton charge density is the
superposition of the charge densities of $m$ quasiexcitons, counted by the
superscript $n$. The first and the second terms in the RHS of Eq. (\ref{119}%
) present the ($n-$th) quasielectron and the ($n-$th) quasihole charge
densities, respectively, of the $n$-th quasiexciton. The former is mainly
localized, within MS, at $y \approx \ell_{0}^{2} \;
k_{xj_{0}}^{(n+\tilde{n})}$ (i.e., almost within the $j_{0}-$th unit cell)
and the latter is mainly localized at $y \approx k_{xi_{0}}^{(n)} \;
\ell_{0}^{2}$ (i.e., almost within the $i_{0}-$th unit cell). Point out, the
RHS of Eq. (\ref{118b}) is calculated from the RHS of Eq. (\ref{118})
without any approximations. Integrating the quasielectron or the quasihole
charge densities, from Eq. (\ref{119}), over $\mathbf{r}$ within the total
area of MS we readily obtain that the total quasielectron or
quasihole charge, within MS, is given as $e/m$ or $-e/m$,
respectively; for given $n-$th quasiexciton. I.e., in the fractional quantum
Hall regime, at $\nu=1/m$, these charges are fractional and have the same
values as quasielectron and quasihole fractionally charged excitations
within the Laughlin model.\cite{laughlin1983} Point out that in our model
the total charge of the quasielectron (the quasihole) is independent from it
"specific" intra-main-strip quantum numbers $n, j_{0}, \tilde{n}$, or $n,
k_{xj_{0}}^{(n+\tilde{n})}$ ($n, i_{0}$, or $k_{xi_{0}}^{(n)}$).

Due to PBC (in particular, the periodicity of
single-electron wave functions) it is clear that the charge density of the $%
n-$th quasiexciton, Eq. (\ref{119}), has its images that are periodic with
the period $L_{x}^{\square }=\sqrt{2m\pi }\ell _{0}$ along the $x-$
direction, for $x$ outside MS; i.e., for $x>L_{x}^{\square
}n_{xs}^{\alpha }$ or $x<L_{x}^{\square }(n_{xs}^{\alpha }-1)$. The same
property of the periodicity holds as well for the charge density of the $n-$%
th quasielectron and of the $n-$th quasihole. Notice, these properties of
the quasielectron and the quasihole periodicity does not have a counterpart
among the properties of the quasiparticles, of fractionally charged
excitations, in the Laughlin model \cite{laughlin1983}, see also, e.g., \cite%
{morf1986,macdonald1986}. In addition, in the present model all $m$
different quasielectrons (quasiholes) and quasiexcitons are strongly
correlated among themselves. Such that $m$ different quasiexcitons (however,
strongly correlated) compose the compound exciton excited-state Eq. (\ref%
{114}), according to Eqs. (\ref{118})-(\ref{119}).

Point out that for the compound spin-exciton wave function $\Psi _{\tilde{N},%
\tilde{N};(m)}^{i_{0},j_{0};\tilde{n},s}$ the form of the spin-exciton
charge density, $\delta \rho _{i_{0},j_{0};\tilde{n}}^{(m),s}(\mathbf{r})$,
coincides with that given by the RHS of Eqs. (\ref{118}), (\ref{118b}) and
Eq. (\ref{119}). With the only difference that here are allowed all possible 
$m$ and $\tilde{n}$: $m=1,\;3,\;5,\ldots $ and $\tilde{n}=0,\pm 1,\ldots
,\pm \ell $. I.e., the compound spin-exciton will have the charge density $%
\delta \rho _{i_{0},j_{0};\tilde{n}}^{(m),s}(\mathbf{r})\equiv 0$, if $%
\tilde{n}=0$ and $n_{ys}^{j_{0}i_{0}}=0$. Notice, for the compound exciton
we have that $\delta \rho _{i_{0},j_{0};\tilde{n}}^{(m)}(\mathbf{r})\neq 0$,
at any allowed $\tilde{n}$, $n_{ys}^{j_{0}i_{0}}$. Hence, $m$ strongly
correlated spin-quasiexcitons, for $m\geq 3$, compose the compound
spin-exciton Eq. (\ref{114s}).

We are interested in calculation of the energy gaps $\Delta E_{i_{0},j_{0};%
\tilde{n}}^{(m)}$ and $\Delta E_{i_{0},j_{0};\tilde{n}}^{(m),s}$ for the
creation of the compound exciton and the compound spin-exciton,
respectively, within MS, cf. with Refs. \cite%
{laughlin1983,morf1986,macdonald1986,bychkov1981, kallin1984,sondhi1993}. We
also call $\Delta E_{i_{0},j_{0};\tilde{n}}^{(m)}$ and $\Delta
E_{i_{0},j_{0};\tilde{n}}^{(m),s}$ as the energy of the compound exciton (or
the energy of the compound exciton excitation) and the energy of the
compound spin-exciton (or the energy of the compound spin-exciton
excitation), respectively. Moreover, we are mainly interested in the
calculation of the \textit{minimal value} of $\Delta E_{i_{0},j_{0};\tilde{n}%
}^{(m)}$, at $m \geq 3$, and $\Delta E_{i_{0},j_{0};\tilde{n}}^{(m),s}$, at $%
m=1$, \textit{that defines the activation gap} at $m \geq 3$ and $m=1$,
respectively. The activation gap is experimentally observable from the
activation behavior of the direct current magnetotransport coefficients
related with dissipation, i.e., typically, the diagonal resistance or
resistivity ($\rho_{yy}$, $\rho_{xx}$) and the diagonal conductance ($%
\sigma_{yy}$). As we will show, for $m=1$ defined in such manner the
activation gap (given by one of the smallest values of $\Delta
E_{i_{0},j_{0};0}^{(1),s}$, however, does not by the smallest one) is much
larger than simply the minimal value of $\Delta E_{i_{0},j_{0};0}^{(1),s}$:
because for the latter gap pertinent transitions does not contribute to any
relevant dissipative kinetic coefficient. Respectively, this gap should be
here discarded.

So far we have assumed UIB. Now, for IJB,
instead of the wave function Eq. (\ref{114}), we obtain the
pertinent compound exciton wave function,
$\Psi_{\tilde{N};(m)}^{i_{0},j_{0};\tilde{n}}$, of the
ground-state Eq. (\ref{18d}) as
\begin{equation}
\Psi_{\tilde{N};(m)}^{i_{0},j_{0};\tilde{n}}= \sum_{n=-\ell}^{\ell }
C_{n}(m) \Phi_{\tilde{N},(m);n}^{i_{0},j_{0};\tilde{n}} (\mathbf{r}%
_{1},\ldots,\mathbf{r}_{\tilde{N}}) ,  \label{114d}
\end{equation}
and instead of the wave function Eq. (\ref{114s}), we obtain the pertinent
compound spin-exciton wave function, $\Psi_{\tilde{N};(m)}^{i_{0},j_{0};%
\tilde{n},s}$, of the ground-state Eq. (\ref{18d}), in the following form
\begin{equation}
\Psi_{\tilde{N};(m)}^{i_{0},j_{0};\tilde{n},s}= \sum_{n=-\ell}^{\ell }
C_{n}(m) \Phi_{\tilde{N},(m);n}^{i_{0},j_{0};\tilde{n},s} (\mathbf{r}%
_{1},\ldots,\mathbf{r}_{\tilde{N}}) .  \label{114sd}
\end{equation}%
It is easy to see that all above results, e.g., Eqs. (\ref{117})- (\ref{119}%
), are valid for the excited-states Eqs. (\ref{114d}), (\ref{114sd}) (and
their ground-state).

\subsection{Energy of the compound exciton}

For UIB, by using the Hamiltonian Eq. (\ref{1}), we calculate the total
energy of the electron-ion system in the compound exciton state Eq. (\ref%
{114}) as 
\begin{equation}
E_{\tilde{N};(m)}^{i_{0},j_{0};\tilde{n}}=\langle \Psi_{\tilde{N},\tilde{N}%
;(m)}^{i_{0},j_{0};\tilde{n}} |\hat{H}_{\tilde{N},\tilde{N}} |\Psi_{\tilde{N}%
,\tilde{N};(m)}^{i_{0},j_{0};\tilde{n}}\rangle ,  \label{120}
\end{equation}%
where in the RHS for the kinetic energy term we, similar to Eq. (\ref{42}),
obtain 
\begin{equation}
\langle \Psi _{\tilde{N},\tilde{N};(m)}^{i_{0},j_{0};\tilde{n}}|\hat{H}%
_{0}|\Psi _{\tilde{N},\tilde{N};(m)}^{i_{0},j_{0};\tilde{n}}\rangle
=\left(\hbar \omega_{c}-|g_{0}| \mu_{B} B\right)\tilde{N}/2 ,  \label{121}
\end{equation}%
where $\mu_{B}$ is the Bohr magneton; here the Zeeman energy is included in $%
\hat{h}_{0}$ explicitly (then the RHS of Eq. (\ref{41}) should be changed by
the RHS of Eq. (\ref{121})).

Point out, for IJB in the RHS of Eq. (\ref{120}) $\Psi _{\tilde{N},\tilde{N}%
;(m)}^{i_{0},j_{0};\tilde{n}}$ is changed on $\Psi _{\tilde{N}%
;(m)}^{i_{0},j_{0};\tilde{n}}$ and $\hat{H}_{\tilde{N},\tilde{N}}$ on $\hat{H%
}_{\tilde{N}}^{JB}$. Then Eq. (\ref{121}) is correct after making the former
change in its LHS.

Using Eqs. (\ref{120}), (\ref{40}), we obtain that the energy of the
excited-state Eq. (\ref{120}) with respect to the energy of the ground-state
Eq. (\ref{40}), $\Delta E_{i_{0},j_{0};\tilde{n}}^{(m)}$, or the energy of
the compound exciton, is given as 
\begin{equation}
\Delta E_{i_{0},j_{0};\tilde{n}}^{(m)}= E_{\tilde{N};(m)}^{i_{0},j_{0};%
\tilde{n}}-E^{(m),eh}_{\tilde{N}} .  \label{122}
\end{equation}%
Eq. (\ref{122}), after using Eqs. (\ref{41}), (\ref{121}), obtains the form 
\begin{equation}
\Delta E_{i_{0},j_{0};\tilde{n}}^{(m)}= \Delta E_{ei,(m)}^{i_{0},j_{0};%
\tilde{n}}+ \Delta E_{ee,(m)}^{i_{0},j_{0};\tilde{n}} ,  \label{123}
\end{equation}%
where in the RHS the term related with the contributions from Eqs. (\ref{41}%
), (\ref{121}) is equal to zero. In addition, it is taken into account that
in the RHS of Eq. (\ref{123}) the term related with the ion-ion (or UIB-UIB
interaction) potential $V_{ii}$, $\Delta E_{ii,(m)}^{i_{0},j_{0};\tilde{n}}=0
$; indeed, in the RHS of Eq. (\ref{120}) the term $\langle \Psi _{\tilde{N},%
\tilde{N};(m)}^{i_{0},j_{0};\tilde{n}}|V_{ii}|\Psi _{\tilde{N},\tilde{N}%
;(m)}^{i_{0},j_{0};\tilde{n}}\rangle$ coincides with the RHS of Eq. (\ref{43}%
).

Point out, for IJB in the RHS of Eq. (\ref{122}) $E^{(m),eh}_{\tilde{N}}$ is
changed on $E^{(m),JB}_{\tilde{N}}$ and in the RHS of Eq. (\ref{123}) $%
\Delta E_{ei,(m)}^{i_{0},j_{0};\tilde{n}}$ is changed on $\Delta
E_{eb,(m)}^{i_{0},j_{0};\tilde{n}}$. It is taken into account that now in
the RHS of Eq. (\ref{123}) the term related with IJB-IJB interaction
potential $V_{bb}$, $\Delta E_{bb,(m)}^{i_{0},j_{0};\tilde{n}}=0$.

In the RHS of Eq. (\ref{123}) the term related with electron-ion potential, 
\begin{eqnarray}
\Delta E_{ei,(m)}^{i_{0},j_{0};\tilde{n}}&=&\langle \Psi_{\tilde{N},\tilde{N}%
;(m)}^{i_{0},j_{0};\tilde{n}} |V_{ei}|\Psi_{\tilde{N},\tilde{N}%
;(m)}^{i_{0},j_{0};\tilde{n}}\rangle  \notag \\
&&-\langle \Psi_{\tilde{N},\tilde{N}}^{(m),eh}|V_{ei}| \Psi_{\tilde{N},%
\tilde{N}}^{(m),eh}\rangle ,  \label{124}
\end{eqnarray}%
i.e., electron-UIB interaction, obtains (cf. with Eq. (\ref{50})) the form 
\begin{eqnarray}
&&\Delta E_{ei,(m)}^{i_{0},j_{0};\tilde{n}}= -\frac{1}{m} \sum_{n=-\ell}^{%
\ell} \sum_{i=1}^{\tilde{N}} \sum_{k=-\infty}^{\infty} \int \int \frac{%
e^{2}\; d \mathbf{r} \; d \mathbf{R} }{\varepsilon |\mathbf{r}- \mathbf{R}-k
L_{x}^{\square} \hat{\mathbf{x}}|}  \notag \\
&&\;\;\;\;\;\; \times [|\varphi_{k_{xj_{0}}^{(n+\tilde{n})}}^{(m)}(\mathbf{r}%
)|^{2}- |\varphi_{k_{xi_{0}}^{(n)}}^{(m)}(\mathbf{r})|^{2}]
|\phi_{n_{ys}^{(i)}}(\mathbf{R})|^{2} ,  \label{125}
\end{eqnarray}%
where, due to the fast convergence of the sum over $k$, in the limits of the
sum $N_{C}$ is substituted by $\infty$.

Point out, for IJB in Eq. (\ref{124}) $\Delta E_{ei,(m)}^{i_{0},j_{0};\tilde{%
n}}$ is changed on $\Delta E_{eb,(m)}^{i_{0},j_{0};\tilde{n}}$, $V_{ei}$ on $%
V_{eb}$, $\Psi _{\tilde{N},\tilde{N};(m)}^{i_{0},j_{0};\tilde{n}}$ is
changed on $\Psi _{\tilde{N};(m)}^{i_{0},j_{0};\tilde{n}}$, and $\Psi_{%
\tilde{N},\tilde{N}}^{(m),eh}$ on $\Psi_{\tilde{N}}^{(m),JB}$. It is easy to
see that
\begin{equation}
\Delta E_{eb,(m)}^{i_{0},j_{0};\tilde{n}}= \Delta E_{ei,(m)}^{i_{0},j_{0};%
\tilde{n}} .  \label{125b}
\end{equation}

Further, in the RHS of Eq. (\ref{123}) the term related with
electron-electron potential, 
\begin{eqnarray}
\Delta E_{ee,(m)}^{i_{0},j_{0};\tilde{n}}&=&\langle \Psi_{\tilde{N},\tilde{N}%
;(m)}^{i_{0},j_{0};\tilde{n}} |V_{ee}|\Psi_{\tilde{N},\tilde{N}%
;(m)}^{i_{0},j_{0};\tilde{n}}\rangle  \notag \\
&&-\langle \Psi_{\tilde{N},\tilde{N}}^{(m),eh}|V_{ee}| \Psi_{\tilde{N},%
\tilde{N}}^{(m),eh}\rangle ,  \label{126}
\end{eqnarray}%
obtains (cf. with Eqs. (\ref{51})-(\ref{57})) the form 
\begin{equation}
\Delta E_{ee,(m)}^{i_{0},j_{0};\tilde{n}}= \Delta
E_{ee,(m)}^{di}(i_{0},j_{0};\tilde{n})+ \Delta E_{ee,(m)}^{xc}(i_{0},j_{0};%
\tilde{n}) ,  \label{127}
\end{equation}%
where the direct-alike (or the Hartree-alike) contribution, cf. with Eq. (%
\ref{56}), is given as 
\begin{eqnarray}
&&\Delta E_{ee,(m)}^{di}(i_{0},j_{0};\tilde{n})= \frac{1}{m}
\sum_{n=-\ell}^{\ell} \sum_{i=1;i \neq i_{0}}^{\tilde{N}}
\sum_{k=-\infty}^{\infty}  \notag \\
&&\times \int d \mathbf{r} \int d \mathbf{r}^{\prime} \frac{e^{2}}{%
\varepsilon|\mathbf{r}- \mathbf{r}^{\prime}- k L_{x}^{\square} \hat{\mathbf{x%
}}|} |\varphi_{k_{xi}^{(n)}}^{(m)}(\mathbf{r})|^{2}  \notag \\
&&\times [|\varphi_{k_{xj_{0}}^{(n+\tilde{n})}}^{(m)}(\mathbf{r}%
^{\prime})|^{2}- |\varphi_{k_{xi_{0}}^{(n)}}^{(m)}(\mathbf{r}%
^{\prime})|^{2}] ,  \label{128}
\end{eqnarray}%
and the exchange-alike (or the Fock-alike) contribution, cf. with Eq. (\ref%
{57}), is given ($m \geq 3$) as 
\begin{eqnarray}
&&\Delta E_{ee,(m)}^{xc}(i_{0},j_{0};\tilde{n})=- \frac{1}{m}
\sum_{n=-\ell}^{\ell} \sum_{i=1; i \neq i_{0}}^{\tilde{N}}
\sum_{k=-\infty}^{\infty}  \notag \\
&&\times \int d \mathbf{r} \int d \mathbf{r}^{\prime} \frac{e^{2}}{%
\varepsilon |\mathbf{r}-\mathbf{r}^{\prime}- k L_{x}^{\square} \hat{\mathbf{x%
}}|} \; \; \varphi _{k_{xi}^{(n)}}^{(m) \ast}(\mathbf{r}) \; \varphi
_{k_{xi}^{(n)}}^{(m)}(\mathbf{r}^{\prime})  \notag \\
&&\times [\varphi _{k_{xj_{0}}^{(n+\tilde{n})}}^{(m)}(\mathbf{r}) \varphi
_{k_{xj_{0}}^{(n+\tilde{n})}}^{(m) \ast}(\mathbf{r}^{\prime}) -\varphi
_{k_{xi_{0}}^{(n)}}^{(m)}(\mathbf{r}) \varphi _{k_{xi_{0}}^{(n)}}^{(m) \ast}(%
\mathbf{r}^{\prime})] .  \label{129}
\end{eqnarray}%
Notice, due to the fast convergence of the sums over $k$ in the RHS of Eqs. (%
\ref{128})-(\ref{129}), in the limits of these sums $N_{C}$ is substituted
by $\infty$.

Point out, for IJB in Eq. (\ref{126}) $\Psi _{\tilde{N},\tilde{N}%
;(m)}^{i_{0},j_{0};\tilde{n}}$ is changed on $\Psi _{\tilde{N}%
;(m)}^{i_{0},j_{0};\tilde{n}}$, and $\Psi_{\tilde{N},\tilde{N}}^{(m),eh}$ on
$\Psi_{\tilde{N}}^{(m),JB}$. Then it is easy to see that Eqs. (\ref{127})-(%
\ref{129}) are obtained again.

Changing in the RHS of Eq. (\ref{125}) $\mathbf{R} \rightarrow \mathbf{r}%
^{\prime}$ and in the RHS of Eq. (\ref{128}) $\mathbf{r} \rightleftarrows 
\mathbf{r}^{\prime}$, $k \rightarrow -k$, we obtain 
\begin{equation}
\Delta E_{ei,(m)}^{i_{0},j_{0};\tilde{n}}+ \Delta
E_{ee,(m)}^{di}(i_{0},j_{0};\tilde{n})= \sum_{i=1}^{2}
E_{i}^{(m)}(i_{0},j_{0};\tilde{n}) ,  \label{131}
\end{equation}%
where 
\begin{eqnarray}
&&E_{1}^{(m)}(i_{0},j_{0};\tilde{n})= \frac{1}{m} \sum_{n=-\ell}^{\ell}
\sum_{k=-\infty}^{\infty} \int \int \frac{e^{2}\; d \mathbf{r} \; d \mathbf{r%
}^{\prime}}{\varepsilon |\mathbf{r}-\mathbf{r}^{\prime}-k L_{x}^{\square} 
\hat{\mathbf{x}}|}  \notag \\
&&\;\;\;\;\;\; \times [|\varphi_{k_{xi_{0}}^{(n)}}^{(m)}(\mathbf{r})|^{2}-
|\varphi_{k_{xj_{0}}^{(n+\tilde{n})}}^{(m)}(\mathbf{r})|^{2}]
|\phi_{n_{ys}^{(i_{0})}}(\mathbf{r}^{\prime})|^{2} ,  \label{132}
\end{eqnarray}%
and 
\begin{eqnarray}
&&E_{2}^{(m)}(i_{0},j_{0};\tilde{n})=-\frac{1}{m} \sum_{n=-\ell}^{\ell}
\sum_{i=1;i \neq i_{0}}^{\tilde{N}} \sum_{k=-\infty}^{\infty} \int d \mathbf{%
r} \int d \mathbf{r}^{\prime}  \notag \\
&&\times \frac{e^{2}}{\varepsilon|\mathbf{r}- \mathbf{r}^{\prime}- k
L_{x}^{\square} \hat{\mathbf{x}}|} [|\varphi_{k_{xj_{0}}^{(n+\tilde{n}%
)}}^{(m)}(\mathbf{r})|^{2}- |\varphi_{k_{xi_{0}}^{(n)}}^{(m)}(\mathbf{r}%
)|^{2}]  \notag \\
&&\times [|\phi_{n_{ys}^{(i)}}(\mathbf{r}^{\prime})|^{2}
-|\varphi_{k_{xi}^{(n)}}^{(m)}(\mathbf{r}^{\prime})|^{2}] .  \label{133}
\end{eqnarray}%
It is shown in Appendix A that Eq. (\ref{132}) is rewritten as
\begin{eqnarray}
E_{1}^{(m)}(i_{0},j_{0}&;&\tilde{n})= \frac{4e^{2}}{\varepsilon
L_{x}^{\square}} \int_{0}^{\infty} \frac{d \eta}{\eta} e^{-\eta^{2}/4}
f_{m}(\eta) S_{m}(\eta)  \notag \\
&& \times \sin^{2}\left(\sqrt{\frac{m \pi}{2}} (n_{ys}^{j_{0}i_{0}}+ \frac{%
\tilde{n}}{m})\eta \right) .  \label{134}
\end{eqnarray}%
Remind that $n_{ys}^{j_{0}i_{0}}=n_{ys}^{(j_{0})}-n_{ys}^{(i_{0})}$.
Further, in Appendix A it is shown that Eq. (\ref{133}) obtains the form
\begin{eqnarray}
E_{2}^{(m)}(i_{0},j_{0}&;&\tilde{n})= \frac{4e^{2}}{\varepsilon
L_{x}^{\square}} \{ \int_{0}^{\infty} \frac{d \eta}{\eta} G_{m}^{v}(\eta) 
\notag \\
&&\times \sin^{2}\left(\sqrt{\frac{m \pi}{2}}(n_{ys}^{j_{0}i_{0}}+ \frac{%
\tilde{n}}{m})\eta\right)  \notag \\
&&-\sum_{k=1}^{\infty} \frac{1}{k} e^{-\pi k^{2}/m} \sin^{2}\left(\frac{\pi
k \tilde{n}}{m}\right) \} ,  \label{135}
\end{eqnarray}%
where it is taken into account that $\sin^{2}\left(\pi k
(n_{ys}^{j_{0}i_{0}}+\tilde{n}/m) \right)= \sin^{2}\left(\pi k \tilde{n}%
/m\right)$; $G_{m}^{v}(\eta)$ is defined in Appendix A. After mutual
cancelling of $E_{1}^{(m)}(i_{0},j_{0};\tilde{n})$ with the pertinent part
of $E_{2}^{(m)}(i_{0},j_{0};\tilde{n})$, we obtain from Eqs. (\ref{134}), (%
\ref{135}) that 
\begin{equation}
\sum_{i=1}^{2} E_{i}^{(m)}(i_{0},j_{0};\tilde{n})= \frac{e^{2}}{\varepsilon
\ell_{0}} F_{di}^{(m)}(n_{ys}^{j_{0}i_{0}};\tilde{n}) ,  \label{136}
\end{equation}%
where 
\begin{eqnarray}
F_{di}^{(m)}(n_{ys}^{j_{0}i_{0}};\tilde{n})&=&2 \sqrt{\frac{2}{m \pi}} \{
\int_{0}^{\infty} \frac{d \eta}{\eta} e^{-\eta^{2}/2}  \notag \\
&& \times \sin^{2}\left(\sqrt{\frac{m \pi}{2}}(n_{ys}^{j_{0}i_{0}}+ \frac{%
\tilde{n}}{m})\eta\right)-\sum_{k=1}^{\infty} \frac{1}{k}  \notag \\
&&\times e^{-\pi k^{2}/m} \; \sin^{2}\left(\frac{\pi k \tilde{n}}{m}\right)
\} .  \label{137}
\end{eqnarray}

In addition, in Appendix A it is shown that from Eq. (\ref{129}) it follows
that 
\begin{equation}
\Delta E_{ee,(m)}^{xc}(i_{0},j_{0};\tilde{n})= \frac{e^{2}}{\varepsilon
\ell_{0}} \left(F_{xc}^{(m)}(n_{ys}^{j_{0}i_{0}};\tilde{n})+2F_{2}(m)\right)
,  \label{138}
\end{equation}%
where we have
\begin{eqnarray}
&&F_{xc}^{(m)}(n_{ys}^{j_{0}i_{0}};\tilde{n})=- \sqrt{\frac{2}{m \pi}} \{
\sum_{k=-\infty}^{\infty} e^{-\pi m (k+\tilde{n}/m)^{2}}  \notag \\
&&\times\int_{0}^{\infty} \frac{d \eta e^{-\eta^{2}/2}}{\sqrt{\eta^{2}+ 2\pi
m(k+\tilde{n}/m)^{2}}} -e^{-\pi m (n_{ys}^{j_{0}i_{0}}+\tilde{n}/m)^{2}} 
\notag \\
&&\times \int_{0}^{\infty} \frac{d \eta e^{-\eta^{2}/2}}{\sqrt{\eta^{2}+
2\pi m(n_{ys}^{j_{0}i_{0}}+\tilde{n}/m)^{2}}} \} .  \label{139}
\end{eqnarray}%
%
%

Using Eqs. (\ref{124})-(\ref{139}), we rewrite the energy of the compound
exciton Eq. (\ref{123}) as follows 
\begin{eqnarray}
\Delta E_{i_{0},j_{0};\tilde{n}}^{(m)}&=&\frac{e^{2}}{\varepsilon \ell_{0}}
\{F_{di}^{(m)}(n_{ys}^{j_{0}i_{0}};\tilde{n})  \notag \\
&&+ F_{xc}^{(m)}(n_{ys}^{j_{0}i_{0}};\tilde{n})+2 F_{2}(m)\} ,  \label{140}
\end{eqnarray}%
where $\tilde{n}=\pm 1,\dots,\pm \ell$, and, remind, $m \geq 3$. 

Now we will study contributions to the RHS of Eq. (\ref{140}). Analytical
and numerical treatment shows that the main contribution to the RHS of Eq. (%
\ref{140}) is given by $F_{di}^{(m)}(n_{ys}^{j_{0}i_{0}};\tilde{n})>0$,
where in turn the main contribution (obviously, positive) is related with
the first term in the curly brackets of the RHS of Eq. (\ref{137}). It is
seen that this integral contribution grows monotonically with the increase
of $|n_{ys}^{j_{0}i_{0}}+\tilde{n}/m|$. In particular, for $%
|n_{ys}^{j_{0}i_{0}}| \gg 1$ it is given as $\approx \sqrt{2/(m \pi)}
[\ln\left(\sqrt{2m \pi} |n_{ys}^{j_{0}i_{0}}|\right)- \left(\ln(2)-\gamma
\right)/2]$, where $\gamma$ is the Euler constant. Moreover, numerical study
shows that both $F_{di}^{(m)}(n_{ys}^{j_{0}i_{0}};\tilde{n})$ and the total
value, always positive, of the RHS of Eq. (\ref{140}) grow monotonically
with the increase of $|n_{ys}^{j_{0}i_{0}}+\tilde{n}/m|$. Point out, these
monotonic increases, of $F_{di}^{(m)}(n_{ys}^{j_{0}i_{0}};\tilde{n})$ and
the RHS of Eq. (\ref{140}), take place starting from the $%
\min\{|n_{ys}^{j_{0}i_{0}}+\tilde{n}/m|\}=1/m$; remind, for $m \geq 3$, $%
|n_{ys}^{j_{0}i_{0}}+\tilde{n}/m| \geq 1/m$. I.e., for $m \geq 3$ the
minimal value of $\Delta E_{i_{0}j_{0};\tilde{n}}^{(m)}$ (here it defines
the activation gap of the compound exciton, $E_{ac}^{(m)}$) is given by the
RHS of Eq. (\ref{140}), for $n_{ys}^{j_{0}i_{0}}=0$ and $\tilde{n}=1$
(notice, $\tilde{n}=-1$ shows the same result), as follows 
\begin{equation}
E_{ac}^{(m)}=\frac{e^{2}}{\varepsilon \ell_{0}}
\left(F_{di}^{(m)}(0;1)+F_{xc}^{(m)}(0;1)+2 F_{2}(m)\right) .  \label{142}
\end{equation}%
Using Eq. (\ref{137}) and Eq. (\ref{139}), we calculate numerically that $%
F_{di}^{(3)}(0;1) \approx 0.104453$, $F_{di}^{(5)}(0;1) \approx 0.025723$, $%
F_{di}^{(7)}(0;1) \approx 0.010725$ and $F_{xc}^{(3)}(0;1) \approx -0.002878$%
, $F_{xc}^{(5)}(0;1) \approx -0.42 \times 10^{-5}$, $F_{xc}^{(7)}(0;1)
\approx -0.63 \times 10^{-8}$. Then substituting these numerical results,
along with given above numerical values of $F_{2}(3), \; F_{2}(5),\;
F_{2}(7) $, in Eq. (\ref{142}) we obtain that dimensionless activation gap
of the compound exciton $\Delta_{ac}^{(m)}=E_{ac}^{(m)}/(e^{2}/\varepsilon
\ell_{0}) $ is given, for $m=3, \; 5, \; 7$, as 
\begin{eqnarray}
&&\Delta_{ac}^{(3)} \approx 0.101596, \;\;\;\; \Delta_{ac}^{(5)} \approx
0.025719,  \notag \\
&&\Delta_{ac}^{(7)} \approx 0.010725 .  \label{143}
\end{eqnarray}%
It is seen that the main contribution to these $\Delta_{ac}^{(m)}$ comes
from the direct-alike term $F_{di}^{(m)}(0;1)$. Point out, this direct-alike
contribution should not be understood as strictly the direct (the Hartree)
term because some important correlations are already taken into account in
the form of many-body electron-ion wave functions, $\Psi_{\tilde{N},\tilde{N}%
}^{(m),eh}$ and $\Psi_{\tilde{N},\tilde{N};(m)}^{eh,i_{0};j_{0}}$, cf. with
the discussion below Eq. (\ref{57}).

Notice, from Eqs. (\ref{137})-(\ref{140}) for the compound exciton energy
larger than $\Delta_{ac}^{(m)}$, however, that is still the smallest one if
not count $\Delta_{ac}^{(m)}$, we obtain that: i) for $m=3$ it is $3.72$
times larger than $\Delta_{ac}^{(3)}$ (it corresponds, e.g., to $\tilde{n}=1$
and $n_{ys}^{j_{0}i_{0}}=-1$), ii) for $m=5$, it is $4.26$ times larger than 
$\Delta_{ac}^{(5)}$ (it corresponds, e.g., to $\tilde{n}=2$ and $%
n_{ys}^{j_{0}i_{0}}=0$), and iii) for $m=7$, it is $4.11$ times larger than $%
\Delta_{ac}^{(7)}$ (it corresponds, e.g., to $\tilde{n}=2$ and $%
n_{ys}^{j_{0}i_{0}}=0$).

Point out, it is shown above that the excitation energy of the compound
exciton for IJB coincides with the excitation energy of the relevant
compound exciton for UIB. In particular, for IJB the results Eqs. (\ref{131}%
)-(\ref{143}) are also valid.

\subsection{Energy of the compound spin-exciton}

For UIB, using the Hamiltonian Eq. (\ref{1}), we calculate the total energy
of the electron-ion system in the compound spin-exciton (remind, here $m
\geq 1$) state Eq. (\ref{114s}) as
\begin{equation}
E_{\tilde{N};(m)}^{i_{0},j_{0};\tilde{n},s}=\langle \Psi_{\tilde{N},\tilde{N}%
;(m)}^{i_{0},j_{0};\tilde{n},s} |\hat{H}_{\tilde{N},\tilde{N}} |\Psi_{\tilde{%
N},\tilde{N};(m)}^{i_{0},j_{0};\tilde{n},s}\rangle ,  \label{145s}
\end{equation}%
where in the RHS for the kinetic energy term we, similar to Eq. (\ref{121}),
obtain 
\begin{eqnarray}
&&\langle \Psi _{\tilde{N},\tilde{N};(m)}^{i_{0},j_{0};\tilde{n},s}|\hat{H}%
_{0}|\Psi _{\tilde{N},\tilde{N};(m)}^{i_{0},j_{0};\tilde{n},s}\rangle  \notag
\\
&&=\left[\hbar \omega_{c}-|g_{0}| \mu_{B} B\right]\tilde{N}/2+ |g_{0}|
\mu_{B} B,  \label{146s}
\end{eqnarray}%
here the Zeeman energy is included in $\hat{h}_{0}$ explicitly (then the RHS
of Eq. (\ref{41}) should be changed by the RHS of Eq. (\ref{121})).

Point out, for IJB in the RHS of Eq. (\ref{145s}) $\Psi _{\tilde{N},\tilde{N}%
;(m)}^{i_{0},j_{0};\tilde{n},s}$ is changed on $\Psi _{\tilde{N}%
;(m)}^{i_{0},j_{0};\tilde{n},s}$ and $\hat{H}_{\tilde{N},\tilde{N}}$ on $%
\hat{H}_{\tilde{N}}^{JB}$. Then Eq. (\ref{146s}) is correct after making the
former change in its LHS.

Using Eqs. (\ref{145s}), (\ref{40}), we obtain that the energy of the
excited-state Eq. (\ref{145s}) with respect to the energy of the
ground-state Eq. (\ref{40}), i.e., the energy of the compound spin-exciton $%
\Delta E_{i_{0},j_{0};\tilde{n}}^{(m),s}$, is given as 
\begin{equation}
\Delta E_{i_{0},j_{0};\tilde{n}}^{(m),s}= E_{\tilde{N};(m)}^{i_{0},j_{0};%
\tilde{n},s}-E^{(m),eh}_{\tilde{N}} .  \label{147s}
\end{equation}%
The energy of the compound spin-exciton, Eq. (\ref{147s}), after using Eqs. (%
\ref{41}), (\ref{146s}), obtains the form
\begin{equation}
\Delta E_{i_{0},j_{0};\tilde{n}}^{(m),s}=|g_{0}| \mu_{B} B + \Delta
E_{ei,(m)}^{i_{0},j_{0};\tilde{n},s}+ \Delta E_{ee,(m)}^{i_{0},j_{0};\tilde{n%
},s} ,  \label{148s}
\end{equation}%
where in the RHS the first term is related with contributions from Eqs. (\ref%
{41}), (\ref{146s}). In addition, it is taken into account that in the RHS
of Eq. (\ref{148s}) the term, related with the ion-ion potential $V_{ii}$, $%
\Delta E_{ii,(m)}^{i_{0},j_{0};\tilde{n},s}=0$. Indeed, in the RHS of Eq. (%
\ref{145s}) the term $\langle \Psi _{\tilde{N},\tilde{N};(m)}^{i_{0},j_{0};%
\tilde{n},s}|V_{ii}|\Psi _{\tilde{N},\tilde{N};(m)}^{i_{0},j_{0};\tilde{n}%
,s}\rangle$ coincides with the RHS of Eq. (\ref{43}).

Point out, for IJB in the RHS of Eq. (\ref{147s}) $E^{(m),eh}_{\tilde{N}}$
is changed on $E^{(m),JB}_{\tilde{N}}$ and in the RHS of Eq. (\ref{148s}) $%
\Delta E_{ei,(m)}^{i_{0},j_{0};\tilde{n},s}$ is changed on $\Delta
E_{eb,(m)}^{i_{0},j_{0};\tilde{n},s}$. It is taken into account that now in
the RHS of Eq. (\ref{148s}) the term related with IJB-IJB interaction
potential $V_{bb}$, $\Delta E_{bb,(m)}^{i_{0},j_{0};\tilde{n},s}=0$.

The treatment shows that in the RHS of Eq. (\ref{148s}) the term related
with electron-ion potential, 
\begin{eqnarray}
\Delta E_{ei,(m)}^{i_{0},j_{0};\tilde{n},s}&=&\langle \Psi_{\tilde{N},\tilde{%
N};(m)}^{i_{0},j_{0};\tilde{n},s} |V_{ei}|\Psi_{\tilde{N},\tilde{N}%
;(m)}^{i_{0},j_{0};\tilde{n},s}\rangle  \notag \\
&&-\langle \Psi_{\tilde{N},\tilde{N}}^{(m),eh}|V_{ei}| \Psi_{\tilde{N},%
\tilde{N}}^{(m),eh}\rangle ,  \label{149s}
\end{eqnarray}%
is equal to the RHS of Eq. (\ref{125}), i.e. 
\begin{equation}
\Delta E_{ei,(m)}^{i_{0},j_{0};\tilde{n},s}= \Delta E_{ei,(m)}^{i_{0},j_{0};%
\tilde{n}} .  \label{150s}
\end{equation}%
%
%
%
Point out, for IJB in Eq. (\ref{149s}) $\Delta E_{ei,(m)}^{i_{0},j_{0};%
\tilde{n},s}$ is changed on $\Delta E_{eb,(m)}^{i_{0},j_{0};\tilde{n},s}$, $%
V_{ei}$ on $V_{eb}$, $\Psi _{\tilde{N},\tilde{N};(m)}^{i_{0},j_{0};\tilde{n}%
,s}$ is changed on $\Psi _{\tilde{N};(m)}^{i_{0},j_{0};\tilde{n},s}$, and $%
\Psi_{\tilde{N},\tilde{N}}^{(m),eh}$ on $\Psi_{\tilde{N}}^{(m),JB}$. It is
easy to see that 
\begin{equation}
\Delta E_{eb,(m)}^{i_{0},j_{0};\tilde{n},s}= \Delta E_{ei,(m)}^{i_{0},j_{0};%
\tilde{n},s} .  \label{150sd}
\end{equation}

Further, in the RHS of Eq. (\ref{148s}) the term related with
electron-electron potential, 
\begin{eqnarray}
\Delta E_{ee,(m)}^{i_{0},j_{0};\tilde{n},s}&=&\langle \Psi_{\tilde{N},\tilde{%
N};(m)}^{i_{0},j_{0};\tilde{n},s} |V_{ee}|\Psi_{\tilde{N},\tilde{N}%
;(m)}^{i_{0},j_{0};\tilde{n},s}\rangle  \notag \\
&&-\langle \Psi_{\tilde{N},\tilde{N}}^{(m),eh}|V_{ee}| \Psi_{\tilde{N},%
\tilde{N}}^{(m),eh}\rangle ,  \label{151s}
\end{eqnarray}%
obtains (cf. with Eqs. (\ref{126})-(\ref{129})) the form 
\begin{equation}
\Delta E_{ee,(m)}^{i_{0},j_{0};\tilde{n},s}= \Delta
E_{ee,(m)}^{di,s}(i_{0},j_{0};\tilde{n})+ \Delta
E_{ee,(m)}^{xc,s}(i_{0},j_{0};\tilde{n}) ,  \label{152s}
\end{equation}%
where the direct-alike (or the Hartree-alike) contribution 
\begin{eqnarray}
\Delta E_{ee,(m)}^{di,s}(i_{0},j_{0};\tilde{n})= \Delta
E_{ee,(m)}^{di}(i_{0},j_{0};\tilde{n}) ,  \label{153s}
\end{eqnarray}%
i.e., $\Delta E_{ee,(m)}^{di,s}(i_{0},j_{0};\tilde{n})$ is given by the RHS
of Eq. (\ref{128}). %
Now the exchange-alike (or the Fock-alike) contribution, cf. with Eq. (\ref%
{129}), is given as 
\begin{eqnarray}
&&\Delta E_{ee,(m)}^{xc,s}(i_{0},j_{0};\tilde{n})= \frac{1}{m}
\sum_{n=-\ell}^{\ell} \sum_{i=1; i \neq i_{0}}^{\tilde{N}}
\sum_{k=-\infty}^{\infty}  \notag \\
&&\times \int d \mathbf{r} \int d \mathbf{r}^{\prime} \frac{e^{2}}{%
\varepsilon |\mathbf{r}-\mathbf{r}^{\prime}- k L_{x}^{\square} \hat{\mathbf{x%
}}|} \; \; \varphi _{k_{xi}^{(n)}}^{(m) \ast}(\mathbf{r}) \; \varphi
_{k_{xi}^{(n)}}^{(m)}(\mathbf{r}^{\prime})  \notag \\
&&\times \varphi _{k_{xi_{0}}^{(n)}}^{(m)}(\mathbf{r}) \varphi
_{k_{xi_{0}}^{(n)}}^{(m) \ast}(\mathbf{r}^{\prime}) .  \label{154s}
\end{eqnarray}%
Notice, Eq. (\ref{154s}) it follows from the RHS of Eq. (\ref{129}) after
formally omitting the first term in the square brackets. Indeed, for the
compound spin-exciton this term vanishes, as it includes both the spin up, $%
|1>$, and the spin down, $|-1>$, spin wave functions.

Point out, for IJB in Eq. (\ref{151s}) $\Psi _{\tilde{N},\tilde{N}%
;(m)}^{i_{0},j_{0};\tilde{n},s}$ is changed on $\Psi _{\tilde{N}%
;(m)}^{i_{0},j_{0};\tilde{n},s}$, and $\Psi_{\tilde{N},\tilde{N}}^{(m),eh}$
on $\Psi_{\tilde{N}}^{(m),JB}$. Then it is easy to see that Eqs. (\ref{152s}%
)-(\ref{154s}) are obtained again.

Comparing Eqs. (\ref{150s})-(\ref{153s}) with Eqs. (\ref{125})-(\ref{137}),
we readily obtain that
\begin{equation}
\Delta E_{ei,(m)}^{i_{0},j_{0};\tilde{n},s}+ \Delta
E_{ee,(m)}^{di,s}(i_{0},j_{0};\tilde{n})= \frac{e^{2}}{\varepsilon \ell_{0}}
F_{di}^{(m)}(n_{ys}^{j_{0}i_{0}};\tilde{n}) .  \label{155s}
\end{equation}

In addition, in Appendix A it is shown that from Eqs. (\ref{154s}) it
follows (cf. with Eq. (\ref{138})) that 
\begin{equation}
\Delta E_{ee,(m)}^{xc,s}(i_{0},j_{0};\tilde{n})= \frac{2e^{2}}{\varepsilon
\ell_{0}} F_{2}(m) ,  \label{156s}
\end{equation}%
where $F_{2}(m)$ is given by Eq. (\ref{90}).

Using Eqs. (\ref{149s})-(\ref{156s}), we rewrite the energy of the compound
spin-exciton Eq. (\ref{148s}), for $m \geq 1$, as follows 
\begin{equation}
\Delta E_{i_{0},j_{0};\tilde{n}}^{(m),s}=|g_{0}| \mu_{B} B + \frac{e^{2}}{%
\varepsilon \ell_{0}} \left(F_{di}^{(m)}(n_{ys}^{j_{0}i_{0}};\tilde{n}%
)+2F_{2}(m)\right) ,  \label{157s}
\end{equation}%
where $\tilde{n}=0, \pm 1,\ldots,\pm \ell$; $%
F_{di}^{(m)}(n_{ys}^{j_{0}i_{0}};\tilde{n})$ is given by Eq. (\ref{137})
both for $m=1$ and $m \geq 3$.

First, we obtain the difference between the energy of the compound
spin-exciton, Eq. (\ref{157s}), and the energy of the compound exciton, Eq. (%
\ref{140}), for the same $m$, $n_{ys}^{j_{0},i_{0}}$, and $\tilde{n}$ (i.e.,
here $m \geq 3$, $\tilde{n}=\pm 1,\ldots,\pm \ell$), as follows 
\begin{equation}
\Delta E_{i_{0},j_{0};\tilde{n}}^{(m),s}- \Delta E_{i_{0},j_{0};\tilde{n}%
}^{(m)}=|g_{0}| \mu_{B} B - \frac{e^{2}}{\varepsilon \ell_{0}}
F_{xc}^{(m)}(n_{ys}^{j_{0}i_{0}};\tilde{n}) ,  \label{158s}
\end{equation}%
where the RHS is always positive as $|g_{0}| \mu_{B} B>0$ and $%
F_{xc}^{(m)}(n_{ys}^{j_{0}i_{0}};\tilde{n}) <0$. Moreover, it is seen (for
given $m \geq 3$, $\tilde{n}=\pm 1,\ldots,\pm \ell$ ) that $%
|F_{xc}^{(m)}(n_{ys}^{j_{0}i_{0}};\tilde{n})|$ is minimal for $%
n_{ys}^{j_{0}i_{0}}=0$ and $\tilde{n}=1$ (for $\tilde{n}=-1$, its value is
the same). I.e., from Eqs. (\ref{140}), (\ref{142}), (\ref{158s}) we obtain
that 
\begin{eqnarray}
&&\min\{\Delta E_{i_{0},j_{0};\tilde{n}}^{(m),s}- \Delta E_{i_{0},j_{0};%
\tilde{n}}^{(m)}\}= \min\{\Delta E_{i_{0},j_{0};\tilde{n}}^{(m),s}-
E_{ac}^{(m)}\}  \notag \\
&&=|g_{0}| \mu_{B} B - \frac{e^{2}}{\varepsilon \ell_{0}} F_{xc}^{(m)}(0;1) ,
\label{159s}
\end{eqnarray}%
where the numerical values of $F_{xc}^{(m)}(0;1)$, for $m=3, \;5$, and $7$,
are given above. Then it is seen that for $m \geq 5$ in the RHS of Eq. (\ref%
{159s}) we typically (estimations are made for conditions relevant to
GaAs-based samples, in particular: $g_{0}=-0.44$, $\varepsilon=12.5$, $%
m^{\ast}/m_{0}=0.067$) obtain that only the first term, due to the ``bare''
Zeeman spin splitting energy, is essential. In addition, even though for $%
m=3 $ the relative role of the bare Zeeman spin splitting energy in the RHS
of Eq. (\ref{159s}) is much smaller than for $m \geq 5$, however, the Zeeman
term is still dominant for $m=3$ as well. Indeed, for $n^{eh} \approx 1.26
\times 10^{11} cm^{-2}$, $B \approx 15.6$T we have (here $e^{2}/\varepsilon
\ell_{0} \approx 200$K, $\hbar \omega_{c} \approx 300$K) that $|g_{0}|
\mu_{B} B \approx \hbar \omega_{c}/68 \approx 4.4$K is more than seven times
greater than $(e^{2}/\varepsilon \ell_{0})|F_{xc}^{(3)}(0;1)| \approx 0.6$K.

Now we will study, both at $m \geq 3$ and $m=1$, the energy of the compound
spin-exciton for $\tilde{n}=0$ and $n_{ys}^{j_{0}i_{0}}=0$. Then from Eq. (%
\ref{137}) it follows that $F_{di}^{(m)}(0;0)=0$, i.e., the result that on
physical grounds is well understood, so we obtain from Eq. (\ref{157s}) that 
\begin{equation}
\Delta E_{i_{0},i_{0};0}^{(m),s}=|g_{0}| \mu_{B} B + \frac{2e^{2}}{%
\varepsilon \ell_{0}} F_{2}(m) ,  \label{160s}
\end{equation}%
where the numerical values of $F_{2}(m)$, for $m=1, \;3, \;5$, and $7$, are
given below Eq. (\ref{90}). In particular, for GaAs-based sample at $m=1$
regime we can make typical estimations in the RHS of Eq. (\ref{160s}) as: $%
|g_{0}| \mu_{B} B \approx 0.015 \hbar \omega_{c}$ and $2F_{2}(1)(e^{2}/%
\varepsilon \ell_{0}) \approx 2F_{2}(1) \hbar \omega_{c} \approx 0.032 \hbar
\omega_{c}$. I.e., the RHS of Eq. (\ref{160s}) is roughly equal to $3|g_{0}|
\mu_{B} B$: so the exchange-alike contribution in Eq. (\ref{160s}) strongly
enhances, at $m=1$ (and $n_{ys}^{j_{0}i_{0}}=0$; $\tilde{n}=0$), the energy
of the compound spin-exciton in comparison to the bare Zeeman spin
splitting. However, for $m \geq 3$ the exchange-alike term in the
RHS of Eq. (\ref{160s}) can be already neglected as it is much smaller than $%
|g_{0}| \mu_{B} B$, for a typical GaAs-based sample: e.g., at $m=3$, $%
e^{2}/\varepsilon \ell_{0} \approx 200$K, $\hbar \omega_{c} \approx 300$K we
have in the RHS of Eq. (\ref{160s}) that $2F_{2}(3)(e^{2}/\varepsilon
\ell_{0})/(|g_{0}| \mu_{B} B) \approx 10^{-3}$.

In addition, it is important to treat the compound spin-exciton energy Eq. (%
\ref{157s}) for $m=1$ (here $\tilde{n}=0$) as well for $n_{ys}^{j_{0}i_{0}}
\neq 0$, here we have that 
\begin{equation}
\Delta E_{i_{0},j_{0};0}^{(1),s}=|g_{0}| \mu_{B} B + \frac{e^{2}}{%
\varepsilon \ell_{0}} \left(F_{di}^{(1)}(n_{ys}^{j_{0}i_{0}};0)+2F_{2}(1)%
\right) ,  \label{161s}
\end{equation}%
where $F_{di}^{(1)}(n_{ys}^{j_{0}i_{0}};0)$ grows monotonically with the
increase of $|n_{ys}^{j_{0}i_{0}}|$. In particular, for $%
|n_{ys}^{j_{0}i_{0}}| \gg 1$ we have that $%
F_{di}^{(1)}(n_{ys}^{j_{0}i_{0}};0) \approx \sqrt{2/\pi} [\ln\left(\sqrt{2\pi%
} |n_{ys}^{j_{0}i_{0}}|\right)- \left(\ln(2)-\gamma \right)/2]$, as the
second term in the RHS of Eq. (\ref{137}) is equal to zero. Notice, $%
F_{di}^{(1)}(n_{ys}^{j_{0}i_{0}};0)= F_{di}^{(1)}(-n_{ys}^{j_{0}i_{0}};0)$,
i.e., $F_{di}^{(1)}(n_{ys}^{j_{0}i_{0}};0)$ is even function over its
argument $n_{ys}^{j_{0}i_{0}}$. In addition, notice that Eq. (\ref{161s}) is
valid also for $n_{ys}^{j_{0}i_{0}}=0$, where $F_{di}^{(1)}(0;0)=0$ and,
respectively, Eq. (\ref{161s}) reduces to the form given by Eq. (\ref{160s}%
), at $m=1$. Point out, the monotonic increase of $%
F_{di}^{(1)}(n_{ys}^{j_{0}i_{0}};0)$, and the RHS of Eq. (\ref{161s}), take
place starting from $n_{ys}^{j_{0}i_{0}}=0$. As we pointed out in the end of
Sec. VI.A, for more details see below discussions, the minimal value of $%
\Delta E_{i_{0},j_{0};0}^{(1),s}$ given by Eq. (\ref{160s}) (or Eq. (\ref%
{161s}), for $n_{ys}^{j_{0}i_{0}}=0$) do not correspond to any relevant
kinetic coefficient (e.g., $\sigma_{yy}$) of the steady state (or direct
current) magnetotransport. So we need to calculate $\Delta
E_{i_{0},j_{0};0}^{(1),s}$, given by Eq. (\ref{161s}), for $%
n_{ys}^{j_{0}i_{0}}=1$ (for $n_{ys}^{j_{0}i_{0}}=-1$ its value is the same),
as 
\begin{equation}
E_{ac}^{(1),s}=|g_{0}| \mu_{B} B + \frac{e^{2}}{\varepsilon \ell_{0}}
\left(F_{di}^{(1)}(1;0)+2F_{2}(1)\right) ,  \label{162s}
\end{equation}%
where, using Eq. (\ref{137}), we calculate numerically that $%
F_{di}^{(1)}(1;0) \approx 1.15194$ 
and, finally, obtain that 
\begin{equation}
E_{ac}^{(1),s}=|g_{0}| \mu_{B} B +1.18431 \frac{e^{2}}{\varepsilon \ell_{0}}
.  \label{163s}
\end{equation}%
Notice that many-body contribution in the RHS of Eq. (\ref{163s}), $\approx
1.18431 e^{2}/\varepsilon \ell_{0}$ is a bit smaller than relevant result of
HFA $\sqrt{\pi/2} \; (e^{2}/\varepsilon \ell_{0})$; also cf. with pertinent
result of Refs. \onlinecite{bychkov1981,kallin1984,sondhi1993,zhang1986}.

Point out that even though the energy of the compound spin-exciton Eq. (\ref%
{160s}), at $m \geq 1$ (and $\tilde{n}=0$, $n_{ys}^{j_{0}i_{0}}=0$), is very
small in respect with the relevant activation gaps $E_{ac}^{(3)}$, $%
E_{ac}^{(5)}$, $E_{ac}^{(7)}$, at $m=3, \;5, \;7$ (given by Eqs. (\ref{142}%
)-(\ref{143})), and $E_{ac}^{(1),s}$, at $m=1$ (given by Eqs. (\ref{162s})-(%
\ref{163s})), it can be seen that such compound spin-excitons, with $\tilde{n%
}=0$ and $n_{ys}^{j_{0}i_{0}}=0$,
will not contribute (to my best knowledge
\cite{balevunpub2}; pertinent complete treatment is
beyond the scope of the
present study) to any pertinent magnetotransport coefficient (e.g., the
diagonal electrical conductance $\sigma_{yy}$) and, respectively, to the
pertinent activation gap. Hence, in the limit of small but nonzero impurity
concentration \cite{morf1986}, I expect that $\sigma_{yy}$ (in the Hall bar
sample it can be related with the electrical resistivity $\rho_{xx}$ Ref. 
\cite{balev1994}) will be thermally activated with the activation energy $%
E_{ac}^{(m)}/2$, at $m \geq 3$, and $E_{ac}^{(1),s}/2$, at $m=1$.

Indeed, typically it is assumed (cf. with Ref. \cite{morf1986}) that, to
obtain the activation gap, the observation of the dissipative conductance $%
\sigma_{yy}$ ($\rho_{yy}$ or $\rho_{xx}$) should be made in the limit of
small but nonzero impurity concentrations. I.e., it is assumed (or implicit 
\cite{morf1986}) that mainly elastic scattering of electrons by impurities
in the ``bulk'' of the channel (e.g., the Hall bar channel) contributes to $%
\sigma_{yy}$; i.e., the contributions related with scattering by edge states
are assumed negligible, cf. with Ref. \cite{balev1994} and references cited
therein. Then, e.g., we can speculate (by taking into account only elastic
scattering between the compound exciton states, the weakness of impurities
scattering potential, etc.) that $\sigma_{yy}$, calculated in the linear
response approximation within MS, should be given (cf., e.g.,
with Ref. \cite{balev1994}), as follows 
\begin{eqnarray}
\sigma_{yy} &\propto& \frac{1}{m} \sum_{n=-\ell}^{\ell} \sum_{i=1}^{\tilde{N}%
} \sum_{\tilde{n}_{0} \geq 1} \int_{-\infty}^{\infty} d\mathbf{q} \left(-%
\frac{\partial f}{\partial E}\right)_{E=\epsilon_{i, \tilde{n}_{0}}^{(m)}} 
\notag \\
&&\times [y_{0}(k_{xi}^{(n+\tilde{n}_{0})})- y_{0}(k_{xi}^{(n-\tilde{n}%
_{0})})]^{2} <U^{2}>_{\mathbf{q}}  \notag \\
&&\times |M(\mathbf{q};k_{xi}^{(n+\tilde{n}_{0})}, k_{xi}^{(n-\tilde{n}%
_{0})})|^{2} G_{i,\tilde{n}_{0};n}(\mathbf{q}),  \label{164s}
\end{eqnarray}%
where $\epsilon_{i,\tilde{n}}^{(m)}=\Delta E_{i,i;\tilde{n}}^{(m)}$, $%
<U^{2}>_{\mathbf{q}}$ is the impurity potential (or only its short-range
part) correlation function, $U(\mathbf{r})$ is the (single-particle)
impurity fluctuation potential within MS (or/and due to a weak
randomness of actual ion distribution), $f(E)$ is the Fermi-Dirac function, $%
G_{i,\tilde{n};n}(\mathbf{q})$ includes the energy $\delta$-function that
expresses the conservation of the energy in the process of scattering
between two compound exciton states $\tilde{n}=\pm \tilde{n}_{0}$ (here it
is essential only $\tilde{n}_{0}=1$) and some effects of impurity potential
(or only its long-range, in comparison with $\ell_{0}$, part). It is crucial
the presence in the RHS of Eq. (\ref{164s}) of the factor $[y_{0}(k_{xi}^{(n+%
\tilde{n}_{0})})- y_{0}(k_{xi}^{(n-\tilde{n}_{0})})]^{2}$, which shows that
a finite contribution to $\sigma_{yy}$ can be given only by processes of
scattering between such two states of the same energy that involve the real
space transition (that leads to usual diffusion alike form for $\sigma_{yy}$%
) of an electron charge between the quantized positions of $%
y_{0}(k_{x\alpha})$ within MS. In addition, it is important that
in Eq. (\ref{164s}) the value of the characteristic matrix element of the
transition, $|M(\mathbf{q};k_{xi}^{(n+1)}, k_{xi}^{(n-1)})|^{2} \sim \exp(-4
\pi/m)$, is not too small. However, the processes of elastic scattering, at $%
m \geq 1$, that involves the spacial transition (quantum diffusion) only of
the electron spin from one spin-exciton state with $\tilde{n}=0$ and $%
n_{ys}^{j_{0}^{(1)}}= n_{ys}^{i_{0}^{(1)}}$ (i.e., $%
n_{ys}^{j_{0}^{(1)},i_{0}^{(1)}}=0$) to any another spin-exciton state with $%
\tilde{n}=0$ and $n_{ys}^{j_{0}^{(2)}}= n_{ys}^{i_{0}^{(2)}} \neq
n_{ys}^{i_{0}^{(1)}}$ does not involve any real space transfer (or quantum
diffusion) of the electron charge and, respectively, does not contribute to $%
\sigma_{yy}$. More general study \cite{balevunpub2} also justifies the above
claim: that at $m \geq 3$ the activation gap is given by the compound
exciton activation gap $E_{ac}^{(m)}$, Eq. (\ref{142}) (at $m=3, \;5, \; 7$,
the dimensionless activation gap is given by Eq. (\ref{143})), and at $m=1$
the activation gap is given by the compound spin-exciton activation gap $%
E_{ac}^{(1),s}$, Eqs. (\ref{162s})-(\ref{163s}).

Point out, it is shown above that the excitation energy of the compound
spin-exciton for IJB coincides with the excitation energy of the relevant
compound spin-exciton for UIB. In particular, for IJB the results Eqs. (\ref%
{155s})-(\ref{163s}) are also valid; for IJB Eq. (\ref{164s}) is also
correct. Notice, except Eq. (\ref{164s}), all expressions of the present
study are given for the zero temperature, $T=0$.

\section{ Quantized Hall conductance}

For UIB, assuming that the Fermi level is located within the finite energy
gap between the ground-state and excited-states, the Hall conductance, $%
\sigma_{H}=-\sigma_{xy}$, can be calculated, within MS, from the
Kubo formula (notice, it readily gives, cf. Ref. \cite{chakraborty1995},
that $\sigma_{yy}=0$) as\cite{niu1985,chakraborty1995}
\begin{equation}
\sigma_{H}=\frac{ie^{2}\hbar}{L_{x}^{\square}L_{y}}\sum_{k(>0)} \frac{%
\langle 0|\tilde{v}_{x}|k\rangle \langle k|\tilde{v}_{y}|0\rangle- \langle 0|%
\tilde{v}_{y}|k\rangle \langle k|\tilde{v}_{x}|0\rangle}{(E_{k}-E_{0})^{2}} ,
\label{165}
\end{equation}%
where $k=0$ and $k=1,2,...$ correspond to the ground-state and excited
states of the Hamiltonian $\tilde{H}=\hat{H}_{\tilde{N}, \tilde{N}%
}+\sum_{i=1}^{\tilde{N}} U(\mathbf{r}_{i})$; $\tilde{H}|k\rangle=E_{k}
|k\rangle$. Here the "ideal" many-body Hamiltonian $\hat{H}_{\tilde{N},%
\tilde{N}}$ is given by Eq. (\ref{1}) and $U(\mathbf{r})$ is a static
fluctuation potential (e.g., due to a weak randomness of actual ion
distribution). The velocity operators $\tilde{v}_{\mu}=\sum_{i=1}^{N}
v_{i\mu} $, $\mu=x, y$; here $v_{ix}=(-i\hbar/m^{\ast})\partial/\partial
x_{i}-\omega_{c}y_{i}$, $v_{iy}=(-i\hbar/m^{\ast})\partial/\partial y_{i}$.
To calculate the RHS of Eq. (\ref{165}), we will use the many-body operator
identities, that generalize the single-electron identities (8a), (8b) of
Ref. \cite{usov1984}, of the following form 
\begin{eqnarray}
&&\tilde{v}_{x}=\frac{\ell_{0}^{2}}{\hbar} \sum_{i=1}^{\tilde{N}}\frac{%
\partial}{\partial y_{i}}V_{ef}(\mathbf{r}_{i}) -\frac{i}{\hbar \omega_{c}} [%
\tilde{v}_{y},\tilde{H}]  \notag \\
&&\tilde{v}_{y}=\; - \frac{\ell_{0}^{2}}{\hbar} \sum_{i=1}^{\tilde{N}}\frac{%
\partial}{\partial x_{i}} V_{ef}(\mathbf{r}_{i}) +\frac{i}{\hbar \omega_{c}}
[\tilde{v}_{x},\tilde{H}] ,  \label{166}
\end{eqnarray}%
where $V_{ef}(\mathbf{r})=U(\mathbf{r})-\sum_{j=1}^{\tilde{N}}
\sum_{k_{1}=-\infty}^{\infty}e^{2}/(\varepsilon |\mathbf{r}-\mathbf{R}_{j}
-k_{1}L_{x}^{\square} \hat{\mathbf{x}}|)$. Notice, that the last term in the
RHS of $V_{ef}(\mathbf{r})$ will give the interaction of an electron at $%
\mathbf{r}$ with the homogeneous ion background. Point out, if in Eq. (\ref%
{166}) to assume that $V_{ef}(\mathbf{r}) \equiv const(\mathbf{r})$, i.e.,
independent of $\mathbf{r}$, than Eq. (\ref{166}) readily follows from Eq.
(5) of Ref. \cite{kohn1961}.

Further, we will neglect by the fluctuation potential $U(\mathbf{r})$, until
it is not stated otherwise. Then taking the ground-state wave function as
given by $\Psi _{\tilde{N},\tilde{N}}^{(m),eh}$, Eq. (\ref{18}), it is
natural to assume that for all many-body wave functions $|k\rangle $ the
part related with ions has the same form, $\dprod\limits_{i=1}^{\tilde{N}%
}\phi _{n_{ys}^{(i)}}(\mathbf{R}_{i})$, as in $\Psi _{\tilde{N},\tilde{N}%
}^{(m),eh}$ (or in Eqs. (\ref{114}), (\ref{114s})). I.e., the ion background
is fixed and has exactly homogeneous the ion charge density. Then from Eq. (%
\ref{166}) we obtain that
\begin{eqnarray}
\langle 0|\tilde{v}_{x}|k\rangle &=&\;-\frac{i(E_{k}-E_{0})}{\hbar \omega
_{c}}\langle 0|\tilde{v}_{y}|k\rangle ,  \notag \\
\langle k|\tilde{v}_{y}|0\rangle &=&\frac{i(E_{0}-E_{k})}{\hbar \omega _{c}}%
\langle k|\tilde{v}_{x}|0\rangle ,  \label{167}
\end{eqnarray}%
where it is taken into account that in the RHS of Eq. (\ref{167}) already
integration over the coordinates of ions gives that 
\begin{equation}
-\frac{\ell _{0}^{2}}{\hbar }\langle 0|\sum_{i=1}^{\tilde{N}}\frac{\partial 
}{\partial y_{i}}\sum_{j=1}^{\tilde{N}}\sum_{k_{1}=-\infty }^{\infty }\frac{%
e^{2}}{\varepsilon |\mathbf{r}_{i}-\mathbf{R}_{j}-k_{1}L_{x}^{\square }\hat{%
\mathbf{x}}|}|k\rangle =0,  \label{168}
\end{equation}%
and 
\begin{equation}
\frac{\ell _{0}^{2}}{\hbar }\langle k|\sum_{i=1}^{\tilde{N}}\frac{\partial }{%
\partial x_{i}}\sum_{j=1}^{\tilde{N}}\sum_{k_{1}=-\infty }^{\infty }\frac{%
e^{2}}{\varepsilon |\mathbf{r}_{i}-\mathbf{R}_{j}-k_{1}L_{x}^{\square }\hat{%
\mathbf{x}}|}|0\rangle =0.  \label{169}
\end{equation}%
In particular, Eqs. (\ref{168})-(\ref{169}) are obtained after using in
their LHS of Eq. (\ref{66}), calculating the matrix elements over the ions
coordinates, $\mathbf{R}_{j}$, the sum over $k_{1}$ and finally the
integrals over $q_{x}$ and $q_{y}$. Indeed, then as the result we have that
the LHS of Eq. (\ref{168}) and Eq. (\ref{169}) tends to zero as $%
L_{x}^{\square }/(\delta \times L_{y})\rightarrow 0$ and $L_{x}^{\square
}/(\delta \times L_{x})\rightarrow 0$, respectively, even for $|k\rangle
=|0\rangle $; for $|k\rangle \neq |0\rangle $ an additional infinitely small
factor will appear in the LHS of Eqs. (\ref{168})-(\ref{169}).

Using Eq. (\ref{167}) and similar relations for $\langle 0|\tilde{v}%
_{y}|k\rangle$, $\langle k|\tilde{v}_{x}|0 \rangle$, from Eq. (\ref{165}) it
follows that 
\begin{equation}
\sigma_{H}=\frac{ie^{2}}{\hbar \omega_{c}^{2} L_{x}^{\square}L_{y}}%
\sum_{k(>0)} \{\langle 0|\tilde{v}_{x}|k\rangle \langle k|\tilde{v}%
_{y}|0\rangle- \langle 0|\tilde{v}_{y}|k\rangle \langle k|\tilde{v}%
_{x}|0\rangle\} ,  \label{170}
\end{equation}%
where, due to the properties (they follow as from Eq. (\ref{167}) so from
direct calculation of these matrix elements) $\langle 0|\tilde{v}%
_{\mu}|0\rangle=0$, $\mu=x,y$, the value of the RHS will not be changed by
adding $k=0$ term to the sum. I.e., in the RHS of Eq. (\ref{170}) we can
change $\sum_{k(>0)}$ on $\sum_{k}$. Then Eq. (\ref{170}) gives, cf. with
Ref. \onlinecite{usov1984}, that 
\begin{equation}
\sigma_{H}=\frac{ie^{2}}{\hbar \omega_{c}^{2} L_{x}^{\square}L_{y}} \langle
0|[\tilde{v}_{x},\tilde{v}_{y}]|0\rangle ,  \label{171}
\end{equation}%
where the many-body commutator can be further simplified as $[\tilde{v}_{x},%
\tilde{v}_{y}]=\sum_{i}^{\tilde{N}}[v_{ix},v_{iy}]$. Further, applying the
single-electron commutator \cite{landau1965} $[v_{jx},v_{jy}]= \; -i\hbar
\omega_{c}/m^{\ast}$, we obtain that $[\tilde{v}_{x},\tilde{v}_{y}]= \;
-i\hbar \omega_{c} \tilde{N}/m^{\ast}$. Using the latter exact result in Eq.
(\ref{171}) we obtain that 
\begin{equation}
\sigma_{H}=\frac{e^{2} \tilde{N}}{\omega_{c} m^{\ast} L_{x}^{\square}L_{y}}= 
\frac{e^{2}}{\omega_{c} m^{\ast} (L_{x}^{\square})^{2}} ,  \label{172}
\end{equation}%
where the last form is obtained by using Eq. (\ref{9}). Finally, using Eq. (%
\ref{13}) in Eq. (\ref{172}) we have 
\begin{equation}
\sigma_{H}=\frac{e^{2}}{\omega_{c} m^{\ast} 2\pi m \ell_{0}^{2}}= \frac{e^{2}%
}{ 2m \pi \hbar} ,  \label{173}
\end{equation}%
i.e., for $m=3, \; 5, \; 7, ....$ the ground-state $\Psi _{\tilde{N},\tilde{N%
}}^{(m),eh}$ corresponds to the fractional quantum Hall effect, $\nu=1/m$.

Now, again neglecting by a weak random potential $U(\mathbf{r})$, we will
calculate the Hall conductance of the quantum Hall system pertinent to the
ground-state $\Psi _{\tilde{N},\tilde{N}}^{(m),eh}$ in a different manner.
Here, similar to Refs. \cite{balev1990,balev1993}, we assume that a static
electric field $\mathbf{E}=E_{H} \hat{\mathbf{y}}$ is applied; it is
implicit that an adiabatic process of turning on of this electric field is
already over. Then the total many-body Hamiltonian $\hat{H}_{\tilde{N},%
\tilde{N}}^{E_{H}}$ it follows from $\hat{H}_{\tilde{N},\tilde{N}}$, Eq. (%
\ref{1}), after changing of $\hat{H}_{0}$ on $\hat{H}_{0}^{E_{H}}$. Where in
the latter $\hat{h}_{0i}$ is changed on $\hat{h}_{0i}-eE_{H}y_{i}$. Then in
Eqs. (\ref{6}), (\ref{7}): $\hbar \omega_{c}(n_{\alpha}+1/2)$ is changed on $%
\hbar \omega_{c}(n_{\alpha}+1/2)-(eE_{H}/m^{\ast}\omega_{c}) [\hbar
k_{x\alpha}+eE_{H}/2\omega_{c}]$, $y_{0}(k_{x\alpha})=\ell_{0}^{2}
k_{x\alpha}$ is changed on $y_{0}^{E_{H}}(k_{x\alpha})=\ell_{0}^{2}
k_{x\alpha}+ eE_{H}/m^{\ast} \omega_{c}^{2}$ and, respectively, $%
\psi_{n_{\alpha};k_{x\alpha}}^{L_{x}^{\square}}(\mathbf{r})$ on $%
\psi_{n_{\alpha};k_{x\alpha}}^{L_{x}^{\square};E_{H}}(\mathbf{r})$. Then $%
\varphi _{k_{xi}^{(n)}}^{(m)}(\mathbf{r})$, Eq. (\ref{15}), is changed on $%
\varphi _{k_{xi}^{(n)}}^{(m);E_{H}}(\mathbf{r})$ and pertinent change should
be done in $\Psi_{\tilde{N},\tilde{N}}^{(m),eh}$, Eq. (\ref{18}),
transforming it to nonequilibrium many-body wave function $\Psi_{\tilde{N},%
\tilde{N}}^{(m),eh;E_{H}}$. Further, the net current, $I_{x}$, in the latter
state is obtained as
\begin{eqnarray}
I_{x}&=&\frac{e}{L_{x}^{\square}} \langle \Psi _{\tilde{N},\tilde{N}%
}^{(m),eh;E_{H}}| \sum_{j=1}^{\tilde{N}} \hat{v}_{jx} |\Psi _{\tilde{N},%
\tilde{N}}^{(m),eh;E_{H}}\rangle  \notag \\
&&=\frac{e}{m L_{x}^{\square}} \sum_{n=-\ell }^{\ell } \sum_{i=1}^{\tilde{N}%
} \int_{L_{x}^{\square}(n_{xs}^{\alpha}-1)}^{L_{x}^{\square}n_{xs}^{\alpha}}
d x \int_{-\infty}^{\infty} dy  \notag \\
&&\times \varphi_{k_{xi}^{(n)}}^{(m);E_{H} \ast}(\mathbf{r}) \hat{v}_{x}
\varphi _{k_{xi}^{(n)}}^{(m);E_{H}}(\mathbf{r}) ,  \label{174}
\end{eqnarray}%
where the matrix element $\langle \varphi _{k_{xi}^{(n)}}^{(m);E_{H}}| \hat{v%
}_{x}|\varphi _{k_{xi}^{(n)}}^{(m);E_{H}}\rangle= \; - eE_{H}/m^{\ast}
\omega_{c}$. Using the latter, Eq. (\ref{174}) gives 
\begin{equation}
\frac{I_{x}}{V_{H}}=-\frac{e^{2} \tilde{N}}{L_{x}^{\square} L_{y} m^{\ast}
\omega_{c}},  \label{175}
\end{equation}%
where $V_{H}=E_{H} \times L_{y}$ is the Hall voltage. As $%
I_{x}/V_{H}=\sigma_{xy}= \; - \sigma_{H}$, from Eq. (\ref{175}) it follow
Eqs. (\ref{172})-(\ref{173}).

Point out, the above treatment of Sec. VII can be readily extended as well
on IJB, i.e., the quantum Hall system at ground-state $\Psi_{\tilde{N}%
}^{(m),JB}$.

Rather similar to Refs. \onlinecite{laughlin1983,niu1985}, we can speculate
that for a weak disorder if the Fermi level still lies in a gap or mobility
gap the Hall conductance should be quantized in agreement with Eq. (\ref{173}%
) as for IJB so for UIB.

\section{Concluding Remarks}

Present study shows (see, in particular, Secs. II C, V C) that
the ground-state and the lowest excited-state
can correspond to partial crystal-like correlation order, Eq. (\ref{DA7}),
among $N$ electrons of MR. As a result, the treatment of
2DES of $N$ electrons within MR is reduced straightforwardly to the study
of 2DES of $\tilde{N}$ electrons ($\tilde{N} \rightarrow \infty$ and
$\tilde{N}/N \rightarrow 0$) localized within MS; with the PBC,
of the period $L_{x}^{\square}$, imposed along $x$.

In particular, present study shows that proper PBC can be totally
relevant to symmetry, periodicity, correlations, and etc. properties
of a sought state. So it will not lead to any
oversimplifications or artificial ``boundary effects''.

In this work, I have presented many-body ground-state wave functions Eq. (%
\ref{18}) and Eq. (\ref{18d}) of the quantum Hall systems (at $\nu =1/m$, $%
m=3,\;5,\ldots $ and $m=1$) for UIB model and IJB model, respectively. For
both these models the charge density of ion background is exactly
homogeneous. However, only for IJB model the ion background is totally
equivalent to the model of ion background used in Ref. \cite{laughlin1983}.
So only the results obtained for IJB can be compared with the results of
Laughlin \cite{laughlin1983} (or studies based on model of Ref. \cite%
{laughlin1983}) directly. For UIB the ground-state
energy is much lower than for IJB due to the difference
between the terms related with ion-ion interaction in these two models. The
\textquotedblleft electron \textquotedblright\ part in the energy (that
includes contributions from electron-electron and electron-ion interactions)
is the same for IJB and UIB. Above it is shown that, per electron, for IJB
the ground-state energy, at $\nu =1/3,\;1/5$ and $1$, is substantially lower
than obtained in well known study \cite{laughlin1983}. It is important that
in the present study (both for IJB and UIB), similar with the model of Ref. 
\cite{laughlin1983}, the ions are localized within the same 2D-plane as
2DES; see also, e.g., Refs. \cite{morf1986,bonsall1977,fano1986}.


Point out that in UIB model I assume ideally periodic distribution of ions
(it follows from the form of the many-body wave functions, e.g., Eqs. (\ref%
{18}), (\ref{114}), and the Hamiltonian Eq. (\ref{1})) with homogeneous ion
density, within 2D-plane. Notice that partly similar model of impurities
that form a regular crystallic lattice (so-called, impurity sublattice)
having a much larger period than the host lattice is widely used to study
the impurity band, etc. of doped semiconductors, see, e.g., Ref. \cite%
{efros1984}.
Notice, definition of very similar model (with UIB model)
of ion background is given by Mahan \cite{mahan1981} as ``One can think
of taking the positive charge of the ions and spreading it uniformly
about unit cell of the crystal''. Indeed, if we have one ion of the charge
$|e|$ within unit cell and will spread it uniformly about our square
unit cell and then take into account that direct interaction of any
ion with itself must be excluded (independently if it ``shape'') we arrive
to UIB model. On the other hand, if we assume that the ion is
spread within rectangular unit cell $L_{x}^{ap} \times L_{y}^{ap}$
of the same area $(L_{x}^{\square})^{2}$ but with, e.g.,
$L_{x}^{ap}/L_{y}^{ap} \rightarrow \infty$ it is easy to see that
we arrive to IJB model. Notice, direct interaction of ion with
itself tends to zero for spreading within above rectangular unit
cell although it is finite for spreading within our square unit cell.

Point out that, due to the quantized according to Eq. (\ref{19})
contributions from the partial many-electron wave functions Eq. (\ref{20}),
for $m \geq 3$ the compound form of the ground-state
wave function Eq. (\ref{18}) leads to the
compound structure of each electron already within MS.
In particular, this
compound structure of the electrons plays important role in the present
treatment of the excited-states. Due to PBC,
the charge density of the excited compound electron and hole (given within
MS as the superposition of $m$, strongly correlated,
quasielectrons and quasiholes, e.g., cf. Eqs. (\ref{118})-(\ref{119})) have
periodical images for $x$ outside MS, with the period $%
L_{x}^{\square}$. Respectively, the same property of the periodicity holds
for the charge density of the compound exciton and spin-exciton or the
charge density of the $n$-th quasiexciton, Eq. (\ref{119}). Point out, these
properties of the quasielectron and the quasihole periodicity are different
from the properties of the fractionally charged elementary excitations of
the Laughlin model \cite{laughlin1983}. Notice, that in the present model,
at $m \geq 3$, it is impossible to create only one quasielectron (quasihole)
without simultaneous excitation another $m-1$ strongly correlated
quasielectrons (quasiholes). Notice, above it is shown that the same
properties of periodicity, compound structure of electrons, holes, compound
excitons and spin-excitons, etc. also are valid for the model with IJB.

Further, from the exact analytical result, Eq. (\ref{142}), for the
activation gap, at $m\geq 3$, pertinent to the compound exciton, I have
obtained numerically that the activation gap at $m=3,\;5,\;7$ is given (in
units of $e^{2}/\varepsilon \ell _{0}$), according to Eq. (\ref{143}), by
(i) $\Delta _{ac}^{(3)}\approx 0.1016$, (ii) $\Delta _{ac}^{(5)}\approx
0.0257$, and (iii) $\Delta _{ac}^{(7)}\approx 0.0107$, respectively. Notice, 
$\Delta _{ac}^{(3)}$ is very close with typically calculated for the
Laughlin liquid pertinent excitation gap \cite%
{haldane1985,morf1986,macdonald1986} $\Delta _{ac,1/3}\approx 0.10\div 0.11$%
. In addition, obtained here $\Delta _{ac}^{(5)}$ is also rather close to
the calculated for the Laughlin liquid pertinent activation gap \cite%
{macdonald1986} $\Delta _{ac,1/5}\approx 0.031$. At $m=1$, from the exact
analytical result, Eq. (\ref{162s}), for the activation gap, pertinent to
the compound spin-exciton, I have obtained it value, Eq. (\ref{163s}), as $%
E_{ac}^{(1),s}\approx |g_{0}|\mu _{B}B+1.1843e^{2}/\varepsilon \ell _{0}$,
where the many-body contribution is a bit smaller than relevant result of
HFA $\sqrt{\pi /2}\;(e^{2}/\varepsilon \ell _{0})$, see also Refs. %
\onlinecite{bychkov1981,kallin1984,sondhi1993,zhang1986}. Notice, for IJB
above it is shown that any activation gap coincides with pertinent gap for
UIB. Moreover, for given $m$ all calculated energies for IJB (of the
ground-state and of the excited-states) can be obtained by upward shift,
on the same value, of the pertinent energies for UIB.

For a detailed comparison of the activation gaps with experiments it is
known that a finite thickness of 2DES should be taken into account as well
as effects of disorder, see, e.g., \cite{willett1988,morf2002} and
references cited therein. In addition, I speculate that many-body effects
similar to those studied in \cite{balev2001} (though for more
"traditional" $\nu =1$ state than the one
presented above) and related with edge states
possibly here also can lead to highly asymmetric pinning of the Fermi level
within the energy gap, as well at fractional $\nu $;
at least in high quality samples with weak disorder
and for very low temperatures, typical
for experiments on the fractional quantum Hall effect. Then, similar with, 
\cite{balev2001} actual activation energy can be substantially smaller than,
e.g., $E_{ac}^{(m)}/2$. However, pertinent calculations should involve
effects due to channel edges and there are beyond the scope of present
study. Other effect that will lead to a decrease of excitation gaps can be
related with a finite separation, along $z-$direction, due to finite spacer
layer of the neutralizing ion background plane from the 2D-plane of 2DES;
we, analogous to Ref. \cite{laughlin1983}, did not study this effect here.

Now we will outline effects of the changing
in Eqs. (\ref{DA7})-(\ref{DA8}) of the period $L_{x}^{\square}$
on the period $L_{x}^{ap}$ of arbitrary value.
Point out, in Eq. (\ref{DA7}) we actually should assume
arbitrary period \cite{balevunpub}
$L_{x}^{ap}=\eta_{a} \times L_{x}^{\square}$
instead of the period $L_{x}^{\square}$;
now a unit cell becomes rectangular
$L_{x}^{ap} \times L_{y}^{ap}$,
where $L_{y}^{ap}=L_{x}^{\square}/\eta_{a}$, etc.
Then we need to find an optimal value of $\eta_{a}$ that
we notate as $\eta_{a}^{\min}(m)$, for which
the trial wave function of a ground-state at $\nu=1/m$ will give
the lowest energy.
Study shows \cite{balevunpub} that
$\eta_{a}^{\min}(m)$ are very close to $1$, at the least, for
$m=1, 3, 5, 7$. The energy very slowly (parabolically) is dependent on
actual small deviations, $|\eta_{a}-\eta_{a}^{\min}(m)| \ll 1$.
Point out, at $m=3,5$ for $\eta_{a} \gg 1$ and $\eta_{a} \ll 1$ this form
of crystal-like correlation order shows that the lowest energy per electron
is substantially higher than relevant ground-state energies
of Ref. \cite{laughlin1983}. In particular, for $\eta_{a} \rightarrow
\infty$ instead of $U^{JB}(m;\eta_{a}=1)$, given for $m=1, 3, 5, 7$
by Eq. (\ref{66d}), we obtain \cite{balevunpub}
$U^{JB}(m;\eta_{a} \rightarrow \infty)=-m^{-1} \sqrt{\pi/8}$;
the latter coincides with the HFA result
$\epsilon_{HF}(\nu=1/m)/(e^{2}/\varepsilon \ell_{0})$.
If $\eta_{a} \rightarrow 0$ then it is seen that \cite{balevunpub}
$U^{JB}(m;\eta_{a})$ becomes positive and divergent (faster than
$\eta_{a}^{-1}$). This result is due to localization of electron
charge along, e.g., few ``lines'' of length $L_{x} \rightarrow \infty$,
of a typical width (along $y$) $\ell_{0}$, infinitely separated
from each other along $y$, within IJB.
In particular,
$U^{JB}(m;\eta_{a} \ll 1)$ gives much higher energy for
$m=1, 3, 5$ than obtained in Ref. \cite{laughlin1983}.
It is seen that $\eta_{a} \ll 1$ correspond
to very short period of PBC (and much stronger
crystal-like correlation order than for $\eta_{a}=1$)
while $\eta_{a} \rightarrow \infty$ correspond to
the (practical) absence of both the PBC and the crystal-like
correlation order.
Notice, we can speculate that, e.g., for Hall bar sample effect of
latteral confinement potential can define optimal value of
$\eta_{a}$ slightly different from $\eta_{a}^{\min}(m)$; however,
anyway it will be very close to $1$, if not equal.
Point out, that some conditions used in well
known studies Refs. \cite{yoshioka83i84,yoshioka84} do not allow
realization of the ground-state, at $\nu=1/3$, of present
type \cite{balevunpub}, with partial crystal-like correlation order.
In particular, in the case assumed as rather favorable in
Refs. \cite{yoshioka83i84,yoshioka84} when the number of electrons within
the rectangular unit cell, $n$, of \cite{yoshioka83i84,yoshioka84}
tends to infinity (then $n$ will correspond to our $\tilde{N}$;
notice, $b$ and $a$ of \cite{yoshioka83i84,yoshioka84}
will correspond to $L_{x}^{ap}$ and $L_{y}$),
while keeping $a/b=n/4$.
Indeed, then for the trial wave function with most relevant
to Refs. \cite{yoshioka83i84,yoshioka84} partial crystal-like correlation
order the energy is higher \cite{balevunpub} than even for the
charge-density wave or Wigner crystal states
\cite{chakraborty1995,yoshioka1983,maki1983,cabo2004,fertig1997},
at $\nu=1/3$.
So it is not too big surprise that numerical study of
Refs. \cite{yoshioka83i84,yoshioka84} (see also Refs.
\cite{chakraborty1995,chakraborty1986}) strongly indicates that
for $n \rightarrow \infty$ in their model the ground-state energy
tends to the result $\approx -0.410$ pertinent to the Laughlin's
trial wave function \cite{levesque1984,morf1986}; also it manifests
a liquidlike ground-state similar to the one of
Ref. \cite{laughlin1983}.

\acknowledgements

This work was supported in part by Universidade Federal do Amazonas,
Universidade Federal de S\~{a}o Carlos, by a grant from Funda\,{c}\~{a}o de
Amparo \`{a} Pesquisa do Estado de Amazonas.

\appendix

\section{Transformation of the compound exciton energy contributions}

Here we will transform the RHS of Eqs. (\ref{129})-(\ref{133}), i.e., the
contributions to the compound exciton energy, to rather simple analytical
forms by carrying out explicitly exact analytical calculations. In
particular, obtained analytical expressions are very suitable even for quite
simple numerical treatment. First, using Eqs. (\ref{66})-(\ref{68}), we
rewrite Eq. (\ref{132}) as 
\begin{eqnarray}
E_{1}^{(m)}(i_{0},j_{0}&;&\tilde{n})= \frac{e^{2}}{2 \pi \varepsilon}
\int_{-\infty}^{\infty} \int_{-\infty}^{\infty} \frac{e^{-q_{y}^{2}
\ell_{0}^{2}/4} \; d q_{x} d q_{y}}{\sqrt{q_{x}^{2}+q_{y}^{2}+\delta^{2}/%
\ell_{0}^{2}}}  \notag \\
&&\times S_{m}(q_{y}\ell_{0}) S_{m}^{2}(q_{x}\ell_{0})
[\sum_{k=-\infty}^{\infty} e^{-ik q_{x} L_{x}^{\square}}]  \notag \\
&& \times f_{m}(q_{y}\ell_{0}) \left(1-e^{i q_{y}
L_{x}^{\square}(n_{ys}^{j_{0}i_{0}}+ \frac{\tilde{n}}{m})} \right) ,
\label{A.1}
\end{eqnarray}%
where the sum over $n$ is carried out. Further, using Eq. (\ref{28}), in Eq.
(\ref{A.1}) the sum over $k$ in the square brackets obtains the form given
by the first line of Eq. (\ref{77}). Then the integral over $q_{x}$ is
calculated with the help of delta-functions and as a result the factor $%
S_{m}^{2}(q_{x}\ell_{0}) \rightarrow S_{m}^{2}(\sqrt{\frac{2 \pi}{m}}M_{x})$%
. The latter there is equal to $1$ for $M_{x}=0$ while it is equal to zero
for any $M_{x} \neq 0$; see also the paragraph below Eq. (\ref{80}).
Therefore we readily arrive from Eq. (\ref{A.1}) to the form given by Eq. (%
\ref{134}).

Further, using Eqs. (\ref{66})-(\ref{68}), we rewrite Eq. (\ref{133}) as
follows
\begin{eqnarray}
E_{2}^{(m)}(i_{0},j_{0}&;&\tilde{n})=- \frac{e^{2}}{2 \pi \varepsilon}
\int_{-\infty}^{\infty} \int_{-\infty}^{\infty} \frac{G_{m}^{v}(q_{y}%
\ell_{0}) \; d q_{x} d q_{y}}{\sqrt{q_{x}^{2}+q_{y}^{2}+\delta^{2}/%
\ell_{0}^{2}}}  \notag \\
&&\times S_{m}^{2}(q_{x}\ell_{0}) \left(1-e^{i q_{y}
L_{x}^{\square}(n_{ys}^{j_{0}i_{0}}+ \frac{\tilde{n}}{m})} \right)  \notag \\
&&\times \sum_{k=-\infty}^{\infty} \sum_{m_{y}=-\infty}^{\infty;m_{y} \neq
0} e^{-ik q_{x} L_{x}^{\square}} e^{-im_{y} q_{y} L_{x}^{\square}} ,
\label{A.2}
\end{eqnarray}%
where $G_{m}^{v}(\eta)=e^{-\eta^{2}/4}[e^{-\eta^{2}/4}- f_{m}(\eta) \;
S_{m}(\eta)]$; $m_{y}=n_{ys}^{(i)}-n_{ys}^{(i_{0})} \neq 0$. The double sums
in the RHS of Eq. (\ref{A.2}) are the same as in the RHS of Eq. (\ref{75}).
Then using for them the form given by Eq. (77) and calculating the integrals
with the help of delta-functions, where they are present, we obtain that
\begin{eqnarray}
E_{2}^{(m)}(i_{0},j_{0}&;&\tilde{n})= \frac{2e^{2}}{\varepsilon
L_{x}^{\square}} \int_{0}^{\infty} \frac{d \eta}{\eta} G_{m}^{v}(\eta) [1 
\notag \\
&&-\cos\left(q_{y} L_{x}^{\square}(n_{ys}^{j_{0}i_{0}}+ \frac{\tilde{n}}{m}%
)\right)]- \frac{2\pi e^{2}}{\varepsilon (L_{x}^{\square})^{2}}  \notag \\
&& \times \sum_{M_{y}=-\infty}^{\infty} \frac{G_{m}^{v}(\sqrt{2\pi/m} \;
M_{y})}{ \sqrt{(2\pi/L_{x}^{\square})^{2} M_{y}^{2}+\delta^{2}/\ell_{0}^{2}}}
\notag \\
&&\times \left(1-e^{-i 2\pi M_{y} (n_{ys}^{j_{0}i_{0}}+ \frac{\tilde{n}}{m}%
)} \right) ,  \label{A.3}
\end{eqnarray}%
where it is taken into account that the factor $S_{m}^{2}(q_{x}\ell_{0})
\rightarrow S_{m}^{2}(\sqrt{\frac{2 \pi}{m}}M_{x})$ is equal to $1$ for $%
M_{x}=0$ while it is zero for any $M_{x} \neq 0$. Now, using that the
contribution from the $M_{y}=0$ term of the sum over $M_{y}$ in the RHS of
Eq. (\ref{A.3}) is exactly equal to zero (it is easy to see as, for $M_{y}=0$%
, both $G_{m}^{v}(0)=0$ and the last factor, within the round brackets, are
exactly equal to zero and the denominator, $\delta/\ell_{0}$, does not equal
to zero even though it is very small), we readily obtain Eq. (\ref{A.3}) in
the form given by Eq. (\ref{135}).

For a transformation of the exchange-alike contribution Eq. (\ref{129}),
first we use in its RHS Eqs. (\ref{66}), (\ref{86}). Then, for $m \geq 3$,
we obtain that
\begin{eqnarray}
\Delta E_{ee}^{xc}(m&;&i_{0},j_{0})=- \frac{e^{2}}{2 \pi \varepsilon}
\sum_{i=1;i\neq i_{0}}^{\tilde{N}} \int_{-\infty}^{\infty} dq_{x}
\int_{-\infty}^{\infty} dq_{y}  \notag \\
&&\frac{e^{-q_{y}^{2}\ell_{0}^{2}/2}}{\sqrt{q_{x}^{2}+q_{y}^{2}+\delta^{2}/%
\ell_{0}^{2}}} \{e^{-(k_{xi}^{(0)}-k_{xj_{0}}^{(\tilde{n})})^{2}%
\ell_{0}^{2}/2}  \notag \\
&&\times S_{m}^{2}\left( (q_{x}+k_{xj_{0}}^{(\tilde{n})}-k_{xi}^{(0)})%
\ell_{0} \right)  \notag \\
&&-e^{-(k_{xi}^{(0)}-k_{xi_{0}}^{(0)})^{2}\ell_{0}^{2}/2} S_{m}^{2}\left(
(q_{x}+k_{xi_{0}}^{(0)}-k_{xi}^{(0)})\ell_{0} \right)\}  \notag \\
&&\times \sum_{k=-\infty}^{\infty} e^{-ik q_{x} L_{x}^{\square}} ,
\label{A.4}
\end{eqnarray}%
where it is used that $(k_{xi}^{(n)}-k_{xj_{0}}^{(n+\tilde{n})})=
(k_{xi}^{(0)}-k_{xj_{0}}^{(\tilde{n})})$ and $%
(k_{xi}^{(n)}-k_{xi_{0}}^{(n)})= (k_{xi}^{(0)}-k_{xi_{0}}^{(0)})$, to carry
out the sum over $n$. Further, the sum over $k$ in Eq. (\ref{A.4}) we will
rewrite, cf. with Eq. (\ref{28}), in the form given by the first line of Eq.
(\ref{77}). Then by using the delta-functions we calculate the integral over 
$q_{x}$ and obtain Eq. (\ref{A.4}) in the following form
\begin{eqnarray}
&&\Delta E_{ee}^{xc}(m;i_{0},j_{0})=- \frac{e^{2}}{\varepsilon
L_{x}^{\square}} \sum_{i=1;i\neq i_{0}}^{\tilde{N}}
\sum_{M_{x}=-\infty}^{\infty} \int_{-\infty}^{\infty} dq_{y}  \notag \\
&&\frac{e^{-q_{y}^{2}\ell_{0}^{2}/2}}{\sqrt{(\frac{2\pi}{L_{x}^{\square}}%
M_{x})^{2}+q_{y}^{2}+\frac{\delta^{2}}{ \ell_{0}^{2}}}}
\{e^{-(k_{xi}^{(0)}-k_{xj_{0}}^{(\tilde{n})})^{2}\ell_{0}^{2}/2}  \notag \\
&&\times S_{m}^{2}\left((k_{xj_{0}}^{(\tilde{n})}-k_{xi}^{(0)}- \frac{2\pi}{%
L_{x}^{\square}}M_{x})\ell_{0} \right)
-e^{-(k_{xi}^{(0)}-k_{xi_{0}}^{(0)})^{2}\ell_{0}^{2}/2}  \notag \\
&&\times S_{m}^{2}\left( (k_{xi_{0}}^{(0)}- k_{xi}^{(0)}-\frac{2\pi}{%
L_{x}^{\square}}M_{x})\ell_{0} \right)\} .  \label{A.5}
\end{eqnarray}%
Point out that Eqs. (\ref{A.4}), (\ref{A.5}) will give the results obtained
from Eq. (\ref{154s}), i.e., for the compound spin-exciton, if formally
neglect by the first term in the curly brackets. Now if take into account
that only for $M_{x}= m(n_{ys}^{(j_{0})}-n_{ys}^{(i)})+\tilde{n}$ the factor
\begin{eqnarray}
&&S_{m}\left((k_{xj_{0}}^{(\tilde{n})}-k_{xi}^{(0)}- \frac{2\pi}{%
L_{x}^{\square}}M_{x})\ell_{0} \right)  \notag \\
&&=\frac{\sin\left(\pi \left(m(n_{ys}^{(j_{0})}-n_{ys}^{(i)})+\tilde{n}-
M_{x} \right) \right)}{\left(\pi \left(m(n_{ys}^{(j_{0})}- n_{ys}^{(i)})+%
\tilde{n}- M_{x} \right) \right)} ,  \label{A.6}
\end{eqnarray}%
in the first term of the curly brackets in the RHS of Eq. (\ref{A.5}), is
equal to $1$ and for any other $M_{x}$ its value is zero, in addition,
treating in a similar manner the second term of these curly brackets we
readily rewrite Eq. (A.5) in the form given by Eqs. (\ref{138}), (\ref{139})
for $m \geq 3$. Respectively, for the compound spin-exciton we obtain from
Eq. (\ref{154s}) the form given by Eq. (\ref{156s}), cf. with Eq. (\ref{138}%
).

\end{document}